\documentclass[twocolumn]{aa}

\usepackage{graphicx}
\usepackage{txfonts}
\usepackage{subcaption}
\usepackage{lscape}
\usepackage{placeins}
\usepackage[hidelinks]{hyperref}
\newcommand{\vsini}{$v\sin i$}

\newcommand{\kms}{\ensuremath{\mathrm{km\,s^{-1}}}}
\newcommand{\Cr}{Collinder\,261}
\newcommand{\msun}{$M_\odot$}
\begin{document}
   \title{Signatures of mass transfer in open-cluster blue straggler stars}
   \subtitle{I. Collinder 261: Evidence from rotation, binarity, and chemical abundances}
   \author{M.\,J.\ Rain\inst{1}\corrauth{mariajoserain@gmail.com}
        \and L. Monaco\inst{2}\email{}
        \and S. Villanova\inst{2}\email{}
        \and G. Carraro\inst{3}\email{}
        }
   \institute{European Southern Observatory, Alonso de C\'ordova 3107, Vitacura, Región Metropolitana, Chile
   \and Universidad Andres Bello, Facultad de Ciencias Exactas, Departamento de F{\'i}sica y Astronom{\'i}a - Instituto de Astrof{\'i}sica, Autopista Concepci\'on-Talcahuano 7100, Talcahuano, Chile
   \and Dipartimento di Fisica e Astronomia, Universit\`{a} di Padova, Vicolo Osservatorio 3, I-35122, Padova, Italy
}
   \date{Received --; accepted --}
  \abstract
   {We present the first dedicated $v\sin i$ survey of blue straggler stars (BSS) in the old open cluster \Cr\ ($t=6.7$~Gyr), based on VLT/FLAMES--GIRAFFE spectroscopy of 40 BSS and 42 turn-off/subgiant (TO/SGB) reference stars. Seventeen of the 39 BSS entering the statistics (44\%) rotate faster than 40~\kms, whereas all TO/SGB reference stars are slow rotators; fast rotation is strongly linked to spectroscopic binarity. TESS light curves identify seven eclipsing or ellipsoidal binaries with $P_{\rm orb}=0.5$--2.1~d, all rotating at the velocity set by tidal synchronisation, so in these systems the fast rotation is maintained by a close post-mass-transfer orbit and does not date the formation event. Differential C, O, Ba, and Fe abundances for the 22 slow rotators reveal that 62\% carry the surface-pollution signature of mass transfer, in two distinct patterns separated by an empty $\sim$1~dex gap in [Ba/Fe]: a Ba-rich group tracing AGB donors and a C/O-enhanced, Ba-normal group tracing donors below the $s$-process activation mass. The BSS iron abundance is consistent with the cluster value; the few apparent Fe excesses are partly explained by differential reddening and do not affect the abundance pattern. The correlated C--O enrichment is opposite to the depletion seen in globular-cluster BSS, indicating binaries wide enough for donors to reach the AGB. With $A^+_{r_h}=0.171\pm0.034$ and $f_{\rm FR-BSS}=44\%\pm8\%$, \Cr\ extends the globular-cluster rotation--density trends into the open-cluster regime: its BSS were assembled by a continuously operating binary channel, and their chemistry records the dynamical environment of the host cluster.}
   \authorrunning{M.J. Rain}
   \titlerunning{Blue straggler $v\sin i$ survey in Collinder~261}
   \maketitle
   \nolinenumbers
\section{Introduction}
\label{sec:intro}
Blue Straggler Stars (BSS) occupy a region of the colour--magnitude diagram (CMD) that should be empty according to standard single-star evolution: they appear bluer and brighter than the MSTO, mimicking a younger stellar population in an old cluster. First identified in M3 by \citet{Sandage1953}, BSS are recognised as a universal feature of old stellar systems \citep{Bailyn1995,Ferraro2023}.
Three formation channels have been proposed. Mass transfer (MT): the primary of a binary accretes mass from an evolved companion, gaining enough material to appear rejuvenated \citep{McCrea1964}. This picture is supported on a population level by \citet{Rain2024}, who found a significant correlation between the number of blue stragglers and the binary fraction of the host cluster in a sample of twelve old open clusters and showed that composite evolutionary sequences from binary evolution models reproduce the observed distributions in
the colour--magnitude diagram. Stellar mergers or stellar triples: close binary evolution or hierarchical-triple dynamics force two stars to coalesce \citep{Leonard1989,Sills2005}. Dynamical collisions: direct stellar encounters in dense environments create BSS with intermediate-to-high rotation \citep{Hills1976,Lombardi2002}.
In GCs, the relative contribution of each channel depends on the dynamical environment: denser cores favour collisions and dynamically-induced mergers, whereas MT operates across all densities \citep{Davies2004,Knigge2009}. In old open clusters, by contrast, the low stellar densities effectively suppress direct collisions; here, BSS are expected to form almost exclusively through MT in binary systems and through mergers driven by close-binary or hierarchical-triple evolution, and recent studies find that mass transfer is the main channel \citep{Leiner2021,Linck2026a}, making these systems particularly clean test beds for binary interaction physics.
Disentangling these formation channels is important for several reasons that extend well beyond BSS themselves. First, BSS are the most accessible laboratory for studying mass transfer in intermediate-age and old stellar populations: the efficiency, stability, and angular-momentum budget of MT govern the production of a wide variety of exotic objects (Type\,Ia supernovae progenitors, cataclysmic variables, symbiotic stars, and subdwarf-B stars) yet these parameters remain poorly constrained observationally \citep{Taam2000,Postnov2014}.
By quantifying the MT-to-merger ratio in different environments and comparing the observed fractions with predictions of binary population-synthesis codes, BSS surveys provide direct empirical constraints on the mass-transfer efficiency, the critical mass-ratio for stable Roche-lobe overflow, and the common-envelope ejection parameter $\alpha_{\rm CE}$ \citep{Iben1993}.
Second, this isolation of MT and binary-driven mergers offers a controlled setting to test MT physics without the complicating effects of the frequent dynamical encounters that dominate GC cores. Third, the MT channel leaves imprints on the companion and on the BSS surface chemistry (C/O depletion, $s$-process enhancement), providing multi-messenger constraints on AGB nucleosynthesis and common envelope evolution that are difficult to obtain by other means \citep{Gosnell2014,Nine2024}. Identifying which BSS formed through MT therefore informs our broader understanding of binary stellar evolution and its end-products across stellar populations.
Stellar rotation helps to separate these channels, together with
radial-velocity (RV) monitoring and chemical abundances. Both
mass-transfer and merger products are expected to spin up when they
form \citep{Sills2005,Glebbeek2008}. Mass-transfer products then lose
angular momentum through magnetic braking or, in close orbits, are
locked by tides to the orbital period, so old mass-transfer products
are usually slow rotators with a companion that can be detected in RV
\citep{Carney2005,Rain2021}. Merger and collision products have no
companion and can stay fast for longer
\citep{Leiner2018,Leiner2019}. Young mass-transfer products can also
rotate fast, however: in M67 and NGC~188 several BSS with hot
white-dwarf companions have projected rotational velocities
($v\sin i$) above 40~\kms\ \citep{Linck2026b}, and stars without a
convective envelope are braked very little. The rotation of a single
star therefore does not identify its formation channel. Combined with
binarity and surface chemistry it does constrain it, and for a
population it traces the time since formation. An advantage of
$v\sin i$ is that it can be measured from a single spectrum.
This diagnostic power has so far been exploited almost exclusively in
globular clusters. There, dedicated $v\sin i$ surveys have revealed
that the fraction of fast-rotating BSS is not random: it correlates
with the density and the dynamical state of the host cluster
\citep{Ferraro2023,Billi2025}, turning BSS rotation into a probe of
the environment that made them. Whether these trends are a property
of dense environments or a universal feature of BSS populations is an
open question, and answering it requires the other end of the density
sequence: the open clusters. Rotation measurements of open-cluster BSS have only begun to appear, and few open clusters have had their BSS populations surveyed in rotation, binarity and surface chemistry together, as done by the WIYN Open Cluster Study for NGC~188 and M67 \citep{Linck2026a, Linck2026b}. This combination is needed to tie the spin of each star to its formation channel, and it is the purpose of this paper.

Collinder\,261 is one of the oldest open clusters of the Galactic disc. Published ages range from about 5 to 11~Gyr, depending on the models, reddening and distance adopted \citep{Mazur1995,Gozzoli1996,Bragaglia2006}, and its metallicity is close to solar \citep{Friel2003,Carretta2005}. \citet{Mazur1995} found 45 short-period eclipsing binaries in the field, 28 of them of W~UMa type. \citet{Rain2020} identified 53 BSS members with \textit{Gaia}\,DR2, one of the largest BSS populations of any open cluster \citep{Rain2021}, and measured RVs at four epochs for nine of them: five are probable close binaries, one is a long-period binary and three show no variation. The short-period variables among its BSS were recently studied with TESS by \citet{Echeveste2026}.

This is the first in a series of papers on the signatures of mass transfer in the blue straggler populations of old open clusters. \Cr\ is a natural starting point, because its age and number of BSS bring it close to the globular-cluster regime. Here we present the projected rotational velocities, the binarity classification, and the differential C, O, Ba, and Fe abundances of its BSS; forthcoming papers will add the spectroscopic orbits and extend the same analysis to the remaining clusters of the programme. \\
\begin{table*}
\centering
\small
\setlength{\tabcolsep}{3.5pt}
\caption{Stellar parameters and projected rotational velocities of the BSS in \Cr.}
\label{tab:bss}
\begin{tabular}{llcccccrrrl}
\hline\hline
ID & Gaia DR3 ID & $T_\mathrm{eff}$ & $\log g$ & $M$ & $R$ & $v\sin i$ & $\Delta$RV & $G$ & SNR & Class \\
 & & (K) & & (\msun) & ($R_\odot$) & (\kms) & (\kms) & (mag) & & \\
\hline
BSS 1 & 5856420729693697408 & $6347\pm66$ & $4.07\pm0.03$ & $1.13\pm0.02$ & $1.62\pm0.06$ & $11\pm6$ & $-3$ & 16.11 & 45 & D \\
BSS 2 & 5856432656849042688 & $6213\pm69$ & $3.98\pm0.03$ & $1.13\pm0.01$ & $1.80\pm0.06$ & $<15$ & $+1$ & 16.02 & 50 & UL \\
BSS 3 & 5856432828647768960 & $7732\pm79$ & $4.10\pm0.02$ & $1.57\pm0.01$ & $1.85\pm0.04$ & $\gtrsim 100$ & $-63$ & 15.00 & 68 & LL, SB1 \\
BSS 4 & 5856437089255328384 & $7152\pm43$ & $4.04\pm0.02$ & $1.46\pm0.01$ & $1.91\pm0.04$ & $26\pm10$ & $+4$ & 15.22 & 72 & D \\
BSS 5 & 5856511271928032256 & $7499\pm129$ & $3.96\pm0.03$ & $1.62\pm0.01$ & $2.21\pm0.08$ & $\gtrsim 100$ & $+16$ & 14.73 & 92 & LL \\
BSS 6 & 5856512680677308160 & $6301\pm57$ & $4.04\pm0.02$ & $1.13\pm0.01$ & $1.68\pm0.04$ & $<26$ & $-3$ & 16.10 & 52 & UL \\
BSS 7 & 5856514879700797696 & $9217\pm267$ & $4.16\pm0.04$ & $1.93\pm0.04$ & $1.91\pm0.09$ & \dots & $+65$ & 14.33 & 90 & NM, SB1 \\
BSS 8 & 5856515601255190272 & $7261\pm103$ & $3.76\pm0.03$ & $1.76\pm0.08$ & $2.90\pm0.12$ & $80\pm22$ & $-37$ & 14.24 & 120 & D, SB1 \\
BSS 9 & 5856515665648722688 & $6783\pm62$ & $4.08\pm0.02$ & $1.34\pm0.02$ & $1.75\pm0.04$ & $63\pm20$ & $-49$ & 15.62 & 64 & D, SB1 \\
BSS 10 & 5856515669974676480 & $6928\pm105$ & $4.14\pm0.02$ & $1.34\pm0.02$ & $1.63\pm0.04$ & $\gtrsim 100$ & $+15$ & 15.70 & 64 & LL \\
BSS 11 & 5856516597687652352 & $6537\pm79$ & $4.06\pm0.04$ & $1.22\pm0.05$ & $1.71\pm0.09$ & $10\pm9$ & $-3$ & 15.89 & 59 & D \\
BSS 12 & 5856517010004497792 & $6783\pm62$ & $4.08\pm0.02$ & $1.34\pm0.02$ & $1.75\pm0.04$ & $\gtrsim 100$ & $-1$ & 15.64 & 63 & LL \\
BSS 13 & 5856517078723954944 & $6028\pm79$ & $3.73\pm0.04$ & $1.32\pm0.03$ & $2.60\pm0.12$ & $<17$ & $+2$ & 15.29 & 78 & UL \\
BSS 14 & 5856517211839051392 & $6406\pm52$ & $3.95\pm0.03$ & $1.33\pm0.05$ & $2.02\pm0.08$ & $17\pm6$ & $+7$ & 15.60 & 72 & D \\
BSS 15 & 5856517284882496128 & $6838\pm69$ & $3.76\pm0.04$ & $1.63\pm0.07$ & $2.79\pm0.14$ & $7\pm6$ & $+1$ & 14.62 & 96 & D \\
BSS 16 & 5856519037229366144 & $6730\pm75$ & $3.68\pm0.02$ & $1.68\pm0.03$ & $3.10\pm0.08$ & $10\pm9$ & $+5$ & 14.44 & 55 & D \\
BSS 17 & 5856527386646253312 & $7273\pm97$ & $3.83\pm0.02$ & $1.68\pm0.02$ & $2.61\pm0.06$ & $\gtrsim 100$ & $+99$ & 14.54 & 107 & LL, SB1 \\
BSS 18 & 5856527455365735680 & $6885\pm96$ & $3.98\pm0.02$ & $1.44\pm0.01$ & $2.03\pm0.05$ & $75\pm25$ & $+59$ & 15.25 & 77 & D, SB1 \\
BSS 19 & 5856527455365740032 & $6299\pm90$ & $3.78\pm0.04$ & $1.43\pm0.05$ & $2.55\pm0.13$ & $\gtrsim 100$ & $+31$ & 15.16 & 86 & LL, SB1 \\
BSS 20 & 5856527455365743488 & $7649\pm98$ & $4.03\pm0.02$ & $1.60\pm0.01$ & $2.02\pm0.05$ & \dots$^{\dagger}$ & $+52$ & 14.81 & 88 & FR, SB1 \\
BSS 21 & 5856527524085202048 & $7032\pm59$ & $4.10\pm0.02$ & $1.39\pm0.01$ & $1.74\pm0.04$ & $\gtrsim 100$ & $+95$ & 15.57 & 63 & LL, SB1$^{\ddagger}$ \\
BSS 22 & 5856527524085206656 & $7002\pm71$ & $3.55\pm0.01$ & $1.80\pm0.07$ & $3.73\pm0.08$ & $13\pm5$ & $+1$ & 13.91 & 123 & D \\
BSS 23 & 5856527558444950656 & $6371\pm59$ & $3.96\pm0.04$ & $1.20\pm0.04$ & $1.90\pm0.09$ & $25\pm9$ & $+3$ & 15.74 & 63 & D \\
BSS 24 & 5856527558444963200 & $6528\pm81$ & $3.98\pm0.04$ & $1.24\pm0.04$ & $1.89\pm0.09$ & $\gtrsim 100$ & $-34$ & 15.67 & 60 & LL, SB1 \\
BSS 25 & 5856527764603421184 & $6226\pm76$ & $3.95\pm0.03$ & $1.16\pm0.01$ & $1.89\pm0.07$ & $13\pm8$ & $-3$ & 15.86 & 54 & D \\
BSS 26 & 5856528348719355648 & $7406\pm76$ & $3.90\pm0.02$ & $1.65\pm0.02$ & $2.39\pm0.06$ & $10\pm8$ & $0$ & 14.62 & 95 & D \\
BSS 27 & 5856528348719366528 & $6834\pm80$ & $4.03\pm0.03$ & $1.39\pm0.01$ & $1.89\pm0.07$ & $13\pm9$ & $+2$ & 15.46 & 68 & D \\
BSS 28 & 5856528383078700288 & $6608\pm92$ & $4.05\pm0.04$ & $1.31\pm0.05$ & $1.79\pm0.09$ & $82\pm44$ & $+165$ & 15.71 & 58 & D, SB1 \\
BSS 29 & 5856528486157941376 & $6378\pm83$ & $4.07\pm0.04$ & $1.14\pm0.02$ & $1.63\pm0.08$ & $36\pm11$ & $+4$ & 16.10 & 51 & D \\
BSS 30 & 5856528520517692032 & $6275\pm103$ & $3.88\pm0.04$ & $1.23\pm0.05$ & $2.11\pm0.11$ & $11\pm8$ & $-2$ & 15.63 & 67 & D \\
BSS 31 & 5856528589237173120 & $6030\pm84$ & $3.69\pm0.02$ & $1.35\pm0.01$ & $2.75\pm0.06$ & $17\pm7$ & $+2$ & 15.22 & 78 & D \\
BSS 32 & 5856528726676132096 & $6501\pm73$ & $4.04\pm0.03$ & $1.29\pm0.05$ & $1.80\pm0.07$ & $<24$ & $+1$ & 15.84 & 58 & UL, SB1 \\
BSS 33 & 5856528761035882496 & $6828\pm77$ & $4.01\pm0.04$ & $1.40\pm0.06$ & $1.94\pm0.10$ & $\gtrsim 100$ & $+175$ & 15.36 & 70 & LL, SB1$^{\ddagger}$ \\
BSS 34 & 5856528962867671936 & $6498\pm83$ & $4.11\pm0.03$ & $1.19\pm0.04$ & $1.59\pm0.06$ & $58\pm20$ & $-1$ & 16.06 & 48 & D, SB1 \\
BSS 35 & 5856529001554008064 & $5984\pm81$ & $3.50\pm0.04$ & $1.54\pm0.04$ & $3.65\pm0.17$ & $24\pm5$ & $+1$ & 14.63 & 99 & D \\
BSS 36 & 5856529242072194816 & $6362\pm42$ & $3.93\pm0.02$ & $1.23\pm0.05$ & $1.99\pm0.06$ & $<45$ & $+7$ & 15.65 & 56 & UL \\
BSS 37 & 5856529276431950080 & $6607\pm77$ & $3.79\pm0.02$ & $1.40\pm0.01$ & $2.49\pm0.06$ & $<16$ & $-1$ & 15.03 & 83 & UL \\
BSS 38 & 5856529276431965696 & $6644\pm80$ & $4.26\pm0.03$ & $1.19\pm0.02$ & $1.34\pm0.05$ & $84\pm57$ & $+16$ & 16.36 & 41 & D, SB1 \\
BSS 39 & 5856529379511194240 & $7499\pm102$ & $3.96\pm0.02$ & $1.62\pm0.02$ & $2.21\pm0.05$ & $\gtrsim 100$ & $-109$ & 14.78 & 87 & LL \\
BSS 40 & 5856533326555127168 & $6378\pm83$ & $4.07\pm0.04$ & $1.14\pm0.02$ & $1.63\pm0.08$ & $10\pm9$ & $+2$ & 16.11 & 52 & D \\
\hline
\end{tabular}
\tablefoot{$T_{\rm eff}$, $\log g$, and mass come from the orthogonal projection of the dereddened CMD position onto PARSEC evolutionary tracks ($[\mathrm{Fe/H}]=-0.22$, $(m-M)_0=12.15$, $A_V=0.93$). The metallicity is from \citet{Friel2003}, and the age, distance, and extinction are those of the PARSEC isochrone that best fits the \textit{Gaia}\,DR3 members \citep{Echeveste2026}. Uncertainties are from the re-projection of the photometric errors and exclude the systematic contribution of unresolved binarity. The radius follows from $\log g$ and $M$, $R=\sqrt{GM/g}$, and its uncertainty includes only the errors on these two quantities. Stars with nearly the same CMD position fall on the same point of the model grid and receive identical parameters (BSS~9 and 12, 29 and 40). The $v\sin i$ uncertainties are $\Delta\chi^2=1$ intervals. D = detection; UL = upper limit ($v\sin i+\sigma$); LL = lower limit (profile too shallow and broad to bound from above at $R=18\,000$); FR = fast rotator without an attributed value (see $^{\dagger}$); NM = not measurable; SB1 = spectroscopic binary (HR18 single-epoch $|\Delta\mathrm{RV}|>20$~\kms, or HR12 $\sigma_{\rm RV}\gtrsim8$~\kms\ and ${\rm ptp}_{\rm RV}>20$~\kms; Sect.~\ref{sec:sb1_rv}). The HR18 offset of BSS~39 is discarded as a cross-correlation artefact (its HR12 velocities are constant). The statistical sample comprises $N=39$ stars (BSS~7 is NM). For the 22 slow rotators, $v\sin i$ comes from a synthesis with the per-star iron of Sect.~\ref{sec:feh} fixed; the fast rotators are fit at the cluster metallicity (Sect.~\ref{sec:vsini_method}). $^{\dagger}$~No Fe\,\textsc{i} line is detected in BSS~20 at ${\rm S/N}\simeq88$ (Sect.~\ref{sec:vsini_results}). $^{\ddagger}$~Radial-velocity membership uncertain (Sect.~\ref{sec:sb1_rv}).}
\end{table*}

\begin{figure}[t]
\centering
\includegraphics[width=0.5\textwidth]{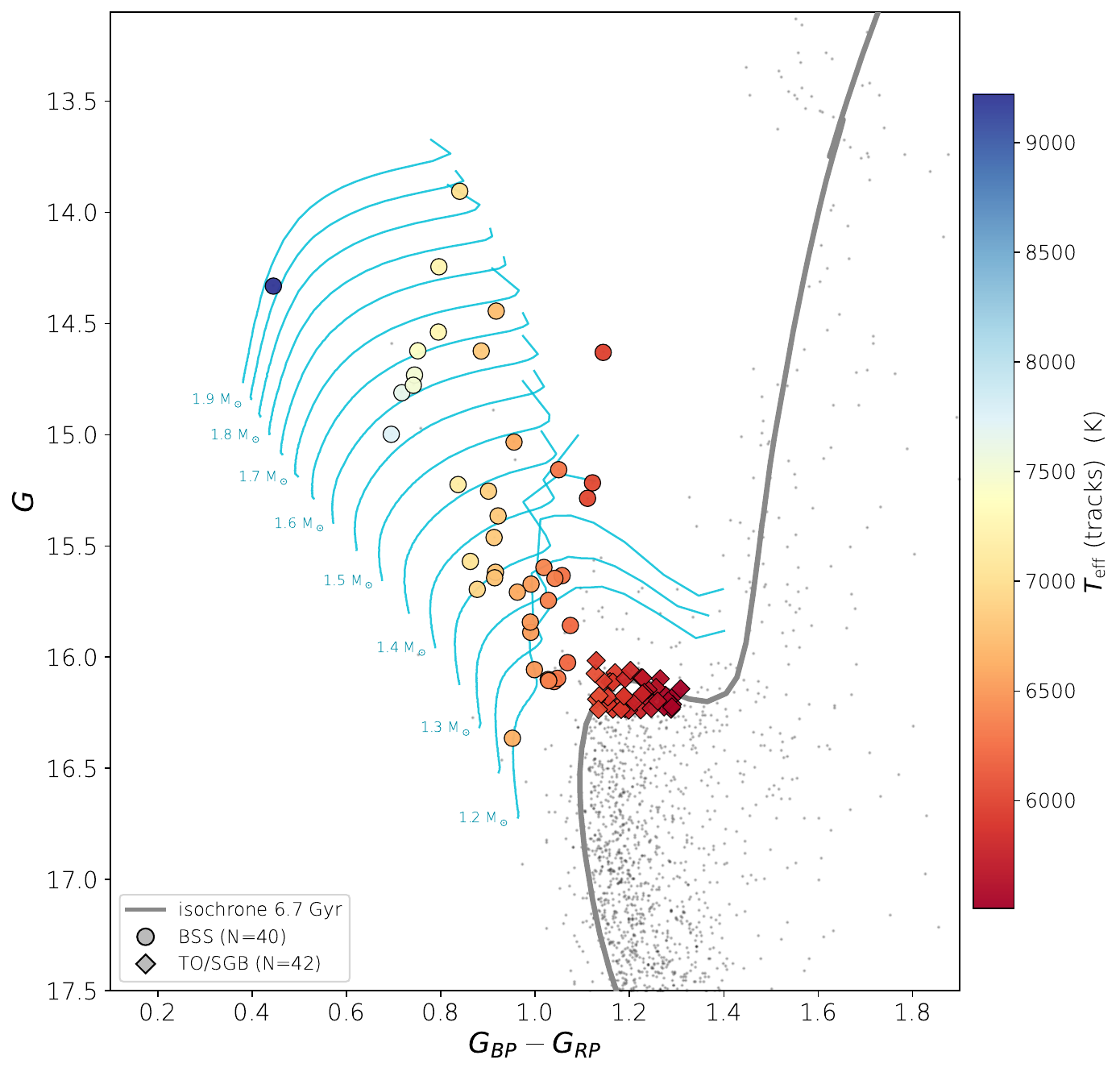}
\caption{\textit{Gaia}\,DR3 colour--magnitude diagram of \Cr\ (grey dots: astrometric members). The 40 BSS (circles) and the 42 TO/SGB reference stars (diamonds) are colour-coded by the effective temperature from the projection onto PARSEC tracks (cyan lines; 1.2--1.95\,\msun, labelled every 0.1\,\msun), computed for $[\mathrm{Fe/H}]=-0.22$, $(m-M)_0=12.15$ and $A_V=0.93$. The grey curve is the 6.7\,Gyr isochrone.}
\label{fig:cmd_tracks}
\end{figure}
This paper is structured as follows. Section~\ref{sec:obs} describes the observations and data reduction. Section~\ref{sec:stellar_params} presents the derivation of the stellar parameters, the radial velocities and binary classification, the projected rotational velocities, and the chemical abundance analysis. The results are discussed in Section~\ref{sec:discussion}, and Section~\ref{sec:conclusions} summarises our conclusions.

\begin{figure*}[ht!]
\centering
\includegraphics[width=\textwidth]{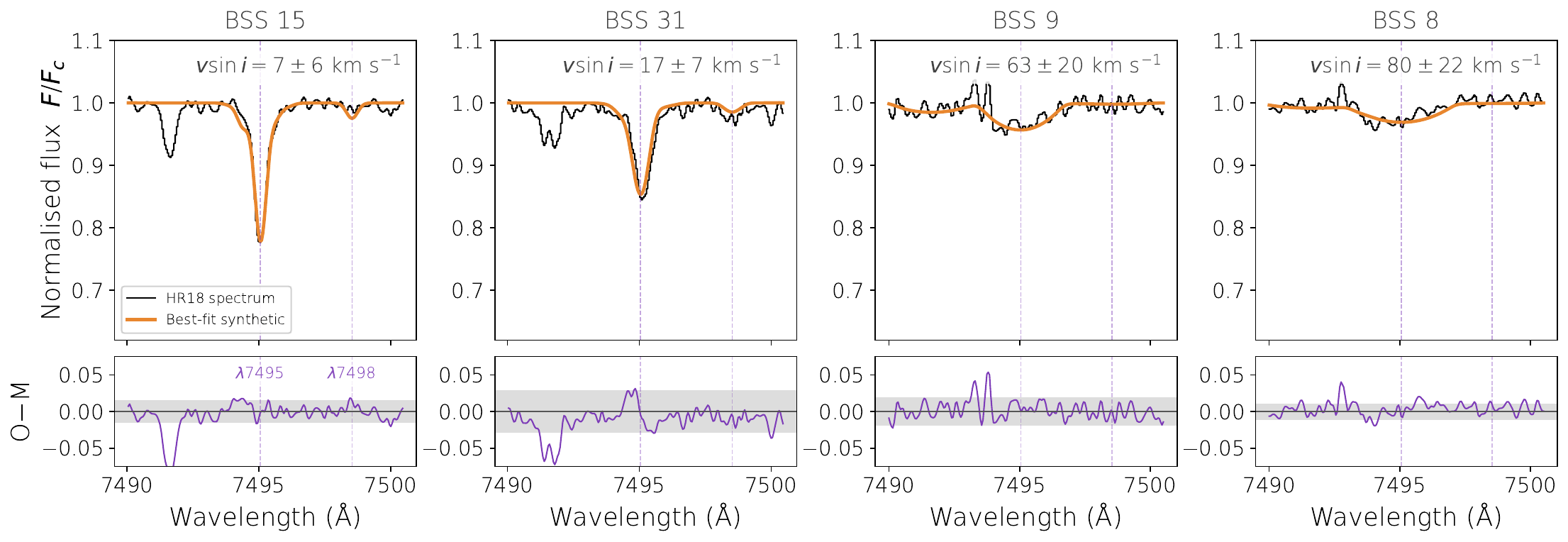}
\caption{Examples of $v\sin i$ fits for four BSS, from the slowest resolved profile (BSS~15) to $v\sin i\approx80$~\kms\ (BSS~8). Observed HR18 spectra (black) and best-fit synthetic profiles (orange); dashed lines mark Fe\,\textsc{i} $\lambda$7495.05 and $\lambda$7498.53. Bottom panels: residuals with the $\pm1\sigma$ noise band. The spectra are smoothed (3 pixels) for display only. All fits are shown in Appendix~\ref{app:vsini_fits}.}
\label{fig:vsini_demo}
\end{figure*}

\begin{figure}[t]
\centering
\includegraphics[width=\columnwidth]{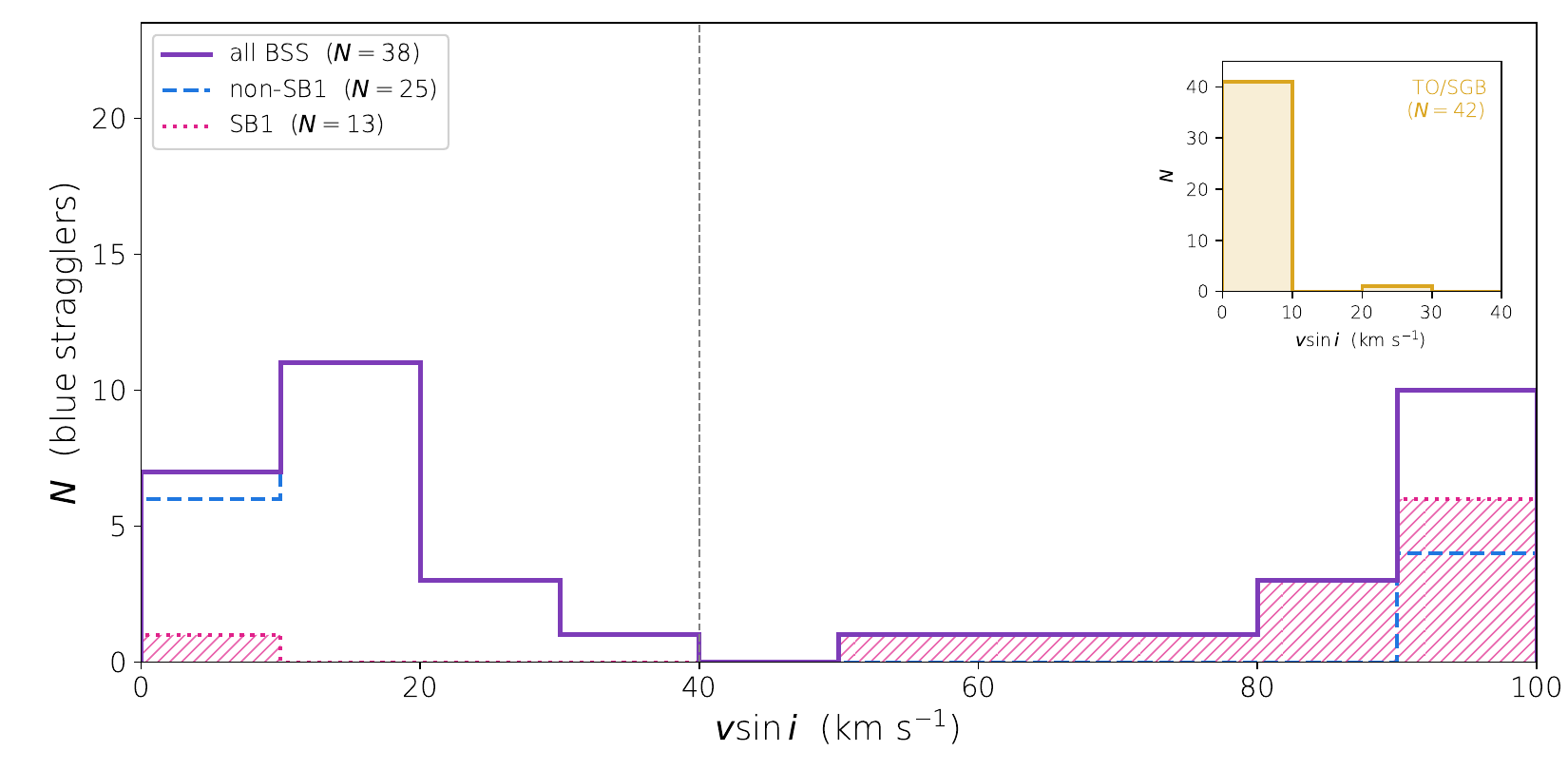}
\caption{Distribution of $v\sin i$ for the 38 BSS with a measurement (purple), split into 13 SB1 (pink, hatched) and 25 non-SB1 stars (blue dashed). Stars with $v\sin i\gtrsim100$~\kms\ are in the last bin. The dashed line marks the 40~\kms\ boundary between slow and fast rotators. Inset: the 42 TO/SGB reference stars.}
\label{fig:vsini_distribution}
\end{figure}

\section{Observations and data reduction}
\label{sec:obs}

The data were collected with the FLAMES--GIRAFFE multi-object spectrograph \citep{Pasquini2002} mounted on the Kueyen unit (UT2) of the ESO Very Large Telescope at Cerro Paranal, under two complementary observing programmes. Programme 0109.D-0492(A) (PI: Rain) provided six HR12 multi-epoch visits between 2022 April~8 and 2022 June~10, with within-programme baselines up to 63\,days. Programme 0114.D-0090(A) (PI: Monaco) provided single-epoch HR18 and HR22 spectroscopy on 2025 March~29. The total temporal coverage of $\sim$\,2.8\,yr between the two programmes enables a long-baseline radial-velocity cross-check that complements the dense multi-epoch monitoring of the 2022 HR12 run. Our sample contains 40 of the 53 BSS candidates of \Cr\ in the catalogue of \citet{Rain2021} (75\%); the candidates we did not observe cover the same range of magnitude and colour as the observed ones. The BSS candidates come from the catalogue of \citet{Rain2021}, which uses the \textit{Gaia}\,DR2 members of \citet{CantatGaudin2018} and \citet{CantatGaudin2020}. These were selected with the UPMASK algorithm from proper motions and parallaxes, down to $G=18$, and only stars with a membership probability of at least 50\% were kept. The BSS were then selected in the CMD between the ZAMS and the equal-mass binary sequence. We test the RV membership of each star in Sect.~\ref{sec:sb1_rv}.
We employed the three GIRAFFE high-resolution setups in a complementary configuration, each delivering a specific kinematic or chemical diagnostic for the same set of BSS. The 2025 observations used a single plate configuration for HR18 and HR22, so the 42 comparison stars, observed with dedicated fibres, have both setups; only the 2022 HR12 monitoring is missing for them. They provide the reference sample against which the BSS rotational velocities and the differential C, O, and Fe abundances are measured (Sect.~\ref{sec:stellar_params}); the lack of HR12 coverage is why no differential barium measurement is possible (Sect.~\ref{sec:chem_method}).
\begin{itemize}
\item HR18 (7462--7883\,\AA, $R\approx18{,}000$) single-epoch grating from the 2025 run. It provides (i) the isolated Fe\,\textsc{i}\,$\lambda$7495.05\,\AA\ line (and Fe\,\textsc{i}\,$\lambda$7507\,\AA\, Fe\,\textsc{i}\,$\lambda$7511\,\AA\, Fe\,\textsc{i}\,$\lambda$7531\,\AA\ for confirmation), used as the rotational-velocity diagnostic for the \vsini\ analysis (Section~\ref{sec:vsini_method}); (ii) the high-S/N continuum required for accurate $\chi^2$ minimisation against synthetic templates; (iii) the O\,\textsc{i}\,$\lambda\lambda$7771--7775\,\AA\ triplet, observed in the ``no-blending'' window and used for the oxygen abundance analysis (Section~\ref{sec:chem_method}); and (iv) the 2025 radial velocity that, combined with the 2022 HR12 RVs, sets a $\sim$1085-day cross-programme baseline for the binary classification.
\item HR12 (5821--6146\,\AA, $R\approx18{,}700$)  The only multi-epoch setup, with six visits during 2022 April to June. HR12 provides the per-star $\sigma_{\rm RV}$ and peak-to-peak ${\rm ptp}_{\rm RV}$ statistics that, combined with the single-epoch HR18 $\Delta\mathrm{RV}$, define our SB1 binarity criterion (Section~\ref{sec:sb1_rv}). The same six epochs are co-added to produce the stacked HR12 spectrum used for the barium abundance measurement at Ba\,\textsc{ii}\,$\lambda$6141.71\,\AA\ (Section~\ref{sec:chem_method}).
\item HR22 (8962--9354\,\AA, $R\approx11{,}500$) single-epoch grating from the 2025 run. It provides the
C\,\textsc{i}\,$\lambda$9094.83\,\AA\ line for the carbon abundance determination, the third element entering the chemical group classification (Section~\ref{sec:chem_groups}).
\end{itemize}
Together, the three setups yield a self-consistent kinematic and chemical characterisation of each BSS: rotation and single-epoch RV from HR18 (2025); multi-epoch RV and Ba from HR12; a 2.8-year cross-programme RV baseline from combining HR12 (2022) with HR18 (2025); and the C abundance from HR22 (2025).
Together they help to separate the formation channels: $v\sin i$ traces the rotational history, the two RV baselines detect the shorter-period spectroscopic binaries, and the C/O/Ba combination identifies the donor type for the MT subset.
Integration times of $\sim$\,2775\,s per field yielded a signal-to-noise ratio per pixel of 42--123 for the BSS sample (median $\approx$\,67) and 30--80 for the comparison stars in the HR18 setup; comparable SNR was achieved in HR12 and HR22 for the BSS sample, sufficient for synthesis-based abundance fitting of the targeted lines at the $\Delta\chi^2=1$ confidence level. Dedicated sky fibres allocated to empty positions were used for background subtraction in all three setups.
Raw frames were reduced with the ESO GIRAFFE pipeline (v2.16), performing bias subtraction, flat-fielding, wavelength calibration from ThAr arc lamps, and spectral extraction. Sky subtraction used the median of dedicated sky fibres. Spectra were normalised with \textsc{iSpec} \citep{Blanco-Cuaresma2014}. Telluric features redward of $\sim$7600\,\AA\ were masked. Radial velocities were measured by cross-correlation against synthetic templates. The binary classification is described in Sect.~\ref{sec:sb1_rv}; spectroscopic binaries are kept in the rotation analysis.

\begin{figure*}
\centering
\includegraphics[width=2\columnwidth]{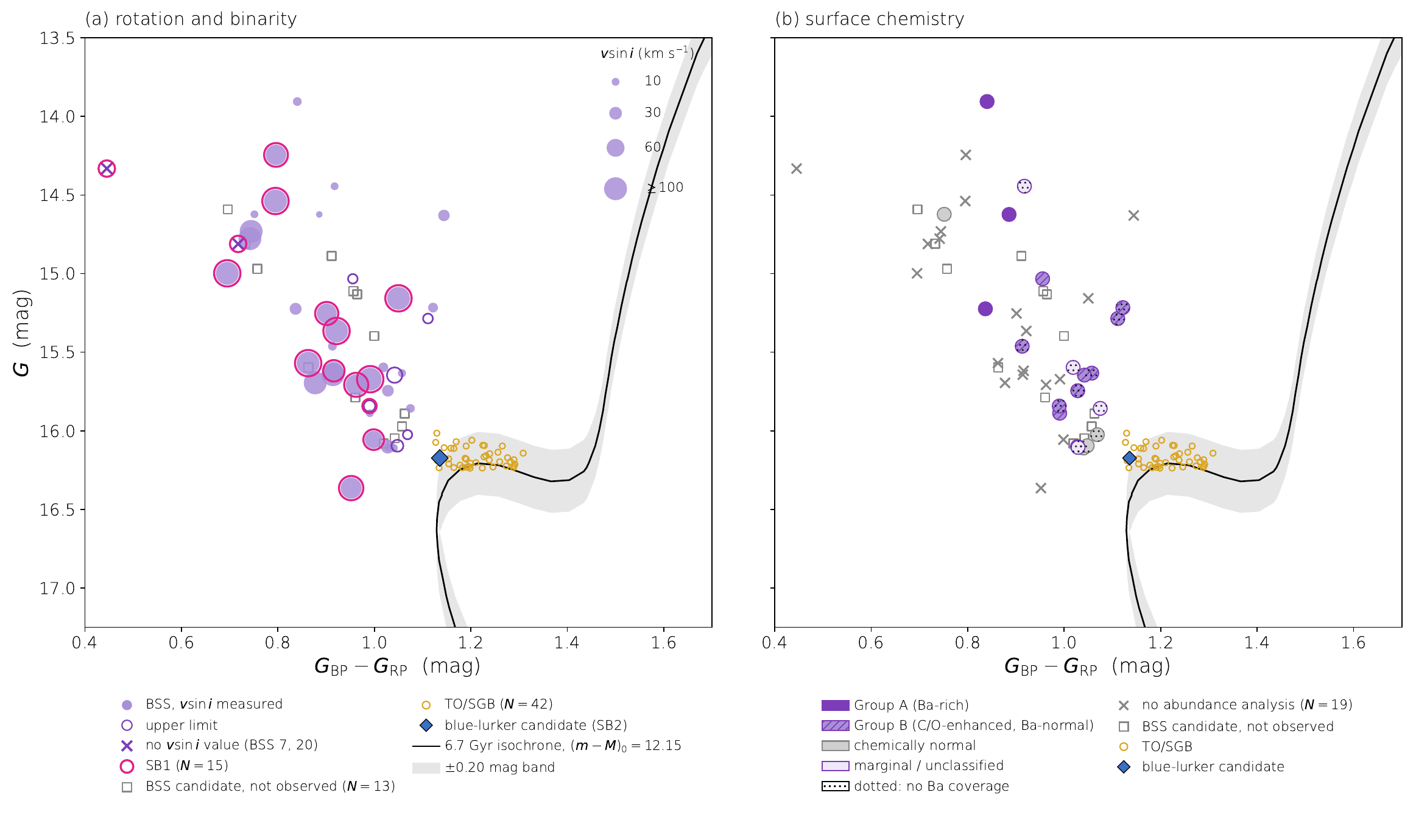}
\caption{\textit{Gaia}\,DR3 CMD of \Cr\ with the 6.7~Gyr isochrone (black line, $\pm0.20$~mag band) and the TO/SGB reference stars (open circles). (a) Rotation and binarity: the symbol area scales with $v\sin i$, and pink rings mark the 15 SB1. (b) Surface chemistry: BSS with measured C and O, marked by chemical group (Sect.~\ref{sec:chem_groups}). In both panels, grey squares are the 13 candidates of \citet{Rain2021} that we did not observe, and the blue diamond is the double-lined reference star (Sect.~\ref{sec:bss_msto}).}
\label{fig:cmd_vsini}
\end{figure*}

\begin{table*}
\centering
\caption{Surface abundances of the 22 slow-rotating BSS.}
\label{tab:chem}
\begin{tabular}{lcccccl}
\hline\hline
BSS & [C/Fe]$_{\rm NLTE}$ & [O/Fe]$_{\rm NLTE}$ &
[Ba/Fe]$_{\rm NLTE}$ & $\Delta$[Fe/H] & SB1 & Group \\
\hline
1  & $-0.20\pm0.11$ & $-0.22\pm0.18$ & \dots            & $+0.40\pm0.11$ & N & normal \\
2  & $-0.10\pm0.11$ & $-0.31\pm0.17$ & \dots            & $+0.30\pm0.11$ & N & normal \\
4  & $+0.20\pm0.26$ & $+0.34\pm0.15$ & $+1.32\pm0.23$   & $+0.45\pm0.11$ & N & A \\
6  & $+0.20\pm0.13$ & $-0.02\pm0.17$ & $-0.38\pm0.32$   & $+0.10\pm0.12$ & N & normal \\
11 & $+0.30\pm0.11$ & $+0.27\pm0.17$ & $\leq-0.62$      & $-0.25\pm0.12$ & N & B \\
13 & $+0.50\pm0.13$ & $+0.80\pm0.15$ & \dots            & $+0.00\pm0.12$ & N & (B) \\
14 & $+0.30\pm0.12$ & $+0.18\pm0.15$ & \dots            & $+0.50\pm0.11$ & N & \dots \\
15 & $+0.40\pm0.12$ & $+0.75\pm0.14$ & $\geq+1.38$      & $+0.55\pm0.11$ & N & A \\
16 & $+0.30\pm0.11$ & $-0.14\pm0.16$ & \dots            & $-0.25\pm0.12$ & N & \dots \\
22 & $+0.90\pm0.15$ & $+1.25\pm0.14$ & $\geq+1.38$      & $+0.35\pm0.11$ & N & A \\
23 & $+0.60\pm0.11$ & $+0.48\pm0.17$ & \dots            & $-0.05\pm0.13$ & N & (B) \\
25 & $+0.30\pm0.11$ & $+0.39\pm0.17$ & \dots            & $-0.15\pm0.14$ & N & \dots \\
26 & $-0.10\pm0.11$ & $+0.02\pm0.16$ & $+0.49\pm0.21$   & $+0.05\pm0.11$ & N & normal$^{a}$ \\
27 & $+0.40\pm0.11$ & $+0.45\pm0.16$ & \dots$^{b}$      & $-0.15\pm0.12$ & N & (B) \\
29 & $+0.70\pm0.21$ & $+0.08\pm0.23$ & $+1.26\pm0.10$   & $+0.20\pm0.14$ & N & A \\
30 & $+0.80\pm0.13$ & $+1.28\pm0.15$ & \dots            & $-0.30\pm0.13$ & N & (B) \\
31 & $+0.50\pm0.12$ & $+0.40\pm0.15$ & \dots            & $+0.00\pm0.11$ & N & (B) \\
32 & $+0.40\pm0.11$ & $+0.47\pm0.19$ & \dots            & $-0.55\pm0.12$ & Y & (B) \\
35 & \dots          & \dots$^{c}$    & \dots            & $+0.00\pm0.11$ & N & \dots \\
36 & $+0.60\pm0.13$ & $+0.78\pm0.16$ & $+0.14\pm0.23$   & $-0.30\pm0.12$ & N & B \\
37 & $+0.70\pm0.13$ & $+0.97\pm0.14$ & $+0.23\pm0.23$   & $-0.10\pm0.11$ & N & B \\
40 & $+0.30\pm0.11$ & $-0.32\pm0.24$ & \dots            & $-0.15\pm0.12$ & N & \dots \\
\hline
\end{tabular}
\tablefoot{Carbon and oxygen are NLTE values anchored to the TO/SGB median (Sect.~\ref{sec:chem_method}); barium is absolute, with the \citet{Korotin2015} NLTE correction. The NLTE correction for C\,\textsc{i}~$\lambda$9094.83 is negligible at these parameters \citep{Amarsi2019} and is set to zero. $\Delta[\mathrm{Fe/H}]$ is the BSS$-$reference differential used as a null test (Sect.~\ref{sec:feh}). SB1 flags spectroscopic binarity (Table~\ref{tab:bss}). Grid-edge Ba values are quoted as limits; ellipses denote lines without coverage or with unusable spectra. Group codes: A = Ba-rich (mass transfer from an AGB donor); B = C/O-enhanced, Ba-normal; (B) = same C/O pattern but no Ba coverage; normal = consistent with the TO/SGB reference; an ellipsis marks marginal ($\Delta[\mathrm{C/Fe}]\approx2\sigma_{\rm MAD}$) or unclassifiable stars. $^{a}$~Normal in C and O but with a moderate Ba excess (Sect.~\ref{sec:chem_groups}). $^{b}$~HR12 stack unusable (corrupted line profile). $^{c}$~Only O measurable, and unreliable (coolest star of the sample); excluded from the analysis.}
\end{table*}

\section{Stellar parameters}
\label{sec:stellar_params}
Effective temperatures, surface gravities, and masses were derived homogeneously for both the BSS and the comparison samples by comparing their positions in the \textit{Gaia}\,DR3 CMD with PARSEC evolutionary tracks \citep{Bressan2012}, following the approach adopted in the reference \vsini\ surveys of globular-cluster BSS (e.g., \citealt{Billi2025}). We retrieved a dense grid of PARSEC models at the cluster metallicity ($[\mathrm{Fe/H}]=-0.22$) spanning initial masses up to $2.0$\,\msun, and orthogonally projected the dereddened CMD position of each star ($(m-M)_0=12.15$, $A_V=0.93$, $A_G/A_V=0.83$, $E(G_{BP}-G_{RP})=0.43\,A_V$) onto the closest model point, adopting the corresponding $T_{\rm eff}$, $\log g$ and stellar mass (Figure~\ref{fig:cmd_tracks}). Since BSS are rejuvenated objects, their projection was performed on the full multi-age track grid rather than on the cluster isochrone, whereas the comparison stars were projected onto the 6.7\,Gyr isochrone itself. Internal uncertainties, estimated by re-projecting each star after perturbing its photometry by the typical errors, are of order $\pm$50--150\,K in $T_{\rm eff}$, $\pm$0.03\,dex in $\log g$ and $\pm$0.02\,\msun\ in mass; these do not include the systematic contribution of unresolved binarity, which shifts the composite CMD position of close pairs.
These uncertainties also do not include differential reddening. The
extinction changes by about 0.15~mag in $A_V$ across our field
(Appendix~\ref{app:diffred}), and this changes the temperature of the
BSS by 260~K (median) and by up to 450~K for the slow rotators at the
edges of the field. We keep the parameters of Table~\ref{tab:bss}
without this correction, and we estimate its effect on our results in
Appendix~\ref{app:diffred}.
The derived parameters span $T_{\rm eff}=5980$--$9220$\,K and $\log g=3.50$--$4.26$ for the BSS, with masses of $1.13$--$1.93$\,\msun\ (all above the turnoff mass ($\approx$1.1\,\msun), as expected for mass-transfer and merger products) and $T_{\rm eff}=5511$--$6076$\,K with $\log g=3.80$--$4.00$ for the comparison sample. At the corresponding masses ($1.05$--$1.10$\,\msun) these gravities imply radii of $1.7$--$2.2\,R_\odot$: the comparison stars lie at the very end of the main sequence and on the early subgiant branch, and we refer to them as the turn-off/subgiant (TO/SGB) reference sample. This is the part of the turn-off region of a cluster as old as \Cr\ that is accessible to FLAMES at useful signal-to-noise ($G\simeq16.0$--$16.3$). One BSS deserves an individual note. BSS~35 ($G=14.63$, $G_{\rm BP}-G_{\rm RP}=1.14$) lies outside the region covered by the track grid: it sits about 0.4~mag blueward of the lower giant branch, in the gap where evolved BSS are expected once they leave the main-sequence band \citep{Ferraro2023}. Its orthogonal projection therefore lands on the terminal, post-main-sequence end of a track rather than inside the grid, and its parameters ($T_{\rm eff}=5984$~K and $\log g=3.50$, the coolest and least compact of the sample) are extrapolations. We keep the star in Table~\ref{tab:bss} but treat these values with caution; consistently, its single abundance measurement is unreliable and was already excluded from the chemical analysis (Table~\ref{tab:chem}). The sample is homogeneous in evolutionary state ($\Delta\log g=0.20$ from end to end), as required by the differential zero point of Sect.~\ref{sec:chem_method}, and the first
dredge-up has not yet begun at these gravities (it sets in near the base of the red
giant branch, $\log g\lesssim3.5$), so their surface C and O are unaltered. At
6.7\,Gyr they rotate at $\lesssim2$\,\kms, an order of magnitude below our detection
threshold, and therefore define the slow-rotation floor against which the BSS are
measured.

\subsection{Radial velocities}
\label{sec:sb1_rv}
Radial velocities were measured by cross-correlation against synthetic templates appropriate for each star's spectral type, and the offset $\Delta\mathrm{RV}=\mathrm{RV}_\star - \langle\mathrm{RV}\rangle_{\rm cluster}$\footnote{$\mathrm{RV}_\star$ is the heliocentric radial velocity of the star measured on the single-epoch HR18 spectrum, and $\langle\mathrm{RV}\rangle_{\rm cluster}=-25.0\pm0.93$~\kms\ is the mean heliocentric velocity of \Cr\ \citep{Rain2020}. A large $|\Delta\mathrm{RV}|$ flags a star that, at the HR18 epoch, is displaced from the systemic velocity of the cluster --- the signature of orbital motion.} was computed relative to the cluster systemic velocity. For slow rotators the measured RVs are insensitive to the template parameters: the cross-correlation peak position shifts by $\ll$1\,\kms\ for template mismatches far larger than our parameter uncertainties, and the multi-epoch variability statistics ($\sigma_{\rm RV}$, ${\rm ptp}_{\rm RV}$) are purely differential and therefore template-independent, so the binarity classification below is unaffected by the parameter revision of Section~\ref{sec:stellar_params}. Of the 40 photometrically-selected BSS candidates listed in Table~\ref{tab:bss}, 38 yielded a $v\sin i$ measurement or upper limit in the rotation analysis of Section~\ref{sec:vsini_method}; BSS~20, for which no value can be attributed but which is nonetheless a fast rotator, and BSS~7, for which no statement is possible, are discussed there. Two complementary criteria flag spectroscopic binarity. First, the single-epoch HR18 cross-correlation gives $|\Delta\mathrm{RV}|>20$\,\kms\ for 13 of the 40 BSS (12 after discarding BSS~39, see below). Second, multi-epoch HR12 monitoring (5--6 epochs per star) provides per-star RV statistics ($\sigma_{\rm RV}$, peak-to-peak ${\rm ptp}_{\rm RV}$) for 27 of the 40 BSS; we classify a star as SB1 when $\sigma_{\rm RV}\gtrsim8$\,\kms\ and ${\rm ptp}_{\rm RV}>20$\,\kms. A star is flagged SB1 if it meets either criterion; no other information enters the flag. Nine BSS satisfy both criteria, three satisfy only the multi-epoch one (BSS~32, 34 and 38, which HR12 catches close to systemic at the single HR18 epoch), and three satisfy only the HR18 criterion because they lack HR12 coverage (BSS~7, 17 and 28). For fast rotators the CCF-based RVs are intrinsically less reliable. We keep BSS~24 as an SB1 because of its large multi-epoch variability (${\rm ptp}_{\rm RV}\approx136$\,\kms), but we discard the single-epoch shift of BSS~39, whose HR12 RVs are constant at the $\sim$1\,\kms\ level and whose apparent $\Delta\mathrm{RV}=-109$\,\kms\ is most likely an artefact of its broad cross-correlation peak. To test how reliable the cross-correlation velocities are for fast rotators, we broadened the HR18 spectra of the TO/SGB stars to different values of $v\sin i$, added noise at the S/N of our BSS spectra (60--90), and measured them against a narrow template, as in our analysis. Up to $v\sin i\approx80$~\kms\ the velocity error is 0.2--3~\kms, with a small positive offset of 2--4~\kms\ at 40--60~\kms\ caused by line blending. Above about 100~\kms\ the cross-correlation peak can jump to a different feature, and the error can reach 100--190~\kms. With a template broadened to the same $v\sin i$ the problem disappears. Single-epoch offsets of the broadest-lined stars are therefore not reliable (for example BSS~17, 28 and 33). For the SB1 classification this matters mostly for four fast rotators whose binarity rests only on their velocities: BSS~19, 20, 21 and 24. The other fast-rotating SB1 are eclipsing or ellipsoidal binaries (Sect.~\ref{sec:tess}) or have $v\sin i<90$~\kms, where our velocities are reliable. A re-analysis with rotationally broadened templates will be part of the orbital study of the SB1 population.

The $\sim$1085-day separation between the two programmes also adds sensitivity at the opposite end: a binary with a period of months to years can appear constant within the 63-day window of the 2022 campaign and still be displaced by several \kms\ three years later. In practice, this does not add new candidates: every star that is constant in the 2022 epochs is also consistent with the cluster velocity in 2025 (Table~\ref{tab:rvstats}), with the single apparent exception of BSS~39, whose offset we discard above as an artefact. The long baseline therefore reveals no long-period system with a large amplitude, consistent with the detectability argument of Sect.~\ref{sec:channels}: typical post-mass-transfer orbits of several hundred days produce velocity amplitudes below our cut, and such systems remain hidden in both windows. Six of our stars were also monitored by \citet{Rain2020} at four epochs between 2012 and 2018. The two classifications agree for all six. The three stars they classified as probable close binaries (BSS~8, 17 and 21) are SB1 here, and the three with constant velocity (BSS~26, 35 and 36) are not. For these three, the RVs stay below our SB1 threshold over about 13~yr. For BSS~21, whose RV membership is uncertain in our data, one of the four velocities of \citet{Rain2020} is at the cluster value. The combined HR18+HR12 cut yields 15 SB1 candidates (38\%).

Four further stars (BSS~12, 22, 26 and 29) show HR12 scatter above the constant level ($\sigma_{\rm RV}=3$--7~\kms) but below the SB1 threshold; we list them as RV-variable in Table~\ref{tab:rvstats} and do not count them as binaries. The SB1 fraction is a lower limit, below the $\sim$50--70\% reported for the BSS populations of old open clusters \citep{Mathieu2009,Geller2012}, because our cuts miss most long-period post-mass-transfer orbits (Sect.~\ref{sec:channels}). The remaining BSS lack HR12 multi-epoch coverage and are unclassified with respect to multi-epoch variability; their HR18 single-epoch $\Delta\mathrm{RV}$ is the only available test, so the true binary fraction is likely still slightly underestimated.

The SB1 candidates are retained in the rotation analysis: their line profiles are dominated by the brighter component, and a $v\sin i$ can still be recovered via the $\chi^2$ surface, although with larger uncertainties, provided any secondary peak is separated by $\gtrsim$\ $1.5\times{\rm FWHM}_{\rm inst}$.  The resulting \vsini\ is, formally, a mixture-dominated estimate for the brighter component and should be interpreted as such.  Excluding the SB1 candidates would remove more than a third of the sample, would bias it against precisely the objects where a mass-transfer origin is most likely to leave a signature on the rotation (Section~\ref{sec:vsini_method}). The reference stars have only the single HR18 epoch and no HR12 monitoring, so their binarity is much less constrained than that of the BSS. The reference sample is therefore not binary-free; its binary fraction is simply unconstrained beyond this one cut, and it should be consistent with the normal binary fraction of the cluster \citep{Mathieu2009,Geller2012}. For our purposes, this is enough, since the reference sample is used as a rotational and chemical baseline, not as a binary census.

We also checked if the BSS are cluster members from their radial
velocities. For the 15 non-SB1 stars with HR12 data, the mean velocity
goes from $-27.5$ to $-20.9$~\kms. The average is $-24.5$~\kms, with a
scatter of 2.0~\kms, in good agreement with the cluster value. The ten
non-SB1 stars with only one HR18 epoch are within 7~\kms\ of the
cluster velocity. The two exceptions are BSS~5 and 10 ($+16$ and
$+15$~\kms), and both are fast rotators. All the stars with a chemical
anomaly (Sect.~\ref{sec:chem_groups}) are among these members.

For the SB1 the test is different, because the mean of a few epochs
is not the systemic velocity of the binary. In this case we checked if
the cluster velocity is inside the range of the velocities we measured
for each star. This is true for nine of the twelve SB1 with HR12 data,
and only marginally true for BSS~38. It is not true for BSS~21 and 33.
All their HR12 velocities are above $-17$ and $-21$~\kms,
respectively. However, both stars are fast rotators. We also note that
all the 13 fast rotators with HR12 data have a mean velocity above the
cluster value, whereas the slow rotators fall on both sides. This agrees
with the positive offset found in our test above, and suggests that the
cross-correlation velocities are biased for very broad lines. For this
reason, the velocities of BSS~21 and 33 alone are not enough to say
that they are field stars. Their Gaia DR3 parallaxes and proper motions
(0.37 and 0.32~mas; $\mu_{\alpha*},\mu_\delta\simeq-6.4,-2.7$ and
$-6.2,-2.7$~mas\,yr$^{-1}$) are also typical of the cluster. We keep
them in the sample, and we mark their membership as uncertain in
Table~\ref{tab:bss}. If we remove the two stars, our results do not
change. The fraction of fast rotators becomes 41\% (15/37), and the
Fisher test gives $p=2\times10^{-5}$.

\begin{figure}[!ht]
\centering
\includegraphics[width=\columnwidth]{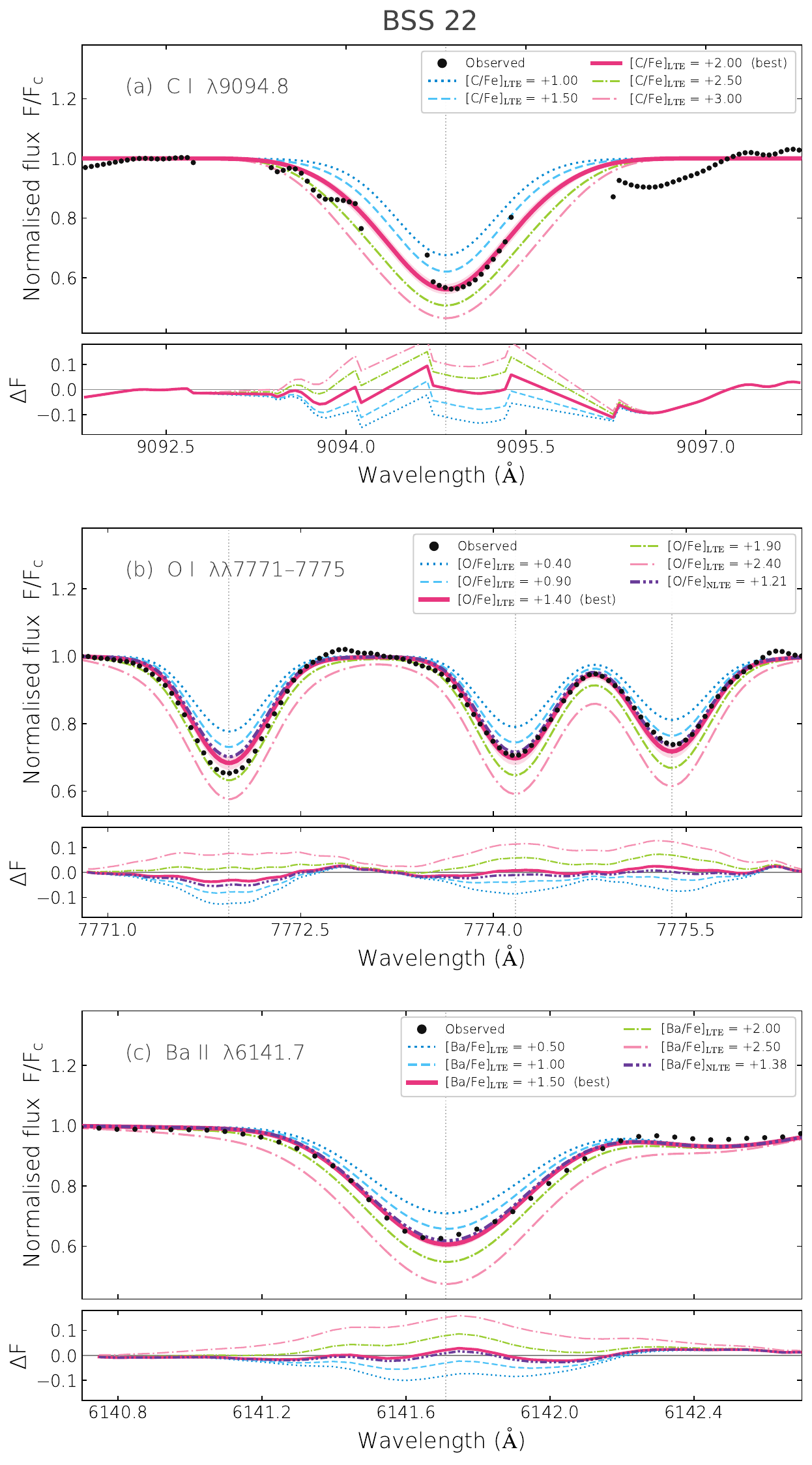}
\caption{Spectral-synthesis fits of BSS~22: (a) C\,\textsc{i}~$\lambda$9094.8, (b) the O\,\textsc{i}~$\lambda\lambda$7771--7775 triplet, and (c) Ba\,\textsc{ii}~$\lambda$6141.7. The observed spectrum (black) is compared with LTE syntheses on a 0.5~dex grid, with the best fit in pink; NLTE abundances are given where they apply. The Ba fit is at the top of the grid and is a lower limit.}
\label{fig:chem_demo}
\end{figure}

\subsection{Projected rotational velocities}
\label{sec:vsini_method}
Projected rotational velocities were measured by direct comparison with synthetic spectra. Synthetic profiles were computed with \textsc{iSpec} \citep{Blanco-Cuaresma2014} using the \textsc{SPECTRUM} radiative transfer code \citep{Gray1994}, MARCS model atmospheres \citep{Gustafsson2008}, and atomic line data from \textsc{linemake} \citep{Placco2021}. For each star, the effective temperature and surface gravity were fixed to the values derived from the CMD projection onto the PARSEC evolutionary tracks (Sect.~\ref{sec:stellar_params}). The synthesis metallicity was set to $[\mathrm{M/H}]=-0.05$, the value that best reproduces the observed Fe\,\textsc{i} line strengths of the TO/SGB reference stars with the same setup, thus absorbing the zero point of the adopted $\log gf$ scale. Microturbulence follows the built-in \textsc{iSpec} calibration, and macroturbulence was fixed by the $T_{\rm eff}$-dependent relation of \citet{Valenti2005} (see also \citealt{Gray2005}), leaving $v\sin i$ as the only free parameter. All syntheses are computed in LTE; the rotation measurement depends on the profile width, not on the absolute line strength, and any residual strength mismatch is absorbed by the metallicity treatment described below.
The fit was performed by $\chi^2$ minimisation over four spectral windows within the HR18 range, centred on the strongest Fe\,\textsc{i} features available ($\lambda$7495+7498, 7507+7511, 7531, and 7583+7586~\AA). Each window was renormalised with a local pseudo-continuum estimated from adjacent line-free sidebands, and the per-pixel noise was taken as the rms of the sideband flux. Residual wavelength shifts (including the orbital shifts of the SB1 candidates) were removed star by star using empirical line-centre offsets measured on the Fe\,\textsc{i}\,$\lambda$7495.05 line prior to the fit. Pixels deviating more than $4\sigma$ above the local continuum (sky-subtraction residuals and cosmic-ray hits, which no photospheric feature can produce) were excluded from the fit; absorption features were never clipped. Uncertainties correspond to $\Delta\chi^2=1$; when the best-fit $v\sin i$ is smaller than its uncertainty we quote an upper limit.

The synthesis metallicity deserves a comment, because the iron abundances of Sect.~\ref{sec:feh} reach up to $+0.55$~dex above the cluster value in a few stars. A template that is too metal-poor produces lines that are too shallow, and the $\chi^2$ fit compensates by broadening them: a control experiment in which the template metallicity is shifted while everything else is held fixed shows a systematic of $\approx$13~\kms\ per dex. We therefore repeated the fit for the 22 slow rotators with the per-star iron of
Sect.~\ref{sec:feh} fixed in the synthesis, and Table~\ref{tab:bss} quotes these values. In practice, the effect is small: the median shift is 0~\kms, the largest is 11~\kms, and no star crosses the 40~\kms\
boundary, so the fast/slow classification and every statistic built on it are unchanged. For the 16 fast rotators with a fitted profile, no per-star iron can be measured because the same broadening that prevents the abundance
analysis also prevents an iron measurement. Rather than assume that the cluster metallicity is adequate for them, we tested it: their fits were repeated with the template metallicity stepped from $-0.05$ to
$+0.60$~dex, the full range spanned by the slow-rotator iron measurements. The well-constrained detections (BSS~8, 9, 18, and 34) prefer the cluster value and change by at most 4~\kms. The remaining profiles are shallow and broad, and there the template strength and the rotation are degenerate: the formal minima drift toward the
metal-rich edge of the grid with very large uncertainties, the
behaviour expected when a washed-out profile no longer constrains the
line strength. In no case, however, does the fitted velocity fall:
the ten lower limits remain lower limits and no star approaches the
40~\kms\ boundary from above, so the classification of every fast
rotator is insensitive to the template metallicity. Note also that the stars with a high measured iron are all slow rotators, and that their excess is not a true metallicity
(Sect.~\ref{sec:feh}). Finally, the synchronised systems provide an external check that no
template choice can affect: their velocities agree with those imposed
by the TESS orbital periods (Sect.~\ref{sec:tess}) at every
metallicity tested (BSS~28, for example, moves only from $82\pm44$ to
$85\pm38$~\kms\ across the full grid, remaining on its
$v_{\rm sync}=83$~\kms\ prediction).

\begin{table}
\centering
\caption{Tidal-synchronisation test for the seven eclipsing or ellipsoidal BSS.}
\label{tab:sync}
\begin{tabular}{lcccccc}
\hline\hline
BSS & $P_{\rm orb}$ & $R$ & $R_{\rm L}$ & $R/R_{\rm L}$ & $v_{\rm sync}$ & $v\sin i$ \\
    & (d) & ($R_\odot$) & ($R_\odot$) & & (\kms) & (\kms) \\
\hline
33 & 0.518 & 1.94 & 1.70 & 1.14 & 190 & $\gtrsim100$ \\
12 & 1.048 & 1.75 & 2.67 & 0.66 &  85 & $\gtrsim100$ \\
28 & 1.090 & 1.79 & 2.71 & 0.66 &  83 & $82\pm44$ \\
3  & 1.128 & 1.85 & 3.01 & 0.62 &  83 & $\gtrsim100$ \\
17 & 1.318 & 2.61 & 3.44 & 0.76 & 100 & $\gtrsim100$ \\
9  & 1.420 & 1.75 & 3.27 & 0.54 &  62 & $63\pm20$ \\
8  & 2.110 & 2.90 & 4.80 & 0.60 &  70 & $80\pm22$ \\
\hline
\end{tabular}
\tablefoot{Orbital periods are from \citet{Echeveste2026} ($P_{\rm orb}=2P_{\rm LS}$); radii are those of Table~\ref{tab:bss}. $R_{\rm L}$ is the Roche-lobe radius of the BSS from \citet{Eggleton1983}, for a companion of 0.25~\msun. $v_{\rm sync}=2\pi R/P_{\rm orb}$ is the equatorial velocity of a synchronised rotator, and therefore an upper limit for $v\sin i$. The stars are ordered by orbital period.}
\end{table}

Two stars require separate treatment. BSS~7 is flagged NM in Table~\ref{tab:bss}: its effective temperature ($\sim$9200~K) lies beyond the MARCS coverage, so no template can be synthesised for it and no statement about its rotation is possible.

BSS~20 is a different case. Despite good S/N ($\simeq$88), no Fe\,\textsc{i} feature is detected in any of the four synthesis windows. This does not conflict with the $\Delta\mathrm{RV}=+52$~\kms\ quoted in Table~\ref{tab:bss}: the radial velocity is obtained by cross-correlating the entire HR18 order against a synthetic template, and the cross-correlation function integrates the combined opacity of hundreds of blended features, so it retains a measurable peak long after any individual line has been broadened below the local noise. The two measurements probe different regimes, and the non-detection of individual lines is itself informative. At $T_{\rm eff}\simeq7650$~K the Fe\,\textsc{i} lines used here are already intrinsically weak, and erasing them entirely at this S/N requires rotation above the ceiling set by our resolution. We therefore count BSS~20 among the fast rotators, flagged FR in Table~\ref{tab:bss}, without attributing a value to it: the star enters the fast fraction and the SB1--rotation statistics, but not the fitted distribution. We note that the star also shows a short-period photometric signal that is not orbital (Sect.~\ref{sec:tess}); the associated line-profile variability may contribute to the smearing, but at the amplitudes typical of such variability, it cannot by itself account for the complete absence of the lines.
Figure~\ref{fig:vsini_demo} illustrates the quality of the fits across the full range of rotation, from the slowest resolved profile (BSS~15) to $v\sin i \sim 80$~\kms\ (BSS~8); the fits for the complete sample are shown in Appendix~\ref{app:vsini_fits}.

\subsubsection{The \texorpdfstring{$v\sin i$}{v sin i} distribution}
\label{sec:vsini_results}
Figure~\ref{fig:vsini_distribution} presents the \vsini\ distribution for the full statistical sample ($N=39$) together with its non-SB1 and SB1 sub-components, and compares it against the 42-star TO/SGB reference. The distribution shows a dominant slow component peaking at 0--20~\kms\ (median $\approx$13~\kms over the 16 slow rotators with a detection; the six upper limits are excluded, since any convention for them shifts the value between 10 and 16~\kms) and an extended fast tail reaching beyond 100~\kms; adopting the usual 40~\kms\ boundary \citep{Leiner2018}, 17 BSS (44\%) are fast rotators and 22 (56\%) are slow. Ten of the fast rotators have profiles that are too shallow and broad to be bounded from above at the HR18 resolution and are quoted as lower limits ($v\sin i \gtrsim 100$~\kms; Table~\ref{tab:bss}), and BSS~20 is counted as fast without a value (Sect.~\ref{sec:vsini_method}), so the median of the fast sub-sample exceeds 100~\kms\, which is roughly an order of magnitude above the slow component, a separation that cannot be produced by measurement noise or projection effects alone (Section~\ref{sec:bimodality}).
A formal Hartigan dip test does not reject unimodality ($p\simeq0.6$--$0.8$, depending on how the ten lower limits are spread over their allowed range; treating them as a point mass at 100~\kms\ is an artefact of the censoring), with a single star occupying the 40--60\,\kms\ interval. The physically meaningful feature is therefore not a gap in the distribution but the presence of a substantial fast-rotating population with no counterpart among the TO/SGB reference stars (0/42): a two-sample KS test rejects a common parent distribution ($D=0.88$, $p\simeq6\times10^{-17}$, with each star entering at its measured value and the censored stars at their quoted bound; any alternative treatment of the limits, including placing every upper limit at zero, gives $D\geq0.72$ and $p\leq1\times10^{-10}$). The fast rotators are strongly concentrated among the SB1 candidates, 13 of the 14 SB1 in the statistical sample are fast (93\%), against 4 of the 25 non-SB1 (16\%). A Fisher exact test rejects independence at $p=4\times10^{-6}$ (odds ratio $\simeq$68).This result does not depend on the fast rotators with the least reliable velocities: if we remove the SB1 flag from BSS~19, 20, 21
and 24 (Sect.~\ref{sec:sb1_rv}), 9 of the 10 SB1 are fast, against 8 of the 29 non-SB1, and the Fisher test still gives $p=9\times10^{-4}$.

\subsubsection{Colour--magnitude diagram}
\label{sec:cmd_results}
Figure~\ref{fig:cmd_vsini} shows the \textit{Gaia}\,DR3 CMD of \Cr\ with the BSS sample (purple, $N=40$) and the TO/SGB reference stars (orange) overplotted on the 6.7~Gyr PARSEC isochrone shifted to the adopted $(m-M)_0=12.15$ and $A_V=0.93$. Symbol areas scale with \vsini; the 15 SB1 candidates are highlighted by a pink edge ring, upper limits are shown as open symbols, and the two stars without a \vsini\ value (BSS~7, NM, and BSS~20, counted as a fast rotator; both SB1) are marked with crosses. Three features are worth highlighting: (i) the fast rotators do not cluster in any particular region of the BSS locus but span the entire strip, from the bright blue end ($G\simeq14.2$, $G_{\rm BP}-G_{\rm RP}\simeq0.7$) down to the faint boundary near the turn-off mass ($G\simeq16.4$);
(ii) the brightest BSS in the sample (BSS~22, $G\simeq13.9$) rotates slowly ($v\sin i = 13\pm5$~\kms) and shows only low-level RV scatter; together with its enhanced [Ba/Fe] (Sect.~\ref{sec:chem_groups}), this is consistent with a long-completed mass-transfer episode in a wide orbit, below our RV detection limit, in which the accretor has already spun down, whereas the fast-rotating SB1s trace more recent binary activity; (iii) the SB1 candidates themselves are distributed across the full CMD region occupied by the BSS, including objects well above the cluster's subgiant branch, consistent with an interpretation as post-MT products with ongoing (wide) companionship rather than primordial
binaries. The TO/SGB sample, as expected, piles up along the isochrone ridge; the only reference star with a $v\sin i$ detection ($26\pm24$~\kms) lies marginally blueward and brighter than the ridge, and its spectrum shows that it is a double-lined binary (Sect.~\ref{sec:bss_msto}).
Point (i) is worth a comparison with globular clusters, where the first $v\sin i$ surveys suggest that fast rotators may be more common among the brighter BSS \citep{Ferraro2023}, although the case of NGC~1851 shows that the evidence is not yet conclusive \citep{Billi2025}. In \Cr\ the fast rotators cover the whole strip. This is what the picture of Sects.~\ref{sec:channels} and \ref{sec:tess} predicts: if much of the fast tail is kept fast by tidal locking, the rotation follows the orbital period and not the luminosity, so no gradient along the strip is expected. In M67, NGC~188 and NGC~6791, \citet{Linck2026b} find three regimes: BSS cooler than about 6200~K are all slow rotators, BSS hotter than about 6750~K are mostly fast, and the stars in between, around the Kraft break, which have thin convective envelopes, show mixed rotation. \Cr\ follows the same pattern. The four BSS cooler than 6215~K are slow, 5 of the 18 in the intermediate range are fast, and 12 of the 17 hotter stars are fast. The five fast rotators of the intermediate range are all SB1, so tides can keep them fast even with a convective envelope. Five of the hot BSS are slow (BSS~4, 15, 22, 26 and 27). \citet{Linck2026b} also find two slow BSS among their hottest stars, and suggest low inclination or further spin-down in stars with large radii. In \Cr\ none of the five is an SB1, but all carry a chemical signature of mass transfer: BSS~4, 15 and 22 are Ba-rich, BSS~27 is C/O-enhanced, and BSS~26 has a moderate Ba excess (Sect.~\ref{sec:chem_groups}).

\subsection{Iron abundance consistency}
\label{sec:feh}
The chemical analysis of Section~\ref{sec:chem_method} assumes a common iron abundance for all targets: any undetected star-to-star spread in [Fe/H] would propagate directly into spurious [X/Fe] anomalies, and BSS are precisely the stars where such a spread is conceivable (composite light from a post-mass-transfer companion, residual field contamination). We therefore verified the iron scale before interpreting any other element.
A classical multi-line analysis is not a fair test with these data: the only setup rich in clean Fe\,\textsc{i} lines (HR12) was observed for the BSS but not for the TO/SGB reference stars, so any HR12-based metallicity would lack a reference measured on the same system. We instead use the one diagnostic available identically for both samples, Fe\,\textsc{i}\,$\lambda$7495.05 (HR18), fitted with the same synthesis pipeline as C and O (Sect.~\ref{sec:chem_method}): $T_{\rm eff}$ and $\log g$ fixed to the track-based parameters, [Fe/H] as the only free parameter. Every systematic shared by the two
samples ($\log gf$, continuum placement, resolution) cancels in the differential $\Delta[\mathrm{Fe/H}]$. The TO/SGB reference stars ($N=42$) define the zero point with a star-to-star scatter of $\sigma_{\rm MAD}=0.15$~dex, a factor of two tighter than with our
previous spectroscopic parameters, an independent confirmation of the CMD-projection parameter scale.
For the 22 slow-rotating BSS we obtain $\langle\Delta[\mathrm{Fe/H}]\rangle = +0.03 \pm 0.06$ (median $0.00$): as a population, the BSS share the cluster iron, validating $[\mathrm{Fe/H}]=-0.22$ for the synthesis grid and securing the abundance anomalies of Sect.~\ref{sec:chem_groups} against any metallicity offset. The distribution is not featureless, however (Table~\ref{tab:chem}).
Seven stars have $\Delta[\mathrm{Fe/H}]\geq+0.20$~dex: the four
Group~A stars, BSS~1, 2 and 14. Mass transfer cannot change the iron
abundance, so these excesses are not real. Light from a white-dwarf
companion cannot explain them either: at 7495~\AA\ a white dwarf
gives less than $10^{-3}$ of the flux of the BSS, and extra continuum
would make the line weaker, not stronger. A temperature that is too
high has the right effect, because Fe\,\textsc{i} lines become weaker
at higher $T_{\rm eff}$, and differential reddening is one source of
such an error (Appendix~\ref{app:diffred}). For BSS~1, 2, 4 and 22
the corrected colour gives a temperature 280--450~K lower, which
brings their $\Delta[\mathrm{Fe/H}]$ close to zero. For BSS~14, 15
and 29 the correction does not help, and with the present data we
cannot explain their excess. This does not affect the abundances of
Sect.~\ref{sec:chem_groups}, which are computed at the cluster
metallicity and do not use these iron values.

\begin{figure}[t]
\centering
\includegraphics[width=\columnwidth]{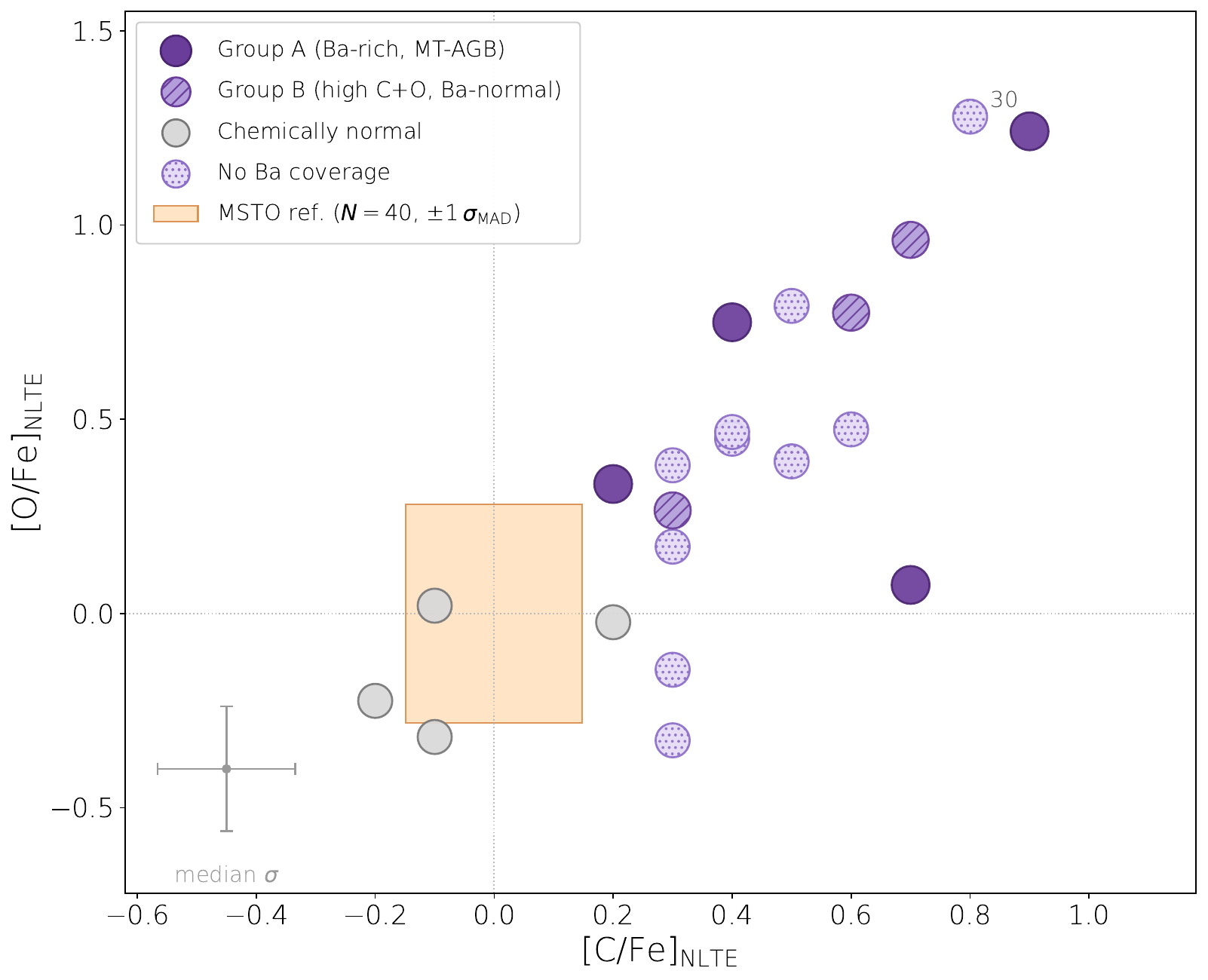}
\caption{[C/Fe]$_{\rm NLTE}$ vs.\ [O/Fe]$_{\rm NLTE}$ for the 21 slow-rotating BSS with measured C and O, relative to the TO/SGB median. The shaded box is the $\pm1\,\sigma_{\rm MAD}$ scatter of the reference sample. Symbols mark the chemical groups (Sect.~\ref{sec:chem_groups}), the grey cross is the median uncertainty.}
\label{fig:c_vs_o}
\end{figure}

\begin{figure}[t]
\centering
\includegraphics[width=\columnwidth]{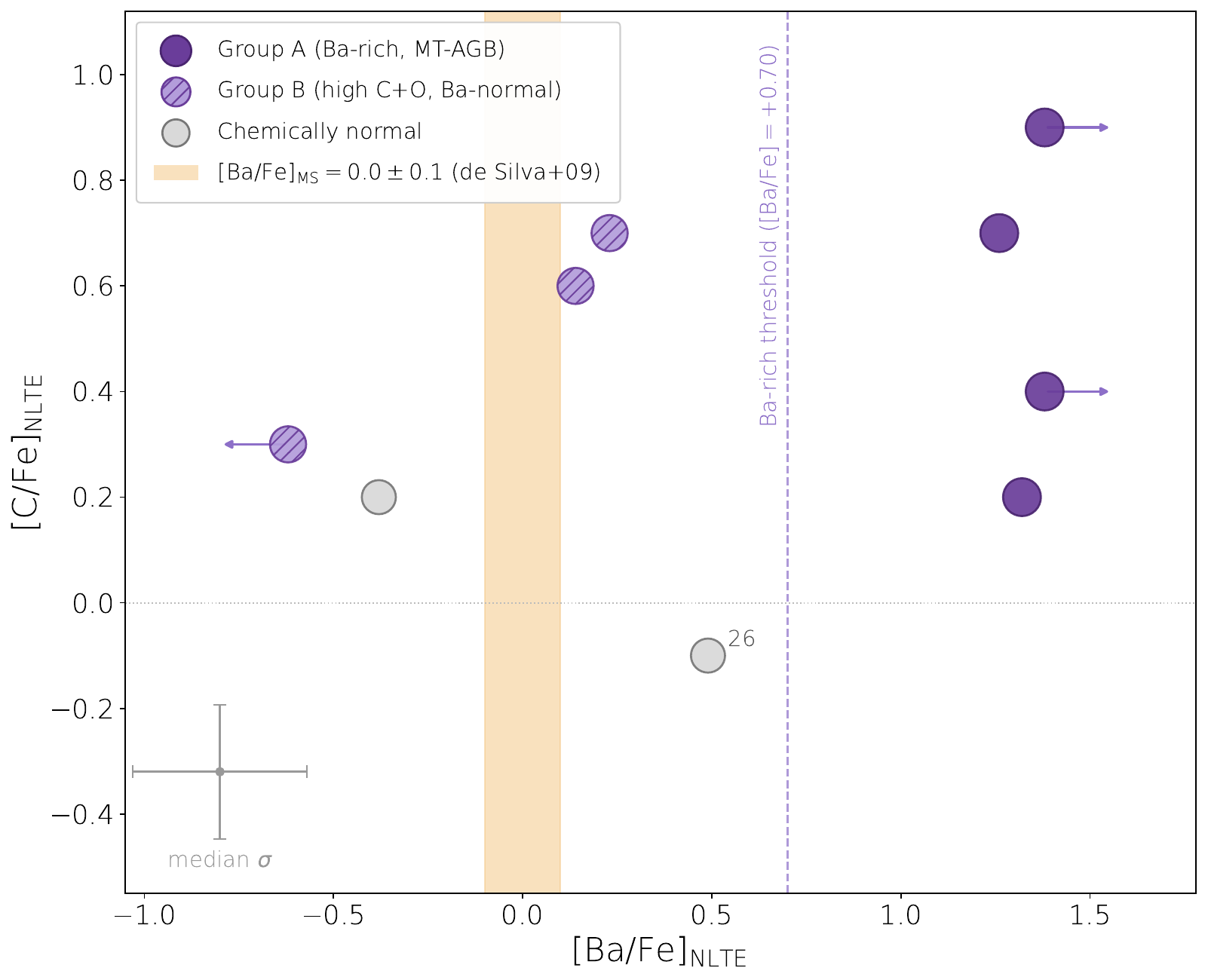}
\caption{[Ba/Fe]$_{\rm NLTE}$ vs.\ [C/Fe]$_{\rm NLTE}$ for the nine slow-rotating BSS with a Ba measurement. The shaded band is the cluster Ba value of \citet{deSilva2009}, and arrows mark grid-edge limits. A gap of about 1~dex separates Group~B from Group~A. BSS~26 is normal in C and O, with a moderate Ba excess.}
\label{fig:c_vs_ba}
\end{figure}
\subsection{Chemical abundances}
\label{sec:chem_method}
Fast-rotating BSS ($v\sin i \gtrsim 40$~\kms) have rotationally broadened lines too diluted for a reliable abundance analysis. For the 22 slow rotators ($v\sin i \leq 40$~\kms) we measured surface abundances of carbon, oxygen, and barium as an independent formation-channel probe, together with iron as a consistency check.
Abundances were derived by spectral synthesis with MOOG \citep{Sneden1973}, using the track-based atmospheric parameters of Section~\ref{sec:stellar_params}, ATLAS9 model atmospheres \citep{Castelli2003} computed at the cluster metallicity $[\mathrm{Fe/H}]=-0.22$ (Sect.~\ref{sec:feh}), and line lists generated with \texttt{linemake}. Four diagnostics were used: C\,\textsc{i}~$\lambda$9094.83 (HR22, $R\simeq11\,500$), the O\,\textsc{i}~$\lambda\lambda$7771--7775 triplet and Fe\,\textsc{i}~$\lambda$7495.05 (both from HR18, $R\simeq18\,000$), and Ba\,\textsc{ii}~$\lambda$6141.71 measured on median stacks of the 5--6 RV-corrected HR12 epochs ($R\simeq18\,700$). For Ba\,\textsc{ii}~$\lambda$6141.71 the synthesis includes hyperfine structure \citep{Gallagher2010} and explicit isotopic splitting, with
the solar mixture of \citet{Asplund2009}: $^{134}$Ba:$^{135}$Ba:$^{136}$Ba:$^{137}$Ba:$^{138}$Ba $= 2.4{:}6.6{:}7.9{:}11.2{:}71.9$. Hyperfine broadening affects only the odd isotopes ($^{135}$Ba, $^{137}$Ba; 17.8\% of the solar mixture), which desaturate the line core; neglecting it would bias the abundances of the Ba-rich stars upward. We note that material polluted by AGB $s$-processing carries a lower odd-isotope fraction ($\sim$11\%; \citealt{Arlandini1999}) than the solar mix: adopting an $s$-only mixture would produce a slightly weaker line at fixed abundance and thus higher inferred [Ba/Fe] for the Group~A stars, so our solar-mixture values (and the grid-edge lower limits of BSS~15 and 22) are conservative.
For each star the spectrum was renormalised with a local continuum estimated from line-free side bands, recentred in wavelength by matching the observed line-core centroid to a synthetic template, and cleaned of sky/telluric emission residuals (pixels $>5$\% above the continuum, plus tellurics already masked by the reduction). The $\chi^2$ statistic was then evaluated over the line cores only, on abundance grids in steps of 0.1~dex for C, O ($-1.0\leq[\mathrm{X/Fe}]\leq+2.4$) and Ba ($-0.5\leq[\mathrm{Ba/Fe}]\leq+1.5$), and 0.05~dex for Fe ($|\Delta[\mathrm{Fe/H}]|\leq0.8$). Stars reaching a grid edge are reported as limits. Uncertainties combine the fit error, rescaled to the observed reduced $\chi^2$ of each star, with the sensitivity to the stellar parameters ($\Delta T_{\rm eff}=\pm100$~K; 0.10--0.13~dex depending on the line), added in quadrature. Figure~\ref{fig:chem_demo} illustrates the three fits for BSS~22, together with the synthesis grids used to bracket the best-fit value.
The MOOG syntheses are computed in LTE, and we correct each element for
departures from LTE before any comparison is made. The
O\,\textsc{i}~$\lambda\lambda$7771--7775 triplet is the diagnostic most
affected: we apply a per-star correction interpolated from the grids of
\citet{Amarsi2016}, which ranges from $-0.14$ to $-0.22$~dex across the
temperatures of our sample. Because the BSS are on average hotter than
the TO/SGB reference stars, this correction does not fully cancel in
the differential abundances, which is why it is applied star by star on
both samples. For barium we adopt the $-0.12$~dex correction computed
by \citet{Korotin2015} for the Ba\,\textsc{ii}~$\lambda$6141 line at
the parameters of our stars. For carbon, the correction to
C\,\textsc{i}~$\lambda$9094 is negligible at these temperatures and
gravities \citep{Amarsi2019}, so its LTE and NLTE values coincide. Iron is used only
differentially, so any residual NLTE effect on
Fe\,\textsc{i}~$\lambda$7495 is common to both samples and cancels in
$\Delta$[Fe/H]. Throughout the paper, abundances labelled NLTE include
these corrections; values shown without the label are LTE.
Because the absolute LTE abundances from single, blended lines at these resolutions carry method systematics that far exceed the star-to-star precision, we work strictly differentially: the identical pipeline was applied to the 42-star TO/SGB reference sample (Section~\ref{sec:obs}; 40 stars with measurable lines), and all C, O, and Fe abundances are quoted as $\Delta[\mathrm{X}] = [\mathrm{X}]_{\rm BSS} - \mathrm{med}([\mathrm{X}]_{\rm ref})$. Any systematic shared by both samples (continuum placement, $\log gf$ scale, resolution, first-order NLTE) cancels in $\Delta[\mathrm{X}]$. The reference-sample
star-to-star scatter, $\sigma_{\rm MAD}=0.15$~dex in $\Delta[\mathrm{C/Fe}]$, $0.22$~dex in $\Delta[\mathrm{O/Fe}]$, and $0.15$~dex in $\Delta[\mathrm{Fe/H}]$, sets the systematic floor of the differentials. Residual differential NLTE effects (the BSS are on average hotter than the reference stars) are below this floor for C \citep{Amarsi2019}; for the O\,\textsc{i} triplet they would reach $\lesssim0.2$~dex, which is why the per-star correction described above is applied to both samples before the differentials are formed. Barium, for which no reference measurement is possible (the TO/SGB fibres lack HR12 coverage), is quoted as absolute $[\mathrm{Ba/Fe}]_{\rm NLTE}$ applying the $-0.12$~dex correction of \citet{Korotin2015}, with the cluster baseline $[\mathrm{Ba/Fe}]=0.0\pm0.1$ of \citet{deSilva2009}.
Iron cannot be altered by mass transfer or collisions, so $\Delta[\mathrm{Fe/H}]$ serves as a null test: a significant apparent offset signals composite light or parameter mismatches rather than chemistry (Sect.~\ref{sec:feh}).
\subsubsection{Chemical group classification}
\label{sec:chem_groups}
Based on the joint pattern of C, O, and Ba, we sort the slow-rotating BSS into three empirical classes: Group A (Ba-rich with enhanced C/O), Group B (C/O-enhanced but Ba-normal), and chemically normal stars, with the remainder marginal or unclassifiable. Table~\ref{tab:chem} lists the individual values, and Figures~\ref{fig:c_vs_o} and \ref{fig:c_vs_ba} show the two diagnostic planes on which the classification is based.
\begin{itemize}

\item Group A ($N=4$: BSS~4, 15, 22, 29). All exceed $[\mathrm{Ba/Fe}]_{\rm NLTE}=+1.2$ (BSS~15 and 22 reach the upper edge of the synthesis grid of Sect.~\ref{sec:chem_method} and are quoted as lower limits, $\gtrsim+1.4$), with enhanced carbon ($\Delta[\mathrm{C/Fe}]=+0.2$ to $+0.9$) and, in BSS~15 and 22, strongly enhanced oxygen. This is the expected signature of accretion from an AGB donor with active third dredge-up ($M_{\rm donor}\gtrsim1.5$~\msun; \citealt{Karakas2014}). All four also show an apparent Fe excess (Sect.~\ref{sec:feh}).
\item Group B ($N=3$ with measured Ba: BSS~11, 36, 37; six more candidates without Ba coverage: BSS~13, 23, 27, 30, 31, 32). They combine $\Delta[\mathrm{C/Fe}]=+0.3$ to $+0.8$ and $\Delta[\mathrm{O/Fe}]$ up to $+1.3$ with $[\mathrm{Ba/Fe}]_{\rm NLTE}\leq+0.35$; an empty $\sim$1~dex gap in [Ba/Fe] separates them from Group~A, so the split is not a threshold effect. We interpret them as accretors from donors below the $s$-process activation mass. The BSS of NGC~188, a cluster of similar age, may belong to the same class: most of them are long-period binaries, some with white-dwarf companions, but none is Ba-enriched \citep{Geller2012,Gosnell2014,Nine2024}. Their cluster-level iron confirms the C/O excesses are not an analysis artefact.
\item Chemically normal ($N=4$: BSS~1, 2, 6, 26), consistent with the TO/SGB reference in both C and O, as expected for collisional or non-polluting channels. BSS~26 is peculiar (normal C/O but $[\mathrm{Ba/Fe}]=+0.49\pm0.21$) and is flagged without a group. No star is C-depleted: the signature of ongoing RGB mass transfer \citep{Stancliffe2007} is absent, in contrast with 47~Tucanae, where \citet{Ferraro2006} found C- and O-depleted BSS and interpreted them as accretors of material from the deep layers of a now-evolved companion.
\item Marginal/unclassified: BSS~14, 16, 25, 40 ($\Delta[\mathrm{C/Fe}]\approx2\sigma_{\rm MAD}$, normal O) and BSS~35 (unreliable single measurement). In total, 13 of the 21 BSS with C and O measurements (62\%) show significant surface pollution, a direct lower limit on the mass-transfer fraction among the slow rotators of Cr~261.

\end{itemize}
\subsubsection{Are the anomalies temperature-driven?}
\label{sec:teff_trends}
Before we can read the groups as accretion signatures, we must rule out
one alternative. Atomic diffusion and radiative levitation change the
surface composition of stars with a thin convective envelope or none at
all, and BSS are the hottest stars of an old cluster.
\citet{Lovisi2012} report the first detection of this effect in blue
stragglers, in the stars of NGC~6397 hotter than 8200~K. In
Population~I main-sequence stars the Am/Fm phenomenon produces similar
anomalies \citep{Preston1974,Richer2000}. If our groups followed
$T_{\rm eff}$ instead of donor type, the chemical argument would not
hold.

Figure~\ref{fig:teff_trends}, in Appendix~\ref{app:teff_trends}, tests this. The Spearman coefficients are $+0.15$
($p=0.51$) for $\Delta[\mathrm{Fe/H}]$, $-0.05$ ($p=0.85$) for
$\Delta[\mathrm{C/Fe}]$ and $+0.04$ ($p=0.87$) for
$\Delta[\mathrm{O/Fe}]$, over the 21--22 slow rotators with
measurements. Iron is the most important test here, because it is the
first element that radiative levitation lifts and it is the element
\citet{Lovisi2012} found altered. It is flat between 5984 and 7406~K,
at the level of our 0.15~dex systematic floor. The fitted slopes are
$+0.18$, $-0.11$ and $+0.08$~dex per 1000~K for Fe, C and O, and all
three agree with zero at 95\% confidence. The small positive slope in
iron comes from the four Group~A stars alone, and if we remove them it
becomes negative ($-0.10$~dex per 1000~K). The slope is therefore set by the excess of these four stars, not by a trend with temperature. Finally, every star in the abundance sample is at least 800~K cooler than the temperature at which the effect has been seen in BSS.

Barium is a different case and needs more care. The nine stars with a Ba
measurement give $\rho_{\rm S}=+0.61$ ($p=0.08$), and three of the four
Group~A stars are in the warmer half of the sample. This correlation
does not pass its own controls. If we remove any single star,
$\rho_{\rm S}$ changes between $+0.52$ and $+0.73$, so we are not
measuring a stable trend. A Mann--Whitney test restricted to the nine
stars that have Ba coverage finds no temperature difference between
Group~A and the rest ($p=0.41$). Two stars are enough to show the
problem. BSS~26 is the hottest star in the abundance sample (7406~K),
but its C and O are normal and its barium is only $+0.49$. BSS~29 is at
6378~K, below the median of the full BSS sample, and reaches $+1.26$. A
mechanism that depends on temperature cannot produce this order.

BSS~26 needs a separate comment, because it is the hottest star of the
abundance sample and the only one with normal C and O together with a
moderate barium excess ($+0.49\pm0.21$). This is why we keep it outside
Group~A and flag it without a group (Sect.~\ref{sec:chem_groups}). Its
barium should indeed be treated with care, but none of our conclusions
depend on it: Group~A requires $[\mathrm{Ba/Fe}]>+1.2$ together with
enhanced carbon, and BSS~26 satisfies neither condition. If part of its
barium excess were produced by diffusion, the star would simply move
from the group of chemically normal BSS to the list of stars we cannot
classify.

The pattern is also wrong for diffusion. In Am/Fm stars diffusion
removes Ca and Sc and enriches the iron peak and the $s$-process
elements, but it does not enrich carbon and oxygen at the same time
\citep{Preston1974,Richer2000}. The correlated C and O excesses of
Sect.~\ref{sec:chem_groups} cannot be produced in this way. Group~B is
the clearest example, with $\Delta[\mathrm{C/Fe}]$ up to $+0.8$~dex and
$\Delta[\mathrm{O/Fe}]$ up to $+1.3$~dex at
$[\mathrm{Ba/Fe}]\leq+0.25$, over the range 6030--6885~K. We therefore
attribute the groups to accreted material. With only nine Ba
measurements we cannot exclude a small diffusive contribution at the hot
end of our range, and the absolute $[\mathrm{Ba/Fe}]$ values of Group~A
should be treated with care until we have HR12 coverage for a larger
part of the sample.

\subsubsection{Where the chemical groups sit in the colour--magnitude diagram}
\label{sec:cmd_groups}
The right-hand panel of Fig.~\ref{fig:cmd_vsini} shows the CMD of \Cr\
with the 21 BSS that have measured C and O marked by chemical group.
The 19 stars without an abundance analysis (the 17 fast rotators and BSS~7, plus
BSS~35, whose single measurement is unreliable) are plotted as grey
crosses.

The groups do not separate. Group~B, the chemically normal stars and
the stars without Ba coverage share the same part of the BSS region,
and a Mann--Whitney test finds no difference between the polluted stars
and the chemically normal ones, either in colour ($p=0.84$) or in
magnitude ($p=0.20$).

Group~A is the one exception. Its four stars have a median colour of
$G_{\rm BP}-G_{\rm RP}=0.86$ against $1.03$ for the other slow rotators
($p=0.026$), whereas their magnitudes are the same ($p=0.30$). This is
not new information. Our temperatures come from the projection of the
CMD position onto the PARSEC tracks (Sect.~\ref{sec:stellar_params}),
so ``Group~A is bluer'' and ``Group~A is hotter''
(Sect.~\ref{sec:teff_trends}) are one measurement seen twice, and the
tests of Sect.~\ref{sec:teff_trends} already show that it does not
produce the barium.

The offset does have a natural reading in terms of mass. The Group~A
pattern needs a donor above the third dredge-up limit
($M\gtrsim1.5$~\msun; \citealt{Karakas2014}), so the progenitor system
had to be more massive than average, and the accretor should be more
massive too. The four Group~A stars have a median mass of $1.54$~\msun\
against $1.30$~\msun\ for the other slow rotators. With four stars the
difference is not significant ($p=0.17$). It is carried by BSS~29,
which is the least massive of the four ($1.14$~\msun) and the same star
we flag in Sect.~\ref{sec:tess}; the other three give $p=0.03$.
Removing a star after seeing the result is not a test, and we quote
that number only to show where the scatter comes from. A larger Ba
sample is what would settle this.

\section{Discussion}
\label{sec:discussion}
\subsection{Two angular momentum histories}
\label{sec:bimodality}
The $v\sin i$ distribution (Fig.~\ref{fig:vsini_distribution}) is strongly asymmetric: a dominant core of slow rotators ($v\sin i\lesssim40$~\kms), sharing the spun-down regime of the TO/SGB reference stars, and an extended fast tail reaching the detection limit. We adopt $v\sin i=40$~\kms\ as the operational boundary between the two regimes; the conclusions below are insensitive to its exact placement ($\pm10$~\kms\ moves individual stars but not the populations). This structure cannot arise from a single spun-down population: measurement uncertainties ($\sim$10~\kms) and projection effects can smear a distribution but cannot generate a 44\% tail extending to $\sim$100~\kms. Following \citet{Ferraro2023}, fast rotation is the signature of recently formed BSS regardless of channel: both mass-transfer and collisional products are born spinning rapidly. While MT products are braked by their convective envelopes on $\tau_{\rm sd}\sim0.3$--1~Gyr \citep{Sills2005,Glebbeek2008}, the braking of collisional products (which develop no convective envelope) remains essentially unconstrained. The two regimes thus separate the population by age since formation. In the low-density environment of \Cr, direct collisions are negligible \citep{Davies2004}, so both must ultimately descend from binary or triple systems. One caveat applies to the rest of this section. The clock only works for stars that are free to spin down. In Sect.~\ref{sec:tess} we find seven BSS whose rotation is fixed by a close companion and we exclude them whenever we use the distribution as an age indicator.

\subsection{Formation channels: mass transfer and tidal spin-up}
\label{sec:channels}
The binary fraction we measure is $f_{\rm SB1}=38\%$ (15/40). This value
is well above what is expected for turn-off stars. In NGC~188, the BSS binary fraction is $76\%\pm19\%$, whereas the turn-off stars of the same cluster give about 20\%, and M67 shows the same difference \citep{Mathieu2009,Geller2012}. Our value is a lower limit, because our RV cuts ($|\Delta\mathrm{RV}|>20$~\kms\ at one epoch, or $\sigma_{\rm RV}\gtrsim8$~\kms\ over a 63-day window) only detect short-period, high-amplitude orbits. The orbits expected after mass transfer from an AGB donor, of several hundred to a few thousand days \citep{Geller2012}, are almost invisible to these cuts: a $1.1+0.5~M_\odot$ pair at 700~d gives $K\simeq9$~\kms. A correction based on an assumed period distribution would therefore be very uncertain. The surface chemistry gives a more direct lower limit. Twelve BSS that are not SB1 carry a chemical signature of mass transfer (Groups A and B, and the C/O-enhanced stars without Ba coverage; Sect.~\ref{sec:chem_groups}). Together with the 15 SB1, at least 27 of the 40 BSS (68\%) are, or have been, in a binary; if we count only the firm Groups A and B, the number is 22 (55\%). Most BSS in \Cr\ therefore have a companion, as the mass-transfer channel requires.

The real fraction may be higher still. Photometric studies of open
clusters show that many members are neither single stars nor simple
binaries. Triples and higher-order systems contribute to the luminosity
function, and if they are ignored both the cluster mass and the
multiplicity fraction are biased \citep{Borodina2021,Malofeeva2023}.

We measure 17 fast rotators, or $44\%\pm8\%$ of the sample. Not all
of them can be used to date the formation of the BSS. In
Sect.~\ref{sec:tess} we find that seven BSS with a photometric orbital
period rotate at the speed set by their orbit (Table~\ref{tab:sync}),
and all seven are in the fast tail. For these stars $v\sin i$ measures
the orbital period, not the time since formation, so the argument of
\citet{Ferraro2023} does not apply to them. If we remove them, 10 of
the remaining 32 BSS are fast rotators, or $31\%\pm8\%$. A constant
formation rate with $\tau_{\rm sd}\sim250$~Myr gives only 10--20\%
fast rotators \citep{Sills2005,Leiner2018}. Even after removing the
tidally locked systems, our fraction is still above this range, so it
needs either a recent increase of the formation rate or longer
spin-down times.

The same observations bear on the hierarchical-triple channel. That
13 of the 14 SB1 are fast rotators could be read as evidence for
triples, in which Kozai--Lidov cycles drive the inner pair to merge
and leave a fast rotator bound to a distant third star
\citep{Perets2009,Naoz2014,Tokovinin2014}. Such a reading would imply
a triple fraction of 13/14 among the BSS, well above the $\sim$10\%
measured for field F and G stars \citep{Tokovinin2014,Moe2017}, and
it invokes a mechanism that is not needed here: tidal synchronisation
already accounts for these stars with close binaries alone. Triples
are certainly present in \Cr, and the systems combining a long period
with a fast $v\sin i$ remain natural candidates for them, but the
rotation data do not call for a third body.

Four fast rotators are left. Seven of the 17 are the synchronised
systems described above, and six have short-period signals that are not
orbital (Sect.~\ref{sec:tess}). The remaining four show no
photometric signal at all. Tides cannot explain these, so their
rotation is probably what a mass-transfer or merger episode produced.
They are the only stars for which the spin-down clock still works, and
four stars are too few to measure a formation rate. Longer TESS
coverage, or the orbital solutions we are now deriving, will show
whether they are also short-period binaries that we see without
eclipses.

\begin{figure*}[ht!]
\centering
\includegraphics[width=0.9\textwidth]{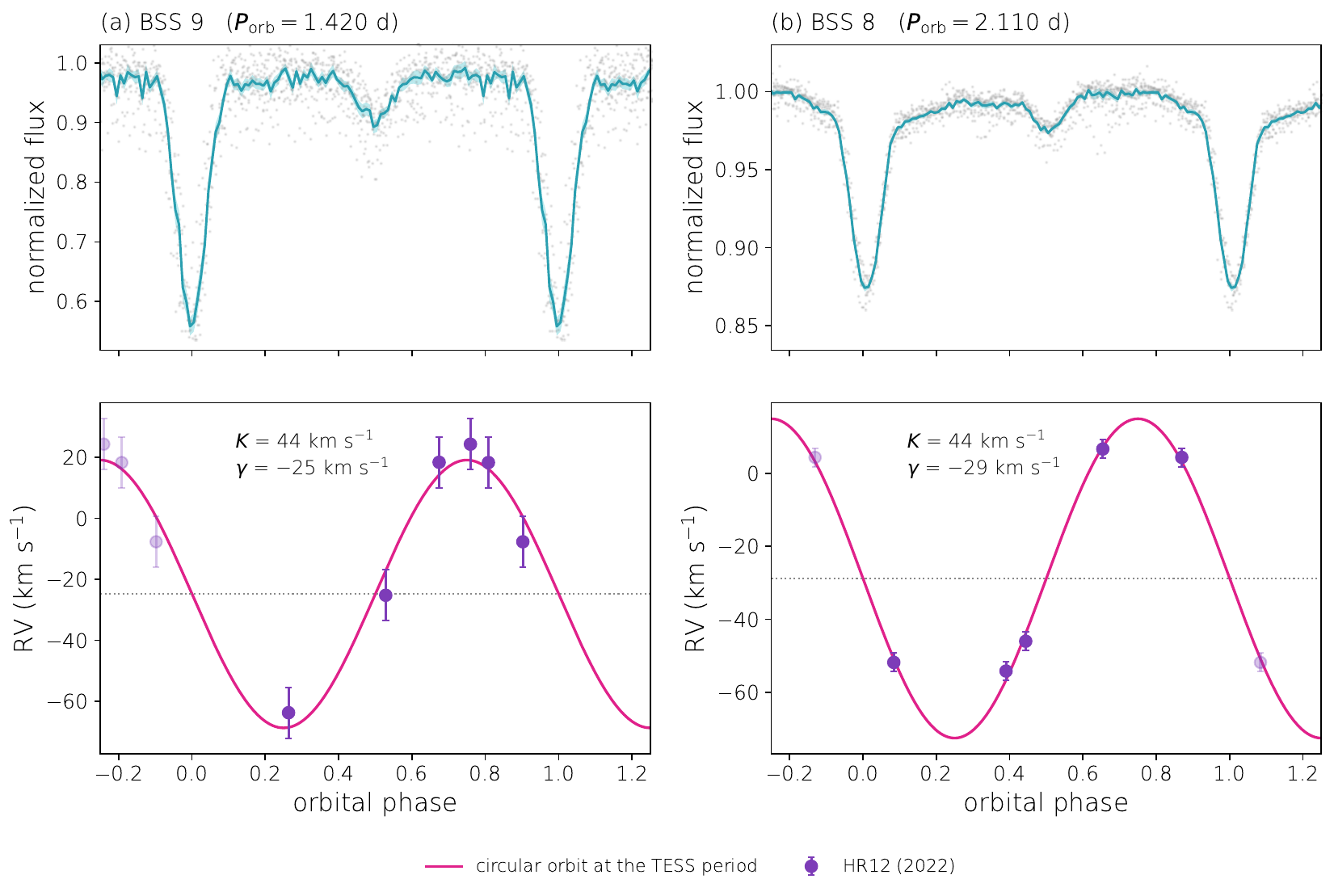}
\caption{TESS/PATHOS light curves (top) and HR12 radial velocities (bottom) of (a) BSS~9 and (b) BSS~8, folded on $P_{\rm orb}=1.420$ and 2.110~d. Phase 0 is the deeper eclipse. The pink line is a circular orbit at the TESS period, placed so that the velocity crosses the systemic value (dotted line) downwards at phase 0. Faded symbols repeat the same epochs.}
\label{fig:tess_rv}
\end{figure*}

\subsection{Short-period variability: pulsators and synchronised binaries}
\label{sec:tess}
We cross-matched the BSS sample with the PATHOS light curves of TESS
Sector~11 \citep{Nardiello_2019,Nardiello_2020}. Sixteen of the 40 BSS
show a significant short-period signal. The full variability analysis
and classification of these signals is the subject of the companion
paper \citep{Echeveste2026}; here we use only the part that matters for
the rotational velocities: eight stars whose light curves show eclipses
or ellipsoidal modulation, so that the photometric period traces the
orbit. The remaining signals are pulsational or instrumental in origin;
none of them traces an orbit, and we exclude those stars from the rest
of this section.

For the eight binaries the orbital periods lie between 0.52 and
2.11~d, once we take into account that an eclipsing or ellipsoidal
light curve repeats twice per orbit ($P_{\rm orb}=2P_{\rm LS}$). Figure~\ref{fig:tess_rv} shows the two clearest cases, BSS~9 and BSS~8, with primary and secondary eclipses at $P_{\rm orb}=1.42$ and 2.11~d. We measured the velocities of their 2022 HR12 spectra by cross-correlation against the spectrum of BSS~15, a constant slow rotator of similar temperature, broadened to the $v\sin i$ of each star. A circular orbit at the TESS period gives $K\simeq44$~\kms\ for both stars, and systemic velocities of $-25$ and $-29$~\kms, close to the cluster value. To align the light curve and the velocities, we place the descending crossing of the systemic velocity at the deeper eclipse, where the BSS is behind its companion in a circular orbit. The mass functions are $f(m)\simeq0.012$ and $0.018~M_\odot$, which for the track masses (1.34 and 1.76~$M_\odot$) and $i\simeq90^\circ$ give companions of about 0.33 and 0.45~$M_\odot$. The light curves also say something about the companions. The primary eclipses are 44\% and 12\% deep and the secondary eclipses are clearly detected, so each companion covers a large part of the BSS disc and contributes a significant fraction of the light: neither is a cool, compact white dwarf. The light curves are of Algol type. With the companion masses above, the BSS lie well inside their Roche lobes ($R/R_{\rm L}\simeq0.55$ and 0.64), but the companions could fill theirs ($R_{\rm L}\simeq1.7$ and 2.4~$R_\odot$), so we cannot exclude that mass transfer from the companion is still going on. With five or six epochs these values are indicative, and the full orbital solutions will be presented in a forthcoming paper.

The TESS light curves can also test the $v\sin i$ of the slow rotators, if star spots modulate the flux at the rotation period. \citet{Echeveste2026} report five tentative variables with $P=3.4$--4.7~d that do not show eclipses (BSS~5, 26, 27, 32 and 39). For BSS~5 and 39 these periods cannot be rotation periods: they give equatorial velocities of 24 and 31~\kms, well below the measured $v\sin i\gtrsim100$~\kms. For BSS~26, 27 and 32 they give $v_{\rm eq}=25$--33~\kms, consistent with the measured $v\sin i$ ($10\pm8$, $13\pm9$ and $<24$~\kms), so rotational modulation is possible. These periods come from a single 27-day sector, however, and they change with the photometric aperture, so we do not use them further. For the other slow rotators no signal is found; for typical inclinations their $v\sin i$ and radii imply periods longer than about 4~d, where a single sector is dominated by instrumental trends.
These seven stars allow a test that $v\sin i$ alone cannot provide. If tides have synchronised the rotation with the orbit, the equatorial velocity must be
\begin{equation}
v_{\rm sync} = \frac{2\pi R}{P_{\rm orb}},
\label{eq:vsync}
\end{equation}
where we take $R$ from the spectroscopic gravity and the track mass
through $R/R_\odot = 10^{(\log g_\odot + \log M/M_\odot - \log g)/2}$.
The test has no free parameters. Table~\ref{tab:sync} compares the prediction with the measurement, and all seven agree.
The agreement holds over a factor of three in $v_{\rm sync}$, the four
lower limits are also consistent, and the measured velocities follow
$P_{\rm orb}$ in the expected sense. The two quantities come from
independent data, one from a light curve taken in 2019 and the other
from an iron line observed six years later, so a chance agreement is
unlikely. At these separations the circularisation and synchronisation
times are much shorter than the age of the cluster
\citep{Zahn1977,Hurley2002}, and locked rotation is what we expect.

The eighth star, BSS~29, does not fit. We measure
$v\sin i=36\pm11$~\kms, whereas its period of 0.982~d predicts
84~\kms. This would need $i\simeq25^{\circ}$, which is not possible
for an eclipsing star. Its five HR12 epochs also have a peak-to-peak
amplitude of only 8.8~\kms, too small for a one-day orbit. In
addition, a source 7\arcsec\ away has the same period, and the two
stars are not separated at the TESS pixel scale (21\arcsec). The
period of BSS~29 is most likely wrong, and we remove the star from
Table~\ref{tab:sync}.

We also checked whether the radial velocities agree with the TESS
periods. For a one- to two-day orbit seen close to edge-on, a
0.25~\msun\ companion gives $K\simeq25$--45~\kms, and five random
epochs should span about $1.6K$. BSS~8 and 9 agree with this
(${\rm ptp}_{\rm RV}=61$ and 88~\kms), and BSS~3 (34~\kms)
agrees with a lighter companion or a lower inclination. BSS~12 and 33
(17 and 26~\kms) are below the prediction, and BSS~17 and 28 (only
HR18) show offsets of $+99$ and $+165$~\kms\ that no plausible
companion can produce. These four are among our broadest-lined stars,
where cross-correlation velocities are least reliable
(Sect.~\ref{sec:sb1_rv}). In BSS~33, which fills its Roche lobe, the
lines of the two stars may also be blended. Their binarity therefore
rests on the light curves. The lower limit of BSS~12
($\gtrsim100$~\kms) is above its $v_{\rm sync}=85$~\kms, but at these
line widths the uncertainty is of the same size as the value. A
spectroscopic follow-up is needed to confirm these periods.

The synchronised systems are post-mass-transfer binaries.
Table~\ref{tab:sync} gives the Roche-lobe radius of each BSS from the
approximation of \citet{Eggleton1983}, for a companion of
0.25~\msun, the typical mass of a helium white dwarf left by mass
transfer on the red giant branch. Stable mass transfer can also leave BSS in orbits of hundreds to thousands of days with a white-dwarf companion \citep{MathieuPols2025}. The short periods of our seven systems point instead to mass transfer that started early, or to a common-envelope phase. BSS~33 is larger than its Roche
lobe ($R/R_{\rm L}=1.14$, and 1.23 for a 0.5~\msun\ companion), so it
is probably a contact system in which mass transfer may still be
active. The other six stars fill 54--82\% of their Roche lobes. They
are detached today, but their orbits are so tight that one of the two
stars must have filled its Roche lobe in the past. Their fast rotation
comes from the orbit that this interaction left, and tides have kept
it since. This is why we cannot use it to date the formation event.

\subsection{Orbital periods compared with M67 and NGC~188}
\label{sec:periods}
In M67 and NGC~188, 25 of the 41 BSS have orbital solutions \citep{Linck2026a}. Only four have periods below 10~d, none has a period between 10 and 100~d, and most of the others have periods of several hundred to a few thousand days, with a median of about 1200~d for the systems with an AGB donor. \citet{Linck2026a} conclude that at least two thirds of these BSS are mass-transfer products, and that mass transfer tends to move the orbits above 100~d.

At first sight \Cr\ looks different, with many short-period systems. Much of this comes from the way the two samples were observed. Our RV cuts detect mainly orbits shorter than about 100~d (Sect.~\ref{sec:channels}), and TESS finds only eclipsing and ellipsoidal systems of a few days, whereas the WIYN survey covers several decades and finds the long orbits. If we count only the systems with $P<10$~d, which both surveys detect, \Cr\ has 7 of 40 BSS ($18\%\pm6\%$), against 4 of 41 ($10\%\pm5\%$) in M67 and NGC~188. The difference is not significant (Fisher test, $p=0.35$).

A real difference may come from the other SB1. Nine of them (BSS~7, 18, 19, 20, 21, 24, 32, 34 and 38) have no TESS period, but our detection limits suggest periods shorter than a few hundred days. If some of them have periods between 10 and 100~d, they fall in the range that is empty in M67 and NGC~188. Their orbits will be presented in a forthcoming paper. The long-period population is probably present in \Cr\ as well: twelve BSS that are not SB1 carry a chemical signature of mass transfer (Sect.~\ref{sec:chem_groups}), as expected for wide post-AGB orbits that our RV data cannot detect.

\subsection{Chemical confirmation}
\label{sec:chem_discussion}
The abundance census (Sect.~\ref{sec:chem_groups}) tests this picture independently. Among the 21 slow rotators with C and O measurements, 62\% show the surface-pollution signature of accreted donor material. This exceeds the SB1 fraction, and only one polluted slow rotator (BSS~32) is an SB1: the chemically polluted stars are the long-period post-mass-transfer binaries that our RV cuts cannot detect (Sect.~\ref{sec:channels}). Mass transfer dominates the spun-down population. Group~A traces AGB donors with active third dredge-up; Group~B (separated by an empty $\sim$1~dex gap in [Ba/Fe]) reveals the sub-$s$-process donor regime. The four chemically normal slow rotators are natural candidates for old, spun-down merger remnants, the aged counterparts of today's fast tail. No star shows the C and O depletion expected for CNO-processed material from an RGB donor \citep{Ferraro2006}. The fraction of Ba-rich stars is similar to that found in other old open clusters. Four of the nine BSS with a Ba measurement belong to Group~A (44\%). \citet{Nine2024} find 15 Ba-enriched BSS out of 35 (43$\pm$11\%) in NGC~7789, NGC~6819 and M67, whereas none is found in NGC~188 \citep[][and references therein]{Nine2024}. In their sample, about 40\% of the Ba-enriched BSS are in spectroscopic binaries with periods up to 5000~d. None of our Ba-rich stars is an SB1, as expected if their orbits are longer than our RV data can detect (Sect.~\ref{sec:channels}).
The C--O plane also marks a sharp contrast with globular clusters.
There, the chemically anomalous BSS show depleted C and O ---
CNO-processed material from an RGB donor, as first found in 47~Tucanae
\citep{Ferraro2006}. In \Cr\ we observe the opposite: a correlated
C--O enrichment sequence capped by $s$-process barium, the
signature of third dredge-up material from AGB donors, with not a
single C-depleted star in the sample. The dichotomy plausibly traces
the dynamical environment: in dense globulars, encounters harden and
shrink binaries, forcing Roche-lobe overflow while the donor is still
on the RGB; in the low-density environment of \Cr, binaries survive
wide enough for the donor to reach the AGB before transferring mass.
The C--O diagram of a BSS population may therefore serve as a
record of the dynamical history of its host cluster. This picture is probably too simple. \citet{Linck2026b} derive the hard--soft period boundary from the velocity-dispersion profile of several globular clusters, and find that in 47~Tuc the BSS of the core should come mostly from main-sequence and RGB interactions, whereas AGB mass transfer is still possible in the outer parts. However, \citet{Ferraro2006} found no difference in radial distribution between the C/O-depleted BSS of 47~Tuc and the others, so the donor stage may not map directly onto C and O. The C and O excesses of Group~B are also not fully explained. In low-mass AGB stars only a few weak third dredge-up episodes take place, and they may raise carbon before much $s$-process material reaches the surface \citep{Karakas2014}, but the large O excesses are harder to explain and we cannot identify their origin with the present data.
\subsection{Rotation and the dynamical state of the cluster}
\label{sec:dynamical}
\begin{figure*}[ht]
\centering
\includegraphics[width=\textwidth]{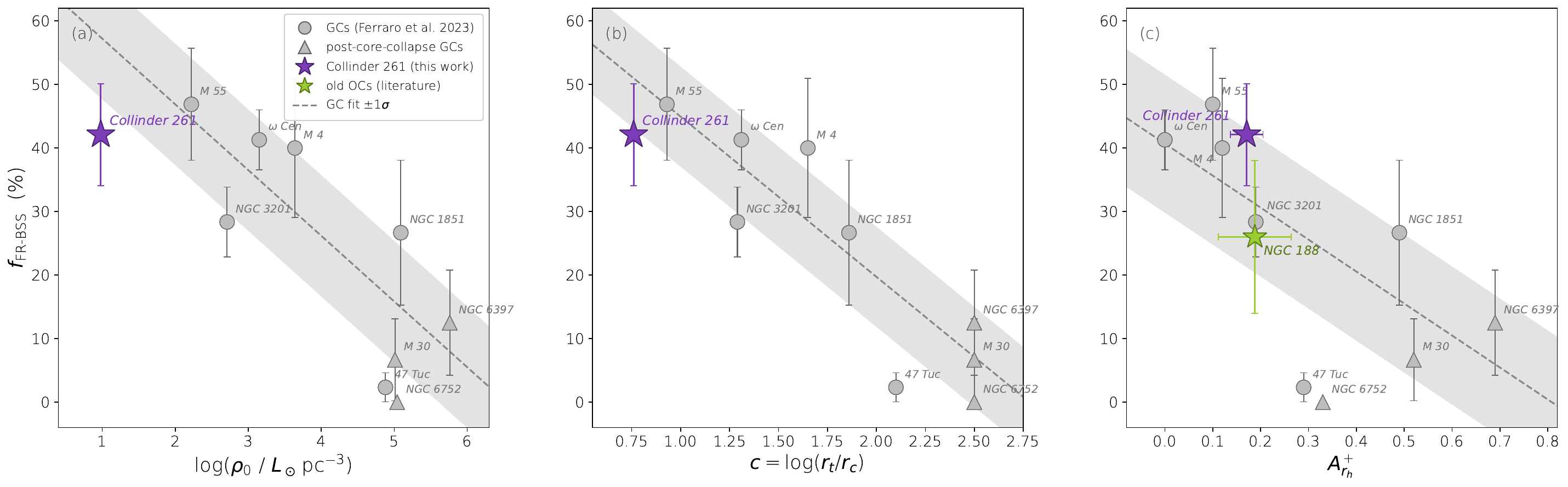}
\caption{Fraction of fast-rotating BSS ($v\sin i\geq40$\,\kms) as a function of (a) central luminosity density, (b) concentration, and (c) $A^+_{r_h}$. Grey symbols: globular clusters from \citet{Ferraro2023} and NGC\,1851 from \citet{Billi2025}; the dashed line and band are the GC fit and its $\pm1\sigma$ dispersion. Purple star: \Cr\ (this work, with $A^+_{r_h}$ from \citealt{Rao2023}). Green star in (c): NGC\,188 \citep{Linck2026b,Rao2023}.}
\label{fig:comparison}
\end{figure*}

For the BSS segregation parameter $A^+_{r_h}$ \citep{Alessandrini2016,Lanzoni2016} we adopt the measurement of \citet{Rao2023}, $A^+_{r_h}=0.171\pm0.034$, obtained from the photometric, bona-fide BSS population of the cluster and a \textit{Gaia}-selected reference sample. We prefer this external value to one based on our own targets, because the FLAMES fibre allocation does not sample the cluster uniformly.\footnote{Measuring $A^+$ with our 40 spectroscopic BSS against the 42 FLAMES TO/SGB stars gives $A^+_{\rm spec}=0.05\pm0.06$, consistent with zero. This is a geometric null, not a physical one: both samples share the same fibre allocation, which prefers the outer regions, so their cumulative radial distributions agree by construction. We quote it only to make this selection effect explicit.} This places \Cr\ in the dynamically intermediate regime (Family\,II of \citealt{Ferraro2012}), corresponding to
several half-mass relaxation times of age and consistent with the pronounced main-sequence mass segregation already reported by \citet{Bragaglia2006}. The non-zero $A^+$ confirms that BSS are more massive ($\langle M\rangle\sim1.3$--$1.6$\,\msun) than the TO/SGB reference (${\sim}1.1$\,\msun) and that two-body relaxation, not
direct collisions, is the dominant kinematic sorting agent.
\Cr\ sits at the low-density extremum of the \citet{Ferraro2023} sample, with central luminosity density $\log\rho_0\ll 2$ (in $L_\odot\,{\rm pc}^{-3}$) and King concentration parameter $c\equiv\log(r_{\rm t}/r_{\rm c})\simeq0.77$. Our fraction of fast-rotating BSS is $f_{\rm FR-BSS}=44\%\pm8\%$\footnote{$f_{\rm FR-BSS}\equiv N(v\sin i \geq v_{\rm crit})/N_{\rm BSS}$ with $v_{\rm crit}=40$\,km\,s$^{-1}$} and falls within the $\pm1\sigma$ band of the GC linear fit in $\log\rho_0$, in $c$, and in $A^+$ (Fig.~\ref{fig:comparison}), demonstrating that the GC trends extend seamlessly into the OC regime. \Cr\ occupies a previously sparsely
populated region of the $A^+$--$f_{\rm FR-BSS}$ plane: dynamically intermediate and rotation-rich. This decoupling has two consequences: (i) direct stellar collisions are negligible in this density regime \citep{Davies2004}, so the high $f_{\rm FR-BSS}$ must come from binary-driven formation, that is mass transfer in close binaries and not collisions (Sect.~\ref{sec:channels}); part of the fast tail is also kept fast by tidal locking rather than by youth (Sect.~\ref{sec:tess}), consistent with the chemistry of the slow population, where 62\% show the pollution signature of mass transfer (Sect.~\ref{sec:chem_discussion}); (ii) the formation channel operates continuously rather than in bursts, because a steady rate predicts only 10--20\% fast rotators for the standard $\tau_{\rm sd}\sim250$\,Myr \citep{Sills2005,Leiner2018}, whereas the observed fraction, 31\% once the tidally locked systems are removed, still requires either a recent enhancement of the formation rate or a longer effective spin-down in BSS --- the latter plausible given the unconstrained braking of merger products
(Sect.~\ref{sec:bimodality}).
Together with the intermediate $A^+_{r_h}=0.17$, which places \Cr\ in the early-to-mid dynamical regime of the \citet{Ferraro2018,Ferraro2023} calibration, this promotes \Cr\ from a peripheral data point to a strategic one: it is the strongest evidence to date that the $f_{\rm FR-BSS}$--$A^+$ relation is a
continuous physical sequence rather than a GC-internal feature. NGC~188 provides a direct test, and the data to run it now exist. Its BSS binary fraction is consistent with 100\%, with long orbital periods \citep{Mathieu2009,Geller2012}, so its population should sit at the slow-rotating end of the sequence. Recent WIYN/Hydra measurements \citep{Linck2026b} give 5 fast rotators out of 19 BSS, or $f_{\rm FR-BSS}=26\%\pm12\%$, and \citet{Rao2023} measure $A^+_{r_h}=0.188\pm0.076$ for the cluster. The point falls on the GC relation in panel~(c) of Fig.~\ref{fig:comparison}. One methodological difference matters for the comparison: \citet{Linck2026b} exclude binaries with known orbital periods below $\sim$60~d, precisely because tides can spin those stars up. Our 44\% includes them. If we exclude our seven synchronised systems in the same way, we are left with 10 fast rotators out of 32, or $f_{\rm FR-BSS}=31\%\pm8\%$ in good agreement with NGC~188. The difference between this number and our full $44\%\pm8\%$ is the tidal contribution identified in Sect.~\ref{sec:tess}, measured in a second, independent way. The same survey reports $44\%\pm17\%$ for M67 and $13\%\pm6\%$ for NGC~6791 \citep{Linck2026b}, so the open-cluster end of the sequence is beginning to fill in.

\subsection{Spatial distribution of fast and slow rotators}
\label{sec:radial_fr_slow}
\begin{figure}[t]
\centering
\includegraphics[width=\columnwidth]{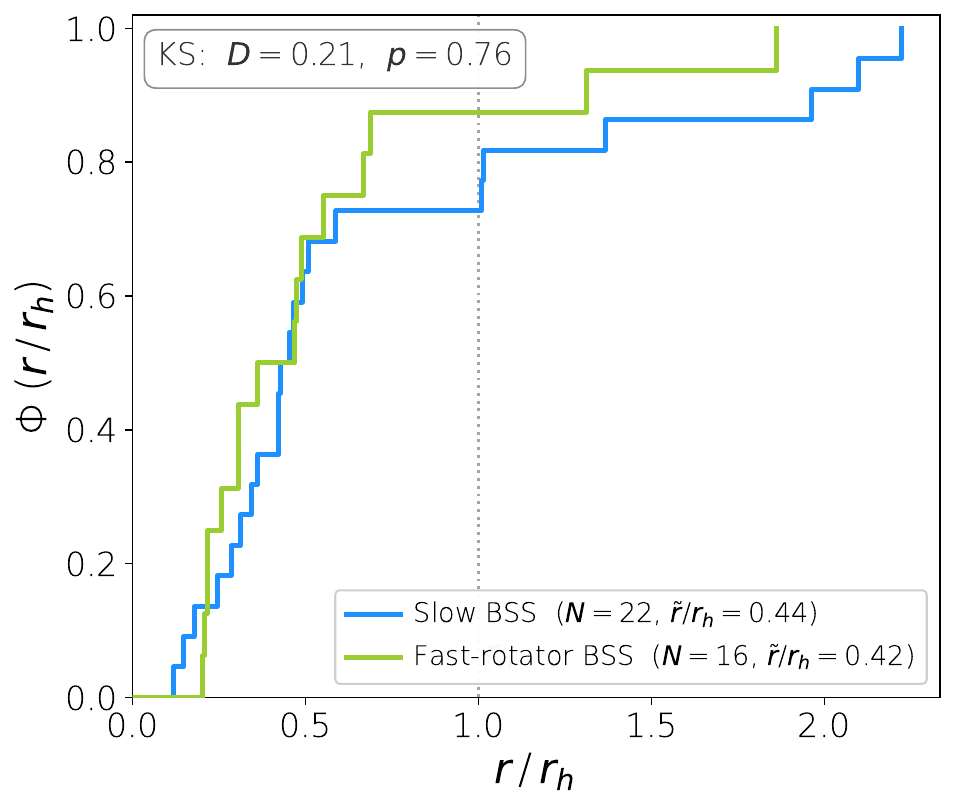}
\caption{Cumulative radial distribution of the fast ($N=16$) and slow ($N=22$) BSS, with the radius in units of the half-light radius ($r_h=5.84'$; \citealt{Gao2023}). A Kolmogorov--Smirnov test gives $D=0.21$ and $p=0.76$.}
\label{fig:radial_fr_slow}
\end{figure}

The cluster-wide $A^+_{r_h}=0.171$ \citep{Rao2023} shows that the BSS as a whole are mass-segregated relative to the photometric reference sample (Sect.~\ref{sec:dynamical}). A complementary question is whether the two rotational regimes of Sect.~\ref{sec:bimodality} (fast and slow rotators) also differ in their radial distributions. Figure~\ref{fig:radial_fr_slow} shows the cumulative distribution  $\Phi(r/r_h)$ of the two sub-samples, where $r_h=5.84'$ is the half-light radius obtained by integration of the King profile of \citet{Gao2023}. The two distributions are statistically indistinguishable (two-sample Kolmogorov--Smirnov test: $D=0.21$, $p=0.76$); the medians are nearly
identical ($\tilde{r}/r_h = 0.42$ versus $0.44$).
This null result is in fact the expected outcome in the framework of Sect.~\ref{sec:channels}. Both rotational regimes ultimately descend from primordial binary (or triple) systems of comparable total mass, which have undergone the same cluster-age dynamical segregation as systems, regardless of when each of them produced its BSS. The rotational split therefore encodes formation epoch, not birthplace: a recently formed fast rotator inherits the radial position its progenitor binary had already reached. A rotation-dependent spatial pattern would only be expected if one channel operated preferentially in a distinct density environment (for instance collisions in a dense core) which the low central density of \Cr\ rules out. The spatial uniformity is thus not evidence against the two-population picture but a consistency check of its binary origin: what segregates in \Cr\ is the binary population itself, whose BSS offspring appear at all radii with whichever rotation their age since formation dictates.

\subsection{The rotational floor of the reference sample}
\label{sec:bss_msto}

One reference star (\textit{Gaia}\,DR3 5856530891339792000) has
$v\sin i=26\pm24$~\kms\ and lies about 0.15~mag bluer and 0.05~mag
brighter than the isochrone ridge. Its spectrum shows that it is a
double-lined binary. The cross-correlation function is broader than
for the other reference stars and asymmetric, and the Fe\,\textsc{i}
lines are split into two components separated by about 25--35~\kms,
on both sides of the cluster velocity (Fig.~\ref{fig:bl_sb2}). The
two components have similar brightness, and the measured $v\sin i$
comes from the blend of the two profiles, not from fast rotation.

\section{Conclusions}
\label{sec:conclusions}

We presented the first dedicated $v\sin i$ survey of BSS in \Cr, combining rotation, multi-epoch radial velocities, and carbon, oxygen, barium, and iron abundances for 39 BSS and 42 TO/SGB reference stars. Our main results are:
\begin{itemize}
    \item[(1)] The BSS rotational distribution splits into a spun-down core and a 44\% fast tail with no counterpart among the TO/SGB reference stars, and fast rotation is strongly linked to binarity (Fisher $p=4\times10^{-6}$).
    \item[(2)] Surface abundances, measured strictly differentially against the TO/SGB reference sample, reveal two mass-transfer signatures: i) a Ba-rich group tracing AGB donors and a ii) C/O-enhanced, Ba-normal group tracing sub-$s$-process donors, separated by an empty 1~dex gap in [Ba/Fe] with 62\% of the slow rotators showing surface pollution.
    \item[(3)] An Fe null test confirms a common metallicity for the BSS; the few apparent Fe excesses are partly due to differential reddening and do not affect the abundance pattern.
    \item[(4)] The C--O enrichment sequence contrasts with the C--O depletion seen in globular clusters, suggesting the C--O diagram traces the dynamical environment of the host cluster.
    \item[(5)] $A^+_{r_h}=0.171\pm0.034$ \citep{Rao2023} and $f_{\rm FR-BSS}=44\%\pm8\%$ place \Cr\ on the extrapolated GC trends of Ferraro et al., extending them into the open-cluster regime and indicating that the engine of the BSS population is a binary channel that operates continuously: mass transfer in close binaries, with tides keeping the shortest-period systems fast.
    \item[(6)] A cross-match with TESS/PATHOS photometry independently confirms the binary nature of the SB1 population: six SB1 are eclipsing or ellipsoidal systems with $P_{\rm orb}=0.5$--2.1~d, and for the two cleanest cases (BSS~9 and BSS~8) the HR12 velocities folded on the TESS period yield circular-orbit amplitudes of $K\simeq44$~\kms\ for both, implying companions of about 0.33 and 0.45~$M_\odot$ in tight post-mass-transfer orbits; their light curves show that neither companion is a cool, compact white dwarf. The full orbital characterisation of the SB1 sample is ongoing and will be presented in a forthcoming paper.
\end{itemize}
A natural next step is to confront the observational picture assembled here with quantitative binary-evolution models. The joint constraints we now have for each star (rotation, radial-velocity variability, and surface C, O and Ba) are precisely the quantities that population-synthesis calculations of mass transfer predict, and their combination breaks degeneracies that any single diagnostic leaves open. \citet{Rain2024} have shown that composite evolutionary sequences built from binary-evolution models reproduce the observed BSS distributions in the colour--magnitude diagrams of old open clusters, and that the size of the BSS
population scales with the binary content of the host cluster. Extending that approach to the individual level (tracking the expected surface rotation and chemical pattern along each mass-transfer sequence, and comparing star by star rather than population by population) would allow the formation channel of each Collinder~261 straggler to be tested rather than inferred. Work in this direction is in progress \citep{Echeveste2026}. A complementary observational step is a dedicated spectroscopic monitoring campaign to derive orbital solutions for the SB1 systems identified here, since orbital periods and mass functions provide
the most direct discriminant between case~A, case~B and wind-accretion scenarios. The same combination of diagnostics used here (rotation, multi-epoch radial velocities, and differential surface abundances measured against a turn-off reference sample observed in the same configuration) will be applied to the remaining clusters of our FLAMES programme, so that the trends established for \Cr\ can be tested across the range of ages, metallicities and densities that the open-cluster regime spans.

\begin{acknowledgements}
Based on observations collected at the European Southern Observatory
under ESO programmes 0109.D-0492(A) (PI: Rain) and 0114.D-0090(A)
(PI: Monaco).
L.M. gratefully acknowledges support from ANID--FONDECYT Regular
Project No.~1251809.
This work has made use of data from the ESA mission \textit{Gaia}
(\url{https://www.cosmos.esa.int/gaia}), processed by the
\textit{Gaia} Data Processing and Analysis Consortium (DPAC,
\url{https://www.cosmos.esa.int/web/gaia/dpac/consortium}). Funding
for the DPAC has been provided by national institutions, in particular
the institutions participating in the \textit{Gaia} Multilateral
Agreement.
This paper includes data collected by the TESS mission, obtained from
the MAST data archive at the Space Telescope Science Institute (STScI);
we use the PATHOS light curves of \citet{Nardiello_2019,Nardiello_2020}.
Funding for the TESS mission is provided by the NASA Explorer Program.
This research made use of
iSpec \citep{Blanco-Cuaresma2014},
SPECTRUM \citep{Gray1994},
MOOG \citep{Sneden1973},
PARSEC \citep{Bressan2012},
and Astropy \citep{AstropyCollaboration2022}.
\end{acknowledgements}
\bibliographystyle{bibtex/aa}
\bibliography{bibtex/biblio}

@ARTICLE{CantatGaudin2020,
  author  = {{Cantat-Gaudin}, T. and {Anders}, F. and {Castro-Ginard}, A. and et al.},
  title   = {Painting a portrait of the Galactic disc with its stellar clusters},
  journal = {\aap},
  year    = 2020,
  volume  = {640},
  pages   = {A1},
  doi     = {10.1051/0004-6361/202038192}
}

@ARTICLE{Bressan2012,
  author  = {{Bressan}, A. and {Marigo}, P. and {Girardi}, L. and et al.},
  title   = {PARSEC: stellar tracks and isochrones with the PAdova and TRieste Stellar Evolution Code},
  journal = {\mnras},
  year    = 2012,
  volume  = {427},
  pages   = {127},
  doi     = {10.1111/j.1365-2966.2012.21948.x}
}

@ARTICLE{Bragaglia2006,
  author  = {{Bragaglia}, A. and {Tosi}, M.},
  title   = {The Bologna Open Cluster Chemical Evolution Project: Midterm Results from the Photometric Sample},
  journal = {\aj},
  year    = 2006,
  volume  = {131},
  pages   = {1544},
  doi     = {10.1086/499537}
}

@ARTICLE{Carretta2005,
  author  = {{Carretta}, E. and {Bragaglia}, A. and {Gratton}, R.~G. and {Tosi}, M.},
  title   = {High-resolution spectroscopy of the old open cluster Collinder 261: abundances of iron and other elements},
  journal = {\aap},
  year    = 2005,
  volume  = {441},
  pages   = {131},
  doi     = {10.1051/0004-6361:20052875}
}

@ARTICLE{deSilva2009,
  author  = {{De Silva}, G.~M. and {Gibson}, B.~K. and {Lattanzio}, J. and {Asplund}, M.},
  journal = {\aap},
  year    = 2009,
  volume  = {500},
  pages   = {L25},
  doi     = {10.1051/0004-6361/200912279}
}

@ARTICLE{Gao2023,
  author  = {{Gao}, X.},
  title   = {Spatial structure and dynamical state of the old open cluster Collinder 261 based on a clustering method},
  journal = {\pasj},
  year    = 2023,
  volume  = {75},
  pages   = {82}
}

@ARTICLE{Pasquini2002,
  author  = {{Pasquini}, L. and {Avila}, G. and {Blecha}, A. and et al.},
  title   = {Installation and Commissioning of FLAMES, the VLT Multifibre Facility},
  journal = {The Messenger},
  year    = 2002,
  volume  = {110},
  pages   = {1}
}

@ARTICLE{AstropyCollaboration2022,
  author  = {{Astropy Collaboration} and {Price-Whelan}, A.~M. and {Lim}, P.~L. and et al.},
  title   = {The Astropy Project: Sustaining and Growing a Community-oriented Open-source Project and the Latest Major Release (v5.0) of the Core Package},
  journal = {\apj},
  year    = 2022,
  volume  = {935},
  pages   = {167},
  doi     = {10.3847/1538-4357/ac7c74}
}

@ARTICLE{Rain2024,
       author = {{Rain}, M.~J. and {Pera}, M.~S. and {Perren}, G.~I. and
                 {Benvenuto}, O.~G. and {Panei}, J.~A. and {De Vito}, M.~A. and
                 {Carraro}, G. and {Villanova}, S.},
        title = "{Binary origin of blue straggler stars in Galactic star clusters}",
      journal = {\aap},
         year = 2024,
        month = may,
       volume = {685},
        pages = {A33},
          eid = {A33},
          doi = {10.1051/0004-6361/202347499},
       adsurl = {https://ui.adsabs.harvard.edu/abs/2024A&A...685A..33R}
}

@ARTICLE{Borodina2021,
       author = {{Borodina}, O.~I. and {Carraro}, G. and {Seleznev}, A.~F. and
                 {Danilov}, V.~M.},
        title = "{Unresolved Multiple Stars and Galactic Clusters' Mass Estimates}",
      journal = {\apj},
         year = 2021,
        month = feb,
       volume = {908},
       number = {1},
        pages = {60},
          eid = {60},
          doi = {10.3847/1538-4357/abd562}
}

@ARTICLE{Malofeeva2023,
       author = {{Malofeeva}, A.~A. and {Mikhnevich}, V.~O. and {Carraro}, G. and
                 {Seleznev}, A.~F.},
        title = "{Unresolved Binaries and Multiples in the Intermediate Mass Range
                  in Open Clusters: Pleiades, Alpha Per, Praesepe, and NGC 1039}",
      journal = {\aj},
         year = 2023,
        month = feb,
       volume = {165},
       number = {2},
        pages = {45},
          eid = {45},
          doi = {10.3847/1538-3881/aca6f0},
       eprint = {2211.12745}
}

@ARTICLE{Sandage1953,
  author  = {{Sandage}, A.~R.},
  title   = {The color-magnitude diagram for the globular cluster M 3},
  journal = {\aj},
  year    = 1953,
  volume  = {58},
  pages   = {61},
  doi     = {10.1086/106822}
}

@ARTICLE{McCrea1964,
  author  = {{McCrea}, W.~H.},
  title   = {Extended main-sequence of some stellar clusters},
  journal = {\mnras},
  year    = 1964,
  volume  = {128},
  pages   = {147},
  doi     = {10.1093/mnras/128.2.147}
}

@ARTICLE{Hills1976,
  author  = {{Hills}, J.~G. and {Day}, C.~A.},
  title   = {Stellar Collisions in Globular Clusters},
  journal = {Astrophys.~Lett.},
  year    = 1976,
  volume  = {17},
  pages   = {87}
}

@ARTICLE{Leonard1989,
  author  = {{Leonard}, P.~J.~T.},
  title   = {Stellar collisions in globular clusters and the blue straggler problem},
  journal = {\aj},
  year    = 1989,
  volume  = {98},
  pages   = {217}
}

@ARTICLE{Bailyn1995,
  author  = {{Bailyn}, C.~D.},
  title   = {Blue Stragglers and Other Stellar Anomalies: Implications for the Dynamics of Globular Clusters},
  journal = {\araa},
  year    = 1995,
  volume  = {33},
  pages   = {133},
  doi     = {10.1146/annurev.aa.33.090195.001033}
}

@ARTICLE{Lombardi2002,
  author  = {{Lombardi}, J.~C., Jr. and {Warren}, J.~S. and {Rasio}, F.~A. and et al.},
  title   = {Stellar Collisions and the Interior Structure of Blue Stragglers},
  journal = {\apj},
  year    = 2002,
  volume  = {568},
  pages   = {939},
  doi     = {10.1086/339060}
}

@ARTICLE{Davies2004,
  author  = {{Davies}, M.~B. and {Piotto}, G. and {de Angeli}, F.},
  title   = {Blue straggler production in globular clusters},
  journal = {\mnras},
  year    = 2004,
  volume  = {349},
  pages   = {129},
  doi     = {10.1111/j.1365-2966.2004.07474.x}
}

@ARTICLE{Carney2005,
  author  = {{Carney}, B.~W. and {Latham}, D.~W. and {Laird}, J.~B.},
  title   = {Metal-poor Field Blue Stragglers: More Evidence for Mass Transfer},
  journal = {\aj},
  year    = 2005,
  volume  = {129},
  pages   = {466},
  doi     = {10.1086/426566}
}

@ARTICLE{Sills2005,
  author  = {{Sills}, A. and {Adams}, T. and {Davies}, M.~B.},
  title   = {Blue stragglers as stellar collision products: the angular momentum question},
  journal = {\mnras},
  year    = 2005,
  volume  = {358},
  pages   = {716},
  doi     = {10.1111/j.1365-2966.2005.08809.x}
}

@ARTICLE{Ferraro2006,
  author  = {{Ferraro}, F.~R. and {Sabbi}, E. and {Gratton}, R. and et al.},
  title   = {Discovery of Carbon/Oxygen-depleted Blue Straggler Stars in 47 Tucanae: The Chemical Signature of a Mass-Transfer Formation Process},
  journal = {\apjl},
  year    = 2006,
  volume  = {647},
  pages   = {L53},
  doi     = {10.1086/507327}
}

@ARTICLE{Glebbeek2008,
  author  = {{Glebbeek}, E. and {Pols}, O.~R. and {Hurley}, J.~R.},
  title   = {Evolution of stellar collision products in open clusters. I. Blue stragglers in N-body models of M 67},
  journal = {\aap},
  year    = 2008,
  volume  = {488},
  pages   = {1007},
  doi     = {10.1051/0004-6361:200809930}
}

@ARTICLE{Knigge2009,
  author  = {{Knigge}, C. and {Leigh}, N. and {Sills}, A.},
  title   = {A binary origin for `blue stragglers' in globular clusters},
  journal = {\nat},
  year    = 2009,
  volume  = {457},
  pages   = {288},
  doi     = {10.1038/nature07635}
}

@ARTICLE{Mathieu2009,
  author  = {{Mathieu}, R.~D. and {Geller}, A.~M.},
  title   = {A binary star fraction of 76 per cent and unusual orbit parameters for the blue stragglers of NGC 188},
  journal = {\nat},
  year    = 2009,
  volume  = {462},
  pages   = {1032},
  doi     = {10.1038/nature08568}
}

@ARTICLE{Perets2009,
  author  = {{Perets}, H.~B. and {Fabrycky}, D.~C.},
  title   = {On the Triple Origin of Blue Stragglers},
  journal = {\apj},
  year    = 2009,
  volume  = {697},
  pages   = {1048},
  doi     = {10.1088/0004-637X/697/2/1048}
}

@ARTICLE{Geller2012,
  author  = {{Geller}, A.~M. and {Mathieu}, R.~D.},
  title   = {WIYN Open Cluster Study. XLVIII. The Hard-binary Population of NGC 188},
  journal = {\aj},
  year    = 2012,
  volume  = {144},
  pages   = {54},
  doi     = {10.1088/0004-6256/144/2/54}
}

@ARTICLE{Ferraro2012,
  author  = {{Ferraro}, F.~R. and {Lanzoni}, B. and {Dalessandro}, E. and et al.},
  title   = {Dynamical age differences among coeval star clusters as revealed by blue stragglers},
  journal = {\nat},
  year    = 2012,
  volume  = {492},
  pages   = {393},
  doi     = {10.1038/nature11686}
}

@ARTICLE{Gosnell2014,
  author  = {{Gosnell}, N.~M. and {Mathieu}, R.~D. and {Geller}, A.~M. and et al.},
  title   = {Detection of White Dwarf Companions to Blue Stragglers in the Open Cluster NGC 188},
  journal = {\apjl},
  year    = 2014,
  volume  = {783},
  pages   = {L8},
  doi     = {10.1088/2041-8205/783/1/L8}
}

@ARTICLE{Moe2017,
       author = {{Moe}, Maxwell and {Di Stefano}, Rosanne},
        title = "{Mind Your Ps and Qs: The Interrelation between Period (P) and
                  Mass-ratio (Q) Distributions of Binary Stars}",
      journal = {\apjs},
         year = 2017,
        month = jun,
       volume = {230},
       number = {2},
        pages = {15},
          eid = {15},
          doi = {10.3847/1538-4365/aa6fb6}
}

@ARTICLE{Alessandrini2016,
  author  = {{Alessandrini}, E. and {Lanzoni}, B. and {Ferraro}, F.~R. and {Miocchi}, P. and {Vesperini}, E.},
  title   = {Investigating the Mass Segregation Process in Globular Clusters with Blue Straggler Stars: The Impact of Dark Remnants},
  journal = {\apj},
  year    = 2016,
  volume  = {833},
  pages   = {252},
  doi     = {10.3847/1538-4357/833/2/252}
}

@ARTICLE{Lanzoni2016,
  author  = {{Lanzoni}, B. and {Ferraro}, F.~R. and {Alessandrini}, E. and et al.},
  title   = {Refining the Dynamical Clock for Star Clusters},
  journal = {\apjl},
  year    = 2016,
  volume  = {833},
  pages   = {L29},
  doi     = {10.3847/2041-8213/833/2/L29}
}

@ARTICLE{Ferraro2018,
  author  = {{Ferraro}, F.~R. and {Lanzoni}, B. and {Raso}, S. and et al.},
  title   = {The Hubble Space Telescope UV Legacy Survey of Galactic Globular Clusters. XV. The Dynamical Clock: Reading Cluster Dynamical Evolution from the Segregation Level of Blue Straggler Stars},
  journal = {\apj},
  year    = 2018,
  volume  = {860},
  pages   = {36},
  doi     = {10.3847/1538-4357/aac01c}
}

@ARTICLE{Leiner2018,
  author  = {{Leiner}, E. and {Mathieu}, R.~D. and {Gosnell}, N.~M. and {Sills}, A.},
  title   = {Observations of Spin-down in Post-mass-transfer Stars and the Possibility for Blue Straggler Gyrochronology},
  journal = {\apjl},
  year    = 2018,
  volume  = {869},
  pages   = {L29},
  doi     = {10.3847/2041-8213/aaf4ed}
}

@ARTICLE{Leiner2019,
  author  = {{Leiner}, E. and {Mathieu}, R.~D. and {Vanderburg}, A. and {Gosnell}, N.~M. and {Smith}, J.~C.},
  title   = {Blue Lurkers: Hidden Blue Stragglers on the M67 Main Sequence Identified from Their Kepler/K2 Rotation Periods},
  journal = {\apj},
  year    = 2019,
  volume  = {881},
  pages   = {47},
  doi     = {10.3847/1538-4357/ab2bf8}
}

@ARTICLE{Ferraro2023,
  author  = {{Ferraro}, F.~R. and {Mucciarelli}, A. and {Lanzoni}, B. and et al.},
  title   = {Fast rotating blue stragglers prefer loose clusters},
  journal = {Nature Communications},
  year    = 2023,
  volume  = {14},
  pages   = {2584},
  doi     = {10.1038/s41467-023-38153-w}
}

@ARTICLE{Rao2023,
  author  = {{Rao}, K.~K. and {Vaidya}, K. and {Bhattacharya}, S. and {Agrawal}, P.},
  title   = {Determination of dynamical ages of open clusters through the A$^{+}$ parameter},
  journal = {\mnras},
  year    = 2023,
  volume  = {508},
  pages   = {4919},
  doi     = {10.1093/mnras/stab2894}
}

@ARTICLE{Nine2024,
  author  = {{Nine}, A.~C. and {Mathieu}, R.~D. and {Schuler}, S.~C. and {Milliman}, K.~E.},
  title   = {WIYN Open Cluster Study. XC. Barium Surface Abundances of Blue Straggler Stars in the Open Clusters NGC 7789 and M67},
  journal = {\apj},
  year    = 2024,
  volume  = {970},
  pages   = {187},
  doi     = {10.3847/1538-4357/ad534b}
}

@ARTICLE{Billi2025,
  author  = {{Billi}, A. and {Monaco}, L. and {Ferraro}, F.~R. and et al.},
  title   = {Fast-rotating blue straggler stars in the globular cluster NGC 1851},
  journal = {\aap},
  year    = 2026,
  volume  = {705},
  pages   = {A36},
  eprint  = {2511.10109},
  doi     = {10.1051/0004-6361/202557251}
}

@ARTICLE{Rain2020,
  author  = {{Rain}, M.~J. and {Carraro}, G. and {Ahumada}, J.~A. and et al.},
  title   = {A Study of the Blue Straggler Population of the Old Open Cluster Collinder 261},
  journal = {\aj},
  year    = 2020,
  volume  = {159},
  pages   = {59},
  doi     = {10.3847/1538-3881/ab5f0b}
}

@ARTICLE{Rain2021,
  author  = {{Rain}, M.~J. and {Ahumada}, J.~A. and {Carraro}, G.},
  title   = {A new, Gaia-based, catalogue of blue straggler stars in open clusters},
  journal = {\aap},
  year    = 2021,
  volume  = {650},
  pages   = {A67},
  doi     = {10.1051/0004-6361/202040072}
}

@ARTICLE{Lovisi2012,
  author  = {{Lovisi}, L. and {Mucciarelli}, A. and {Lanzoni}, B. and et al.},
  title   = {Chemical and Kinematical Properties of Blue Straggler Stars and Horizontal Branch Stars in NGC 6397},
  journal = {\apj},
  year    = 2012,
  volume  = {754},
  pages   = {91},
  doi     = {10.1088/0004-637X/754/2/91}
}

@ARTICLE{Iben1993,
  author  = {{Iben}, I., Jr. and {Livio}, M.},
  title   = {Common envelopes in binary star evolution},
  journal = {\pasp},
  year    = 1993,
  volume  = {105},
  pages   = {1373},
  doi     = {10.1086/133321}
}

@ARTICLE{Taam2000,
  author  = {{Taam}, R.~E. and {Sandquist}, E.~L.},
  title   = {Common Envelope Evolution of Massive Binary Stars},
  journal = {\araa},
  year    = 2000,
  volume  = {38},
  pages   = {113},
  doi     = {10.1146/annurev.astro.38.1.113}
}

@ARTICLE{Naoz2014,
  author  = {{Naoz}, S. and {Farr}, W.~M. and {Lithwick}, Y. and {Rasio}, F.~A. and {Teyssandier}, J.},
  title   = {Secular dynamics in hierarchical three-body systems},
  journal = {\mnras},
  year    = 2013,
  volume  = {431},
  pages   = {2155},
  doi     = {10.1093/mnras/stt302}
}

@ARTICLE{Tokovinin2014,
  author  = {{Tokovinin}, A.},
  title   = {From Binaries to Multiples. II. Hierarchical Multiplicity of F and G Dwarfs},
  journal = {\aj},
  year    = 2014,
  volume  = {147},
  pages   = {87},
  doi     = {10.1088/0004-6256/147/4/87}
}

@ARTICLE{Postnov2014,
  author  = {{Postnov}, K.~A. and {Yungelson}, L.~R.},
  title   = {The Evolution of Compact Binary Star Systems},
  journal = {Living Reviews in Relativity},
  year    = 2014,
  volume  = {17},
  pages   = {3},
  doi     = {10.12942/lrr-2014-3}
}

@ARTICLE{Sneden1973,
  author  = {{Sneden}, C.},
  title   = {The nitrogen abundance of the very metal-poor star HD 122563},
  journal = {\apj},
  year    = 1973,
  volume  = {184},
  pages   = {839},
  doi     = {10.1086/152374}
}

@ARTICLE{Gray1994,
  author  = {{Gray}, R.~O. and {Corbally}, C.~J.},
  title   = {The calibration of MK spectral classes using spectral synthesis. 1: The effective temperature calibration of dwarf stars},
  journal = {\aj},
  year    = 1994,
  volume  = {107},
  pages   = {742}
}

@BOOK{Gray2005,
  author    = {{Gray}, D.~F.},
  title     = {The Observation and Analysis of Stellar Photospheres},
  edition   = {3rd},
  publisher = {Cambridge University Press},
  address   = {Cambridge},
  year      = 2005
}

@ARTICLE{Arlandini1999,
  author  = {{Arlandini}, C. and {K{\"a}ppeler}, F. and {Wisshak}, K. and et al.},
  title   = {Neutron Capture in Low-Mass Asymptotic Giant Branch Stars: Cross Sections and Abundance Signatures},
  journal = {\apj},
  year    = 1999,
  volume  = {525},
  pages   = {886},
  doi     = {10.1086/307938}
}

@INPROCEEDINGS{Castelli2003,
  author    = {{Castelli}, F. and {Kurucz}, R.~L.},
  title     = {New Grids of ATLAS9 Model Atmospheres},
  booktitle = {Modelling of Stellar Atmospheres},
  series    = {IAU Symposium},
  volume    = {210},
  editor    = {{Piskunov}, N. and {Weiss}, W.~W. and {Gray}, D.~F.},
  publisher = {ASP},
  address   = {San Francisco, CA},
  year      = 2003,
  pages     = {A20}
}

@ARTICLE{Valenti2005,
  author  = {{Valenti}, J.~A. and {Fischer}, D.~A.},
  title   = {Spectroscopic Properties of Cool Stars (SPOCS). I. 1040 F, G, and K Dwarfs from Keck, Lick, and AAT Planet Search Programs},
  journal = {\apjs},
  year    = 2005,
  volume  = {159},
  pages   = {141},
  doi     = {10.1086/430500}
}

@ARTICLE{Stancliffe2007,
  author  = {{Stancliffe}, R.~J. and {Glebbeek}, E. and {Izzard}, R.~G. and {Pols}, O.~R.},
  title   = {Carbon-enhanced metal-poor stars and thermohaline mixing},
  journal = {\aap},
  year    = 2007,
  volume  = {464},
  pages   = {L57},
  doi     = {10.1051/0004-6361:20066891}
}

@ARTICLE{Gustafsson2008,
  author  = {{Gustafsson}, B. and {Edvardsson}, B. and {Eriksson}, K. and et al.},
  title   = {A grid of MARCS model atmospheres for late-type stars. I. Methods and general properties},
  journal = {\aap},
  year    = 2008,
  volume  = {486},
  pages   = {951},
  doi     = {10.1051/0004-6361:200809724}
}

@ARTICLE{Asplund2009,
  author  = {{Asplund}, M. and {Grevesse}, N. and {Sauval}, A.~J. and {Scott}, P.},
  title   = {The Chemical Composition of the Sun},
  journal = {\araa},
  year    = 2009,
  volume  = {47},
  pages   = {481},
  doi     = {10.1146/annurev.astro.46.060407.145222}
}

@ARTICLE{Gallagher2010,
  author  = {{Gallagher}, A.~J. and {Ryan}, S.~G. and {Garc{\'\i}a P{\'e}rez}, A.~E. and {Aoki}, W.},
  title   = {The barium isotopic mixture for the metal-poor subgiant star HD 140283},
  journal = {\aap},
  year    = 2010,
  volume  = {523},
  pages   = {A24},
  doi     = {10.1051/0004-6361/201014970}
}

@ARTICLE{Blanco-Cuaresma2014,
  author  = {{Blanco-Cuaresma}, S. and {Soubiran}, C. and {Heiter}, U. and {Jofr{\'e}}, P.},
  title   = {Determining stellar atmospheric parameters and chemical abundances of FGK stars with iSpec},
  journal = {\aap},
  year    = 2014,
  volume  = {569},
  pages   = {A111},
  doi     = {10.1051/0004-6361/201423945}
}

@ARTICLE{Karakas2014,
  author  = {{Karakas}, A.~I. and {Lattanzio}, J.~C.},
  title   = {The Dawes Review 2: Nucleosynthesis and Stellar Yields of Low- and Intermediate-Mass Single Stars},
  journal = {\pasa},
  year    = 2014,
  volume  = {31},
  pages   = {e030},
  doi     = {10.1017/pasa.2014.21}
}

@ARTICLE{Korotin2015,
  author  = {{Korotin}, S.~A. and {Andrievsky}, S.~M. and {Hansen}, C.~J. and et al.},
  title   = {Grid of theoretical NLTE equivalent widths of four Ba II lines and barium abundance in cool stars},
  journal = {\aap},
  year    = 2015,
  volume  = {581},
  pages   = {A70},
  doi     = {10.1051/0004-6361/201526558}
}

@ARTICLE{Amarsi2016,
  author  = {{Amarsi}, A.~M. and {Asplund}, M. and {Collet}, R. and {Leenaarts}, J.},
  title   = {Non-LTE oxygen line formation in 3D hydrodynamic model stellar atmospheres},
  journal = {\mnras},
  year    = 2016,
  volume  = {455},
  pages   = {3735},
  doi     = {10.1093/mnras/stv2608}
}

@ARTICLE{Amarsi2019,
  author  = {{Amarsi}, A.~M. and {Nissen}, P.~E. and {Skúladóttir}, {\'A}.},
  title   = {Carbon, oxygen, and iron abundances in disk and halo stars. Implications of 3D non-LTE spectral line formation},
  journal = {\aap},
  year    = 2019,
  volume  = {630},
  pages   = {A104},
  doi     = {10.1051/0004-6361/201936265}
}

@ARTICLE{Placco2021,
  author  = {{Placco}, V.~M. and {Sneden}, C. and {Roederer}, I.~U. and et al.},
  title   = {Linemake: An Atomic and Molecular Line List Generator},
  journal = {Research Notes of the American Astronomical Society},
  year    = 2021,
  volume  = {5},
  pages   = {92},
  doi     = {10.3847/2515-5172/abf651}
}

@ARTICLE{Nardiello_2020,
       author = {{Nardiello}, D. and {Piotto}, G. and {Deleuil}, M. and {Malavolta}, L. and {Montalto}, M. and {Bedin}, L.~R. and {Borsato}, L. and {Granata}, V. and {Libralato}, M. and {Manthopoulou}, E.~E.},
        title = "{A PSF-based Approach to TESS High quality data Of Stellar clusters (PATHOS) - II. Search for exoplanets in open clusters of the Southern ecliptic hemisphere and their frequency}",
      journal = {\mnras},
         year = 2020,
        month = jul,
       volume = {495},
       number = {4},
        pages = {4924-4942},
          doi = {10.1093/mnras/staa1465},
archivePrefix = {arXiv},
       eprint = {2005.12281},
 primaryClass = {astro-ph.EP},
       adsurl = {https://ui.adsabs.harvard.edu/abs/2020MNRAS.495.4924N}
}

@ARTICLE{Nardiello_2019,
       author = {{Nardiello}, D. and {Deleuil}, M. and {Mantovan}, G. and {Malavolta}, L. and {Lacedelli}, G. and {Libralato}, M. and {Bedin}, L.~R. and {Borsato}, L. and {Granata}, V. and {Piotto}, G.},
        title = "{A PSF-based Approach to TESS High quality data Of Stellar clusters (PATHOS) - IV. Candidate exoplanets around stars in open clusters: frequency and age-planetary radius distribution}",
      journal = {\mnras},
         year = 2021,
        month = aug,
       volume = {505},
       number = {3},
        pages = {3767-3784},
          doi = {10.1093/mnras/stab1497},
archivePrefix = {arXiv},
       eprint = {2105.09952},
 primaryClass = {astro-ph.EP},
       adsurl = {https://ui.adsabs.harvard.edu/abs/2021MNRAS.505.3767N}
}

@ARTICLE{Echeveste2026,
  author  = {{Echeveste}, M. and {Rain}, M.~J. and {Carraro}, G. and
             {Benvenuto}, O.~G. and {Panei}, J.~A. and {De Vito}, M.~A.},
  title   = "{Binary evolution models and short-period variability of the
              blue straggler population of Collinder 261}",
  journal = {\aap},
  year    = 2026,
  volume  = 712,
  pages   = {A126}
}

@ARTICLE{Zahn1977,
  author  = {{Zahn}, J.-P.},
  title   = "{Tidal friction in close binary stars}",
  journal = {Astronomy \& Astrophysics},
  year    = 1977, volume = 57, pages = {383-394},
  adsurl  = {https://ui.adsabs.harvard.edu/abs/1977A&A....57..383Z}
}

@ARTICLE{Hurley2002,
  author  = {{Hurley}, Jarrod R. and {Tout}, Christopher A. and {Pols}, Onno R.},
  title   = "{Evolution of binary stars and the effect of tides on binary populations}",
  journal = {MNRAS}, year = 2002, volume = 329, number = 4, pages = {897-928},
  doi     = {10.1046/j.1365-8711.2002.05038.x}
}

@ARTICLE{Preston1974,
  author  = {{Preston}, G.~W.},
  title   = "{The chemically peculiar stars of the upper main sequence}",
  journal = {ARA\&A}, year = 1974, volume = 12, pages = {257-277},
  doi     = {10.1146/annurev.aa.12.090174.001353}
}

@ARTICLE{Richer2000,
  author  = {{Richer}, J. and {Michaud}, G. and {Turcotte}, S.},
  title   = "{The Evolution of AmFm Stars, Abundance Anomalies, and Turbulent Transport}",
  journal = {ApJ}, year = 2000, volume = 529, number = 1, pages = {338-356},
  doi     = {10.1086/308274}
}

@ARTICLE{Linck2026a,
  author  = {{Linck}, E. and {Mathieu}, R.~D.},
  title   = {The Distribution of Blue Straggler Stars in the Color-Magnitude Diagrams of Old Open Clusters},
  journal = {\aj}, year = 2026, volume = 172, pages = {61},
  eprint  = {2605.14187}
}

@ARTICLE{Linck2026b,
  author  = {{Linck}, E. and {Mathieu}, R.~D.},
  title   = {Blue Straggler Stars in Old Open Clusters and the Kraft Break},
  journal = {\apjl}, year = 2026, volume = 1005,
  doi     = {10.3847/2041-8213/ae6e3d}, eprint = {2605.18995}
}

@ARTICLE{Eggleton1983,
  author  = {{Eggleton}, P.~P.},
  title   = {Approximations to the radii of Roche lobes},
  journal = {ApJ},
  year    = 1983,
  volume  = 268,
  pages   = {368-369},
  doi     = {10.1086/160960}
}

@ARTICLE{Leiner2021,
       author = {{Leiner}, Emily M. and {Geller}, Aaron},
        title = "{A Census of Blue Stragglers in Gaia DR2 Open Clusters as a Test of Population Synthesis and Mass Transfer Physics}",
      journal = {\apj},
         year = 2021,
        month = feb,
       volume = {908},
       number = {2},
          eid = {229},
        pages = {229},
          doi = {10.3847/1538-4357/abd7e9},
archivePrefix = {arXiv},
       eprint = {2101.11047},
 primaryClass = {astro-ph.SR},
       adsurl = {https://ui.adsabs.harvard.edu/abs/2021ApJ...908..229L}
}

@ARTICLE{Mazur1995,
       author = {{Mazur}, Beata and {Krzeminski}, Wojciech and {Kaluzny}, Janusz},
        title = "{Discovery of 45 short-period eclipsing binaries in the field of the old open cluster Collinder 261}",
      journal = {\mnras},
         year = 1995,
        month = mar,
       volume = {273},
       number = {1},
        pages = {59-71},
          doi = {10.1093/mnras/273.1.59},
       adsurl = {https://ui.adsabs.harvard.edu/abs/1995MNRAS.273...59M}
}

@ARTICLE{Gozzoli1996,
       author = {{Gozzoli}, E. and {Tosi}, M. and {Marconi}, G. and {Bragaglia}, A.},
        title = "{CCD photometry of the old open cluster Collinder 261.}",
      journal = {\mnras},
         year = 1996,
        month = nov,
       volume = {283},
       number = {1},
        pages = {66-76},
          doi = {10.1093/mnras/283.1.66},
archivePrefix = {arXiv},
       eprint = {astro-ph/9607081},
 primaryClass = {astro-ph},
       adsurl = {https://ui.adsabs.harvard.edu/abs/1996MNRAS.283...66G}
}

@ARTICLE{Friel2003,
       author = {{Friel}, Eileen D. and {Jacobson}, Heather R. and {Barrett}, Elizabeth and {Fullton}, Laura and {Balachandran}, Suchitra C. and {Pilachowski}, Catherine A.},
        title = "{Abundances of Red Giants in the Old Open Cluster Collinder 261}",
      journal = {\aj},
         year = 2003,
        month = nov,
       volume = {126},
       number = {5},
        pages = {2372-2384},
          doi = {10.1086/378600},
       adsurl = {https://ui.adsabs.harvard.edu/abs/2003AJ....126.2372F}
}

@ARTICLE{CantatGaudin2018,
       author = {{Cantat-Gaudin}, T. and {Jordi}, C. and {Vallenari}, A. and {Bragaglia}, A. and {Balaguer-N{\'u}{\~n}ez}, L. and {Soubiran}, C. and {Bossini}, D. and {Moitinho}, A. and {Castro-Ginard}, A. and {Krone-Martins}, A. and {Casamiquela}, L. and {Sordo}, R. and {Carrera}, R.},
        title = "{A Gaia DR2 view of the open cluster population in the Milky Way}",
      journal = {\aap},
         year = 2018,
        month = oct,
       volume = {618},
          eid = {A93},
        pages = {A93},
          doi = {10.1051/0004-6361/201833476},
archivePrefix = {arXiv},
       eprint = {1805.08726},
 primaryClass = {astro-ph.GA},
       adsurl = {https://ui.adsabs.harvard.edu/abs/2018A&A...618A..93C}
}

@ARTICLE{Schlegel1998,
       author = {{Schlegel}, David J. and {Finkbeiner}, Douglas P. and {Davis}, Marc},
        title = "{Maps of Dust Infrared Emission for Use in Estimation of Reddening and Cosmic Microwave Background Radiation Foregrounds}",
      journal = {\apj},
         year = 1998,
        month = jun,
       volume = {500},
       number = {2},
        pages = {525-553},
          doi = {10.1086/305772},
archivePrefix = {arXiv},
       eprint = {astro-ph/9710327},
 primaryClass = {astro-ph},
       adsurl = {https://ui.adsabs.harvard.edu/abs/1998ApJ...500..525S}
}

@ARTICLE{Green2018,
       author = {{Green}, Gregory M. and {Schlafly}, Edward F. and {Finkbeiner}, Douglas and {Rix}, Hans-Walter and {Martin}, Nicolas and {Burgett}, William and {Draper}, Peter W. and {Flewelling}, Heather and {Hodapp}, Klaus and {Kaiser}, Nicholas and {Kudritzki}, Rolf-Peter and {Magnier}, Eugene A. and {Metcalfe}, Nigel and {Tonry}, John L. and {Wainscoat}, Richard and {Waters}, Christopher},
        title = "{Galactic reddening in 3D from stellar photometry - an improved map}",
      journal = {\mnras},
         year = 2018,
        month = jul,
       volume = {478},
       number = {1},
        pages = {651-666},
          doi = {10.1093/mnras/sty1008},
archivePrefix = {arXiv},
       eprint = {1801.03555},
 primaryClass = {astro-ph.GA},
       adsurl = {https://ui.adsabs.harvard.edu/abs/2018MNRAS.478..651G}
}

@ARTICLE{Milone2012,
  author  = {{Milone}, A.~P. and {Piotto}, G. and {Bedin}, L.~R. and others},
  title   = {The ACS survey of Galactic globular clusters. XII. Photometric binaries along the main sequence},
  journal = {\aap},
  year    = 2012,
  volume  = 540,
  pages   = {A16},
  eprint  = {1111.0552}
}

@ARTICLE{MathieuPols2025,
  author  = {{Mathieu}, R.~D. and {Pols}, O.~R.},
  title   = {Blue Stragglers and Friends: Initial Evolutionary Pathways in Close Low-Mass Binaries},
  journal = {\araa},
  year    = 2025,
  volume  = 63,
  pages   = {467-512},
  doi     = {10.1146/annurev-astro-071221-054402}
}

@ARTICLE{Mucciarelli2021,
  author  = {{Mucciarelli}, A. and {Bellazzini}, M. and {Massari}, D.},
  title   = {Exploiting the Gaia EDR3 photometry to derive stellar temperatures},
  journal = {\aap},
  year    = 2021,
  volume  = 653,
  pages   = {A90},
  eprint  = {2106.03882}
}

\clearpage
\begin{appendix}

\clearpage
\onecolumn
\section{Chemical differentials against effective temperature}

\label{app:teff_trends}
\begin{figure*}[h]
\centering
\includegraphics[width=\textwidth]{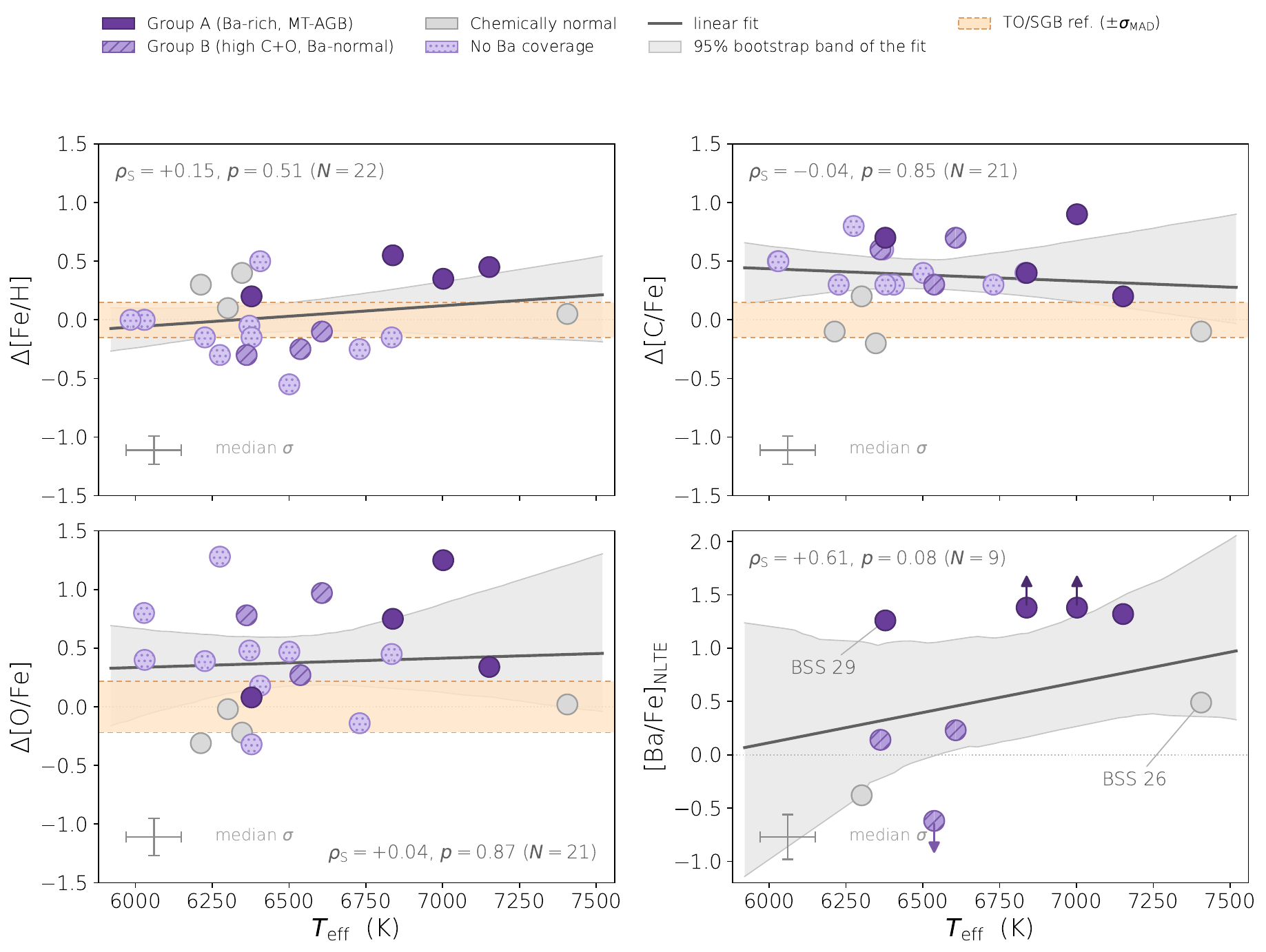}
\caption{Chemical differentials against $T_{\rm eff}$ for the 22 slow-rotating BSS, coloured by chemical group: $\Delta[\mathrm{Fe/H}]$ (top left), $\Delta[\mathrm{C/Fe}]$ (top right), $\Delta[\mathrm{O/Fe}]$ (bottom left), and $[\mathrm{Ba/Fe}]_{\rm NLTE}$ (bottom right). Open symbols with arrows are grid-edge limits, excluded from the fit. The orange band is the $\pm\sigma_{\rm MAD}$ scatter of the reference sample; the line and grey band are a least-squares fit and its 95\% bootstrap interval. Each panel gives the Spearman coefficient.}
\label{fig:teff_trends}
\end{figure*}

Figure~\ref{fig:teff_trends} supports the test of
Sect.~\ref{sec:teff_trends}, where we ask whether the chemical groups
of Sect.~\ref{sec:chem_groups} could be produced by atomic diffusion
and radiative levitation rather than by accretion. We collect it here
because the result is a null one, and the discussion in the main text
does not require the reader to inspect the panels.

The four panels show the three differentials
($\Delta[\mathrm{Fe/H}]$, $\Delta[\mathrm{C/Fe}]$,
$\Delta[\mathrm{O/Fe}]$) and the absolute
$[\mathrm{Ba/Fe}]_{\rm NLTE}$ against the track-based effective
temperature of Sect.~\ref{sec:stellar_params}, for the 22
slow-rotating BSS with abundance measurements. The three differential
panels share a common ordinate so that their amplitudes can be
compared directly, and each is drawn against the
$\pm\sigma_{\rm MAD}$ scatter of the TO/SGB reference sample, which
sets the systematic floor of the measurement
(Sect.~\ref{sec:chem_method}). Fe, C, and O are flat over the full
temperature range, which rules out the radiative-levitation signature
that \citet{Lovisi2012} detect in BSS hotter than 8200~K. The two
labelled stars break the apparent Ba trend: BSS~26 is the hottest star
of the sample and is only moderately Ba-enhanced, whereas BSS~29 is
Ba-rich at a temperature below the sample median.

\onecolumn

\section{Individual \texorpdfstring{$v\sin i$}{v sin i} fits}
\label{app:vsini_fits}
Figure~\ref{fig:vsini_gallery} presents the comparison between the
observed HR18 spectra and the best-fit rotationally broadened
synthetic profiles for the 38 BSS with a $v\sin i$ measurement,
sorted by increasing $v\sin i$, in the primary fitting window
(Fe\,\textsc{i} $\lambda\lambda$7495.05, 7498.53; dashed vertical
lines). The best-fit value or the corresponding upper limit is quoted
inside each panel, and the bottom sub-panels show the
observed$-$model residuals together with the $\pm1\sigma$ noise band.
The observed spectra are boxcar-smoothed (3 pixels) for display only;
the fits always use the unsmoothed data over the four spectral
windows described in Sect.~\ref{sec:vsini_method}. Because each panel
shows only the primary window, the quoted value can look larger than
the displayed line would suggest, in particular where a nearby
residual distorts the blue wing (e.g.\ BSS~29). We checked this
directly: refitting the displayed window alone returns lower central
values with much larger uncertainties (for BSS~29, $24\pm17$~\kms\
against the joint $36\pm11$~\kms), consistent within $1\sigma$ in
every case. The joint four-window fit uses four times the signal and
is the measurement we quote.
\begin{figure}[ht]
\centering
\includegraphics[width=\textwidth]{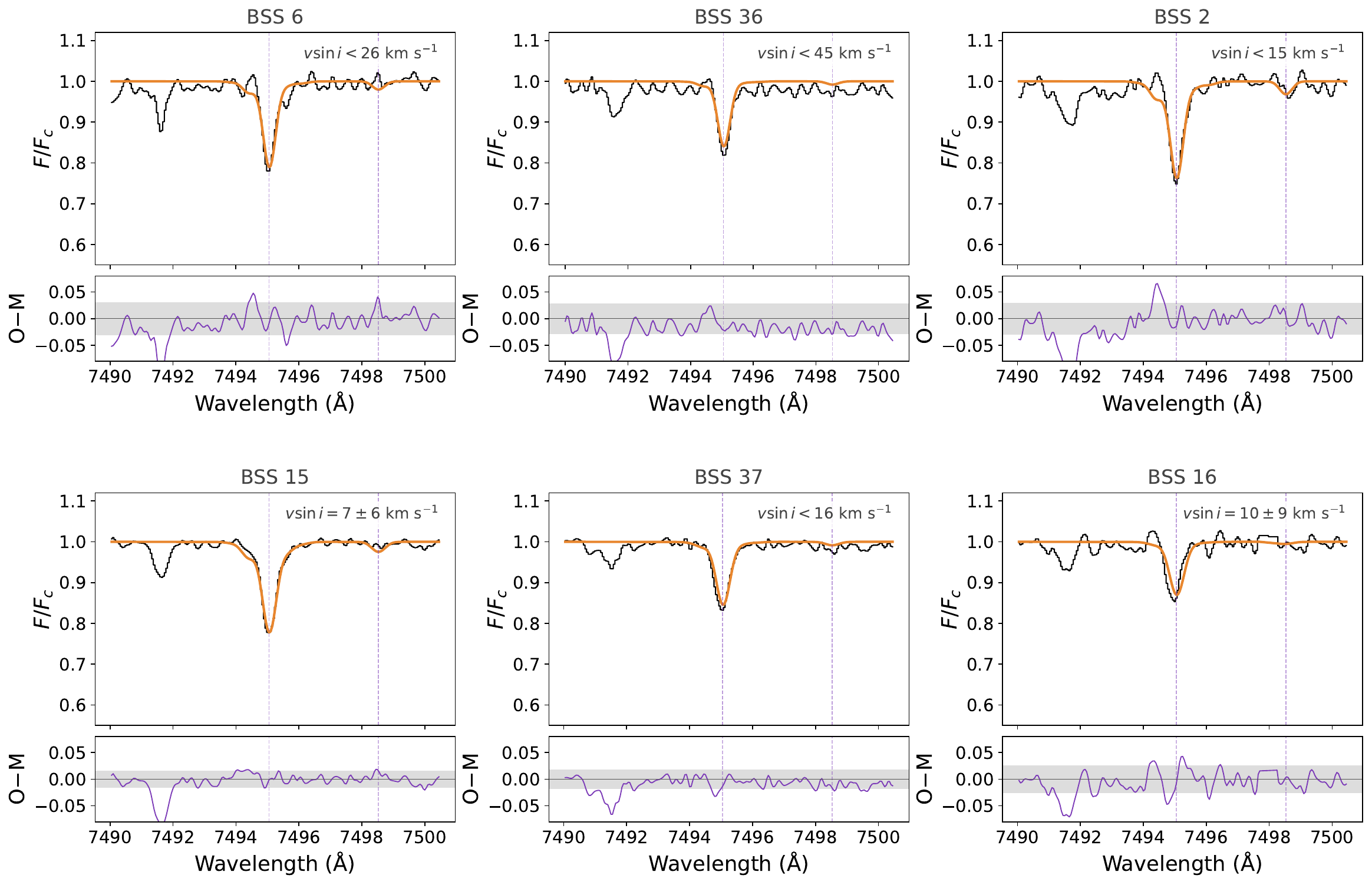}
\caption{Individual $v\sin i$ fits for the BSS, sorted by increasing $v\sin i$. Observed HR18 spectra (black) and best-fit synthetic profiles (orange); dashed lines mark Fe\,\textsc{i} $\lambda\lambda$7495.05, 7498.53. Bottom panels: residuals with the $\pm1\sigma$ noise band.}
\label{fig:vsini_gallery}
\end{figure}
\begin{figure*}[p]
\centering
\ContinuedFloat
\includegraphics[width=\textwidth]{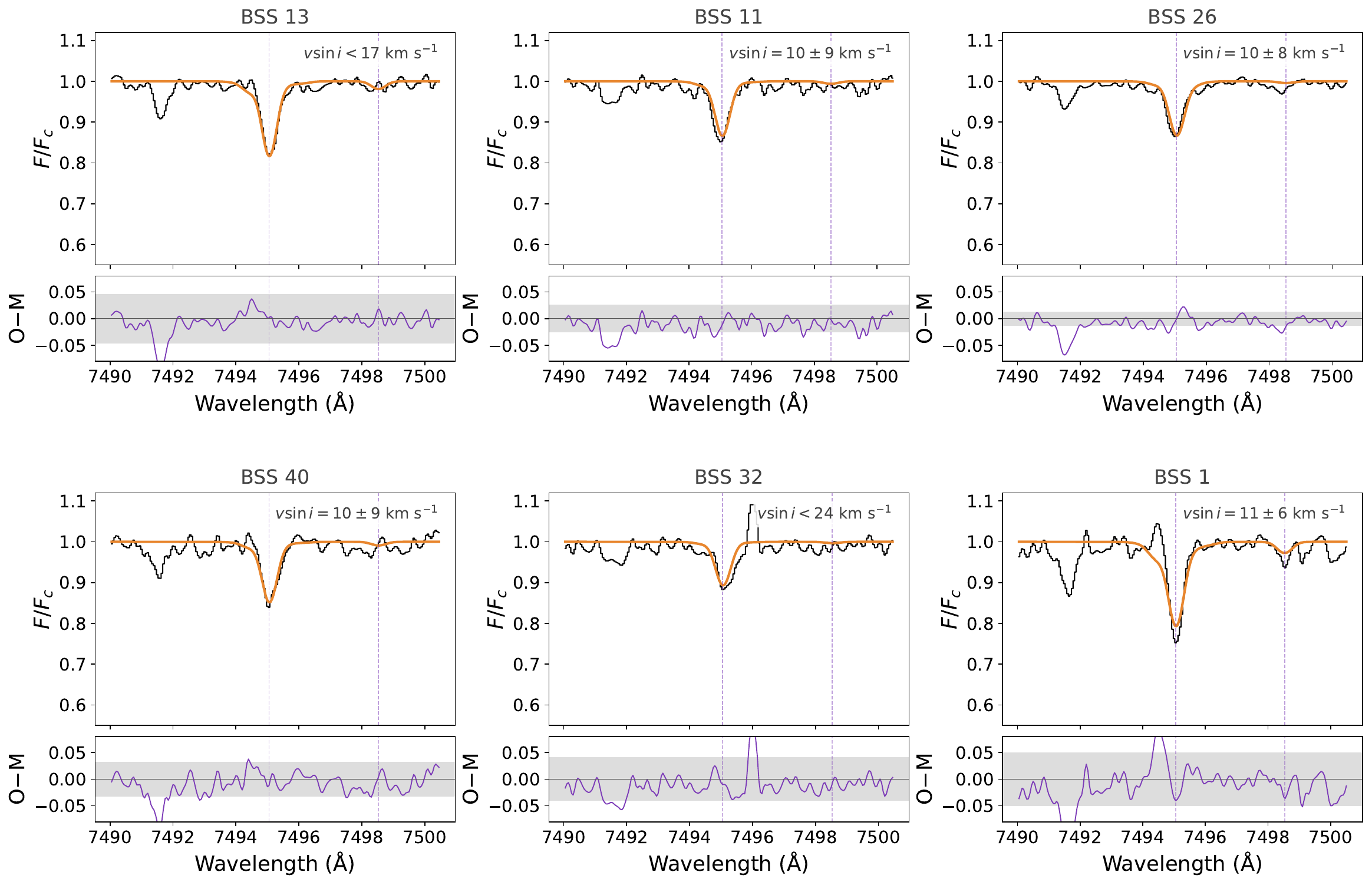}
\includegraphics[width=\textwidth]{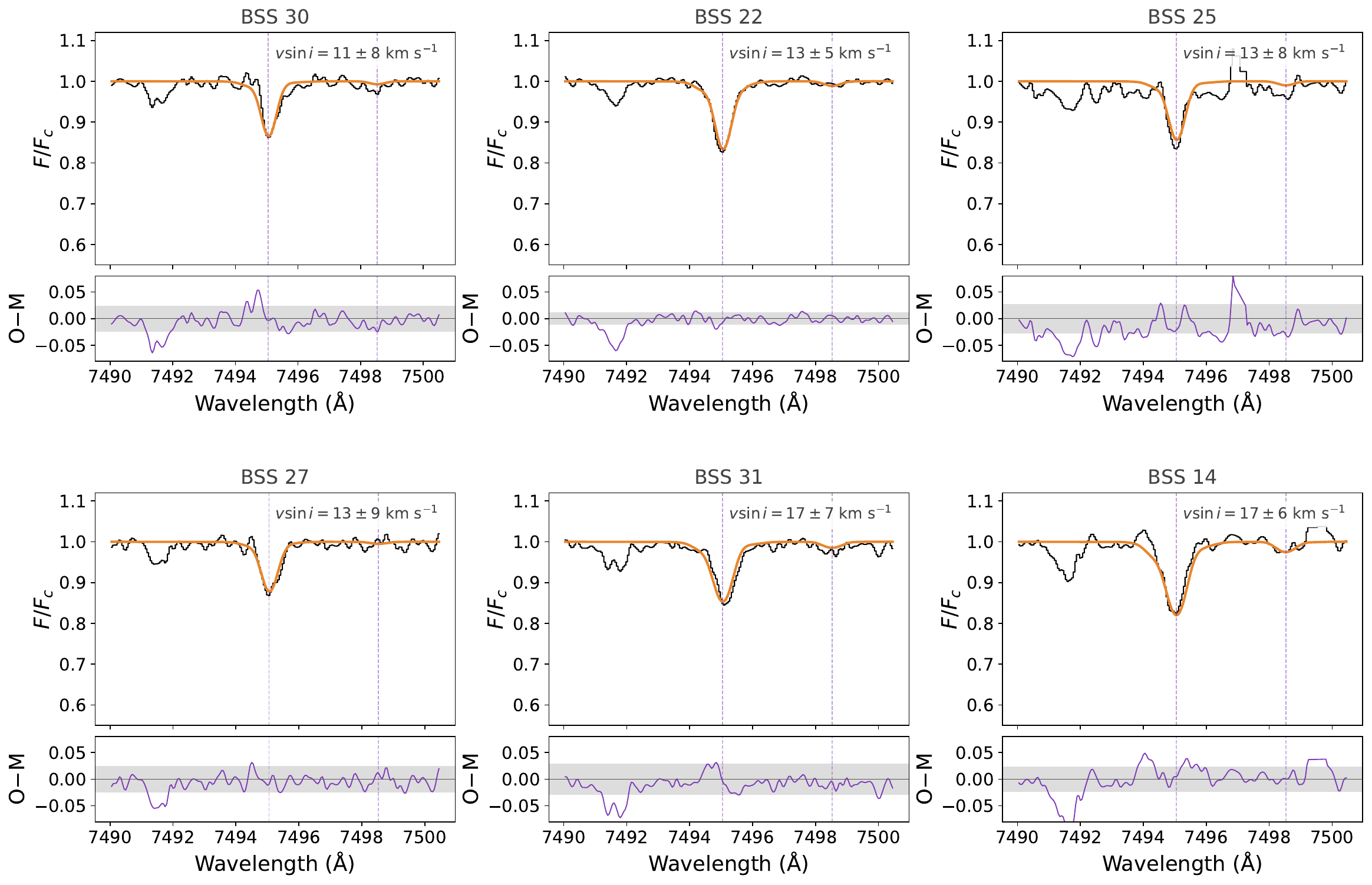}
\caption{Continued.}
\end{figure*}
\begin{figure*}[p]
\centering
\ContinuedFloat
\includegraphics[width=\textwidth]{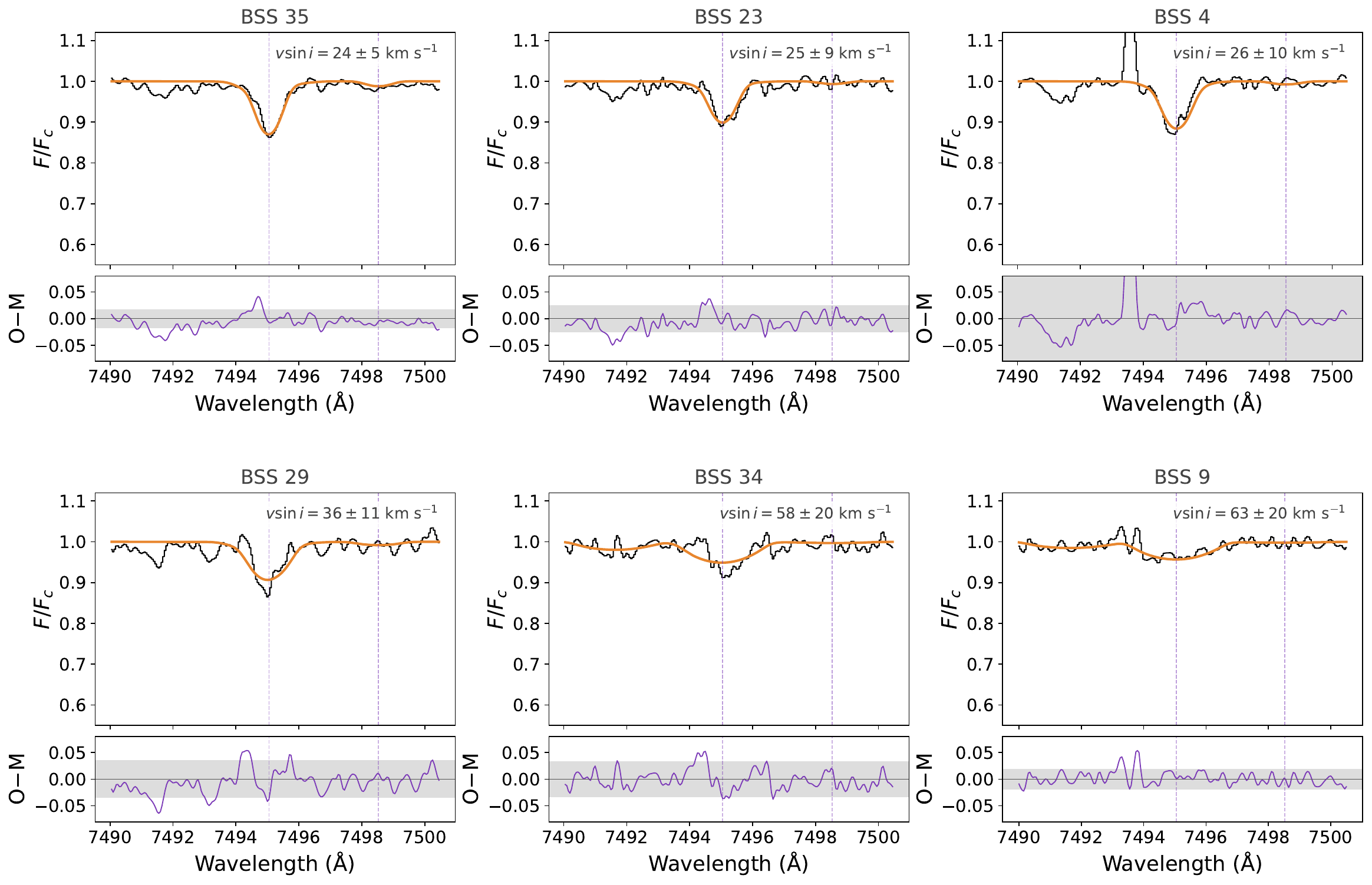}
\includegraphics[width=\textwidth]{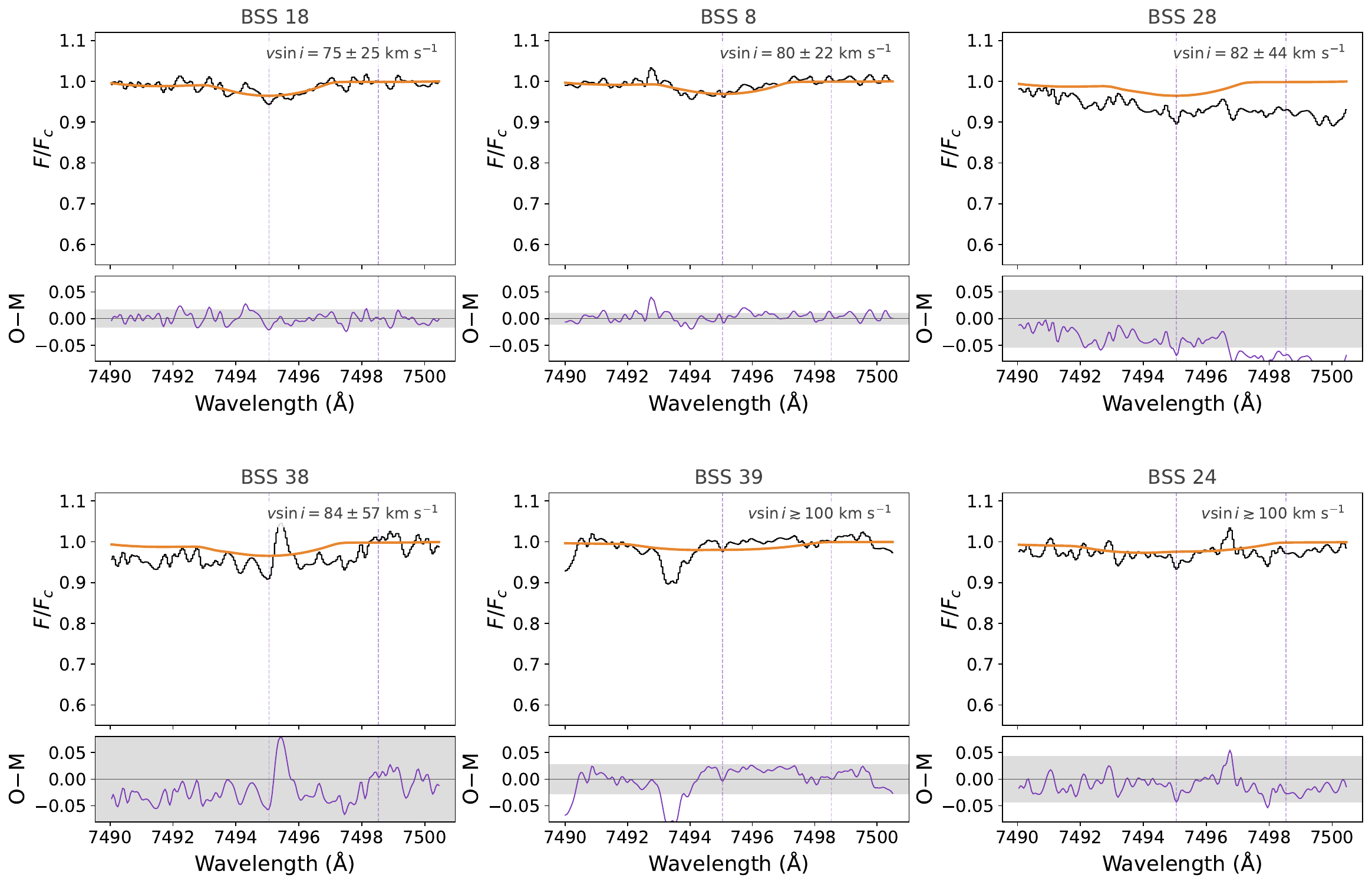}
\caption{Continued.}
\end{figure*}
\begin{figure*}[p]
\centering
\ContinuedFloat
\includegraphics[width=\textwidth]{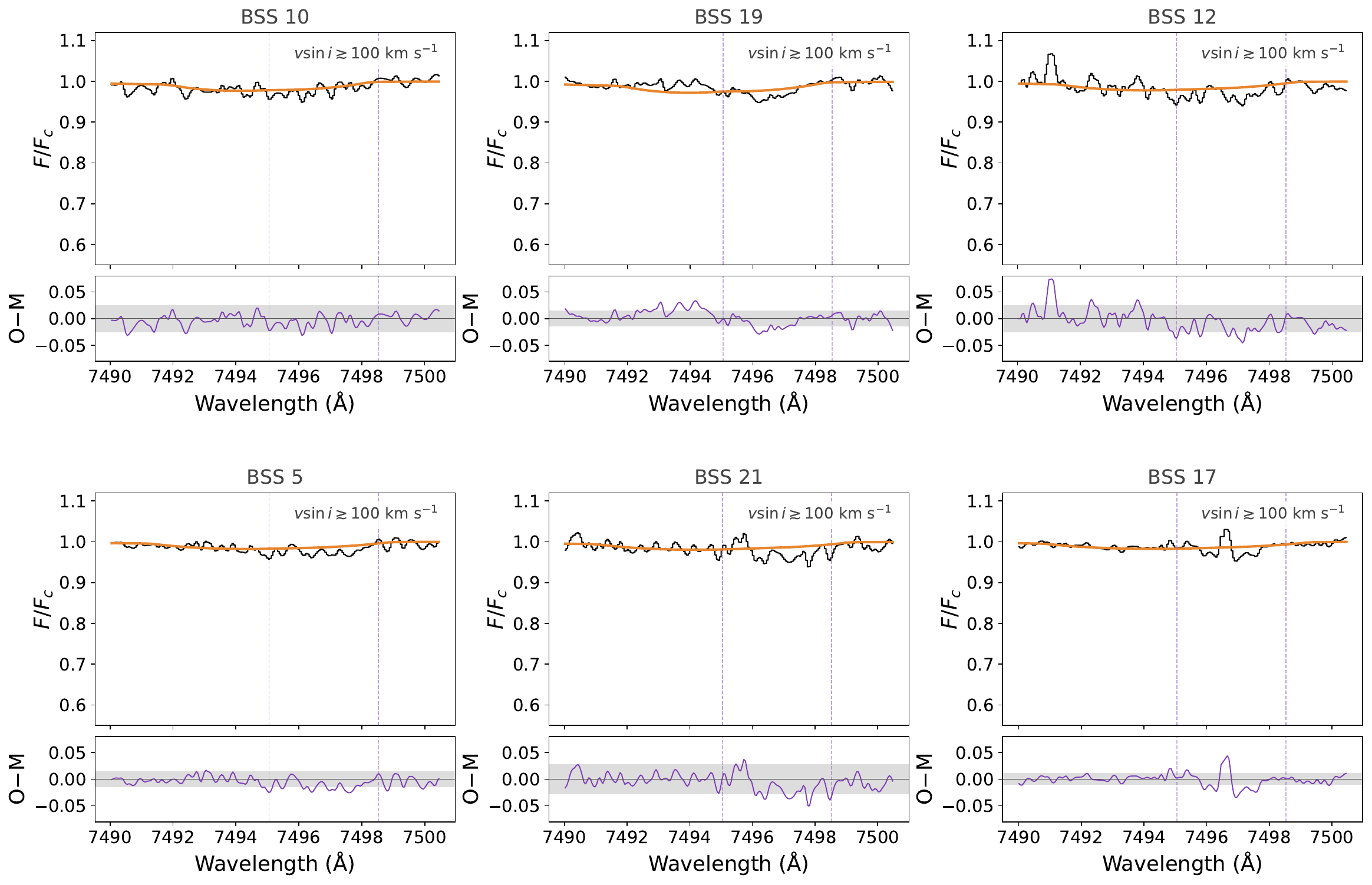}
\includegraphics[width=\textwidth]{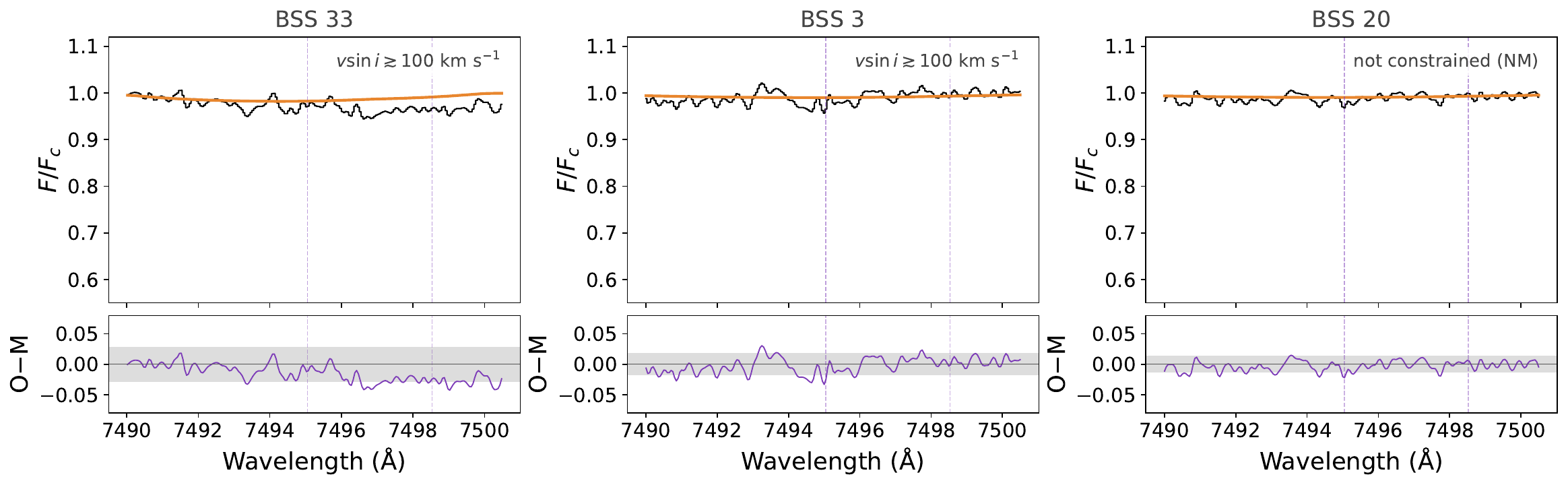}
\caption{Continued.}
\end{figure*}
\clearpage

\section{Radial-velocity statistics of the BSS sample}
\label{app:rvstats}
Table~\ref{tab:rvstats} lists the per-star radial-velocity statistics
underlying the binarity classification of Sect.~\ref{sec:sb1_rv}. For the
27 BSS with HR12 multi-epoch coverage (5--6 visits between 2022 April
and June, baseline $\simeq$63~d) we report the number of epochs, the
mean HR12 radial velocity, the standard deviation $\sigma_{\rm RV}$,
and the peak-to-peak amplitude ptp$_{\rm RV}$; for all stars we also
list the single-epoch HR18 offset $\Delta$RV measured in 2025~March,
which extends the effective baseline to $\sim$2.8~yr. Stars without
HR12 coverage rely on the HR18 criterion alone. The RV class encodes
the multi-epoch variability thresholds adopted in
Sect.~\ref{sec:sb1_rv} (constant: $\sigma_{\rm RV}<3$~\kms;
RV-variable: $3\leq\sigma_{\rm RV}<8$~\kms; SB1-candidate:
$\sigma_{\rm RV}\gtrsim8$~\kms\ and ptp$_{\rm RV}>20$~\kms;
SB1-confirmed: $\sigma_{\rm RV}>20$~\kms\ or ptp$_{\rm RV}>60$~\kms),
and the last column gives the SB1 flag adopted in
Table~\ref{tab:bss}, which follows only from these RV statistics (a star is SB1 if it is an HR12 SB1-candidate or SB1-confirmed, or if $|\Delta\mathrm{RV}_{\rm HR18}|>20$~\kms). The individual heliocentric velocities per epoch are available at the CDS (Table~C.2).
\begin{table*}[h]
\centering
\caption{Multi-epoch radial-velocity statistics of the BSS.}
\label{tab:rvstats}
\begin{tabular}{lccccccc}
\hline\hline
ID & $N_{\rm ep}$ & $\langle\mathrm{RV}\rangle_{\rm HR12}$ &
$\sigma_{\rm RV}$ & ptp$_{\rm RV}$ & $\Delta$RV$_{\rm HR18}$ &
RV class & SB1 \\
 & & (\kms) & (\kms) & (\kms) & (\kms) & & \\
\hline
BSS 1  & \dots & \dots & \dots & \dots & $-3$   & \dots         & N \\
BSS 2  & \dots & \dots & \dots & \dots & $+1$   & \dots         & N \\
BSS 3  & 5     & $-12.4$ & 15.0  & 33.6  & $-63$  & SB1-candidate & Y \\
BSS 4  & 5     & $-23.5$ & 2.1   & 5.2   & $+4$   & constant      & N \\
BSS 5  & \dots & \dots & \dots & \dots & $+16$  & \dots         & N \\
BSS 6  & 6     & $-27.2$ & 0.7   & 2.0   & $-3$   & constant      & N \\
BSS 7  & \dots & \dots & \dots & \dots & $+65$  & \dots         & Y$^{a}$ \\
BSS 8  & 5     & $-28.2$  & 30.9  & 60.9  & $-37$  & SB1-confirmed & Y \\
BSS 9  & 5     & $-5.9$  & 34.1  & 88.0  & $-49$  & SB1-confirmed & Y \\
BSS 10 & \dots & \dots & \dots & \dots & $+15$  & \dots         & N \\
BSS 11 & 5     & $-25.9$ & 2.6   & 6.2   & $-3$   & constant      & N \\
BSS 12 & 5     & $-22.4$ & 6.6   & 16.6  & $-1$   & RV-variable   & N \\
BSS 13 & \dots & \dots & \dots & \dots & $+2$   & \dots         & N \\
BSS 14 & \dots & \dots & \dots & \dots & $+7$   & \dots         & N \\
BSS 15 & 5     & $-23.3$ & 0.4   & 1.2   & $+1$   & constant      & N \\
BSS 16 & \dots & \dots & \dots & \dots & $+5$   & \dots         & N \\
BSS 17 & \dots & \dots & \dots & \dots & $+99$  & \dots         & Y$^{a}$ \\
BSS 18 & 6     & $-9.3$  & 18.1  & 42.3  & $+59$  & SB1-candidate & Y \\
BSS 19 & 6     & $-15.1$ & 16.0  & 38.4  & $+31$  & SB1-candidate & Y \\
BSS 20 & 5     & $-9.9$  & 14.6  & 30.2  & $+52$  & SB1-candidate & Y \\
BSS 21 & 5     & $+15.8$ & 13.1  & 33.0  & $+95$  & SB1-candidate & Y \\
BSS 22 & 5     & $-25.7$ & 4.0   & 10.4  & $+1$   & RV-variable   & N \\
BSS 23 & \dots & \dots & \dots & \dots & $+3$   & \dots         & N \\
BSS 24 & 6     & $+16.8$ & 48.2  & 135.8 & $-34$  & SB1-confirmed & Y \\
BSS 25 & 6     & $-24.9$ & 0.4   & 1.2   & $-3$   & constant      & N \\
BSS 26 & 5     & $-27.5$ & 3.0   & 7.3   & $0$    & RV-variable   & N \\
BSS 27 & 6     & $-25.2$ & 2.6   & 7.4   & $+2$   & constant      & N \\
BSS 28 & \dots & \dots & \dots & \dots & $+165$ & \dots         & Y$^{a}$ \\
BSS 29 & 5     & $-24.9$ & 3.6   & 8.8   & $+4$   & RV-variable   & N \\
BSS 30 & 5     & $-23.8$ & 1.1   & 2.4   & $-2$   & constant      & N \\
BSS 31 & \dots & \dots & \dots & \dots & $+2$   & \dots         & N \\
BSS 32 & 5     & $-29.7$ & 9.9   & 26.3  & $+1$   & SB1-candidate & Y \\
BSS 33 & 5     & $+5.6$  & 10.9  & 26.1  & $+175$ & SB1-candidate & Y \\
BSS 34 & 5     & $-22.4$ & 9.3   & 24.5  & $-1$   & SB1-candidate & Y \\
BSS 35 & \dots & \dots & \dots & \dots & $+1$   & \dots         & N \\
BSS 36 & 5     & $-23.5$ & 0.9   & 2.1   & $+7$   & constant      & N \\
BSS 37 & 5     & $-21.5$ & 1.7   & 4.6   & $-1$   & constant      & N \\
BSS 38 & 6     & $+5.9$  & 17.2  & 50.7  & $+16$  & SB1-candidate & Y \\
BSS 39 & 5     & $-20.9$ & 1.2   & 2.7   & $-109$ & constant      & N$^{b}$ \\
BSS 40 & 6     & $-26.8$ & 2.5   & 6.7   & $+2$   & constant      & N \\
\hline
\end{tabular}
\tablefoot{The RV classes follow the thresholds given in the text.
\tablefoottext{a}{SB1 from the HR18 single-epoch criterion only (no HR12 coverage).}
\tablefoottext{b}{HR18 $\Delta$RV $=-109$~\kms\ discarded: the HR12 epochs are constant at the $\sim$1~\kms\ level (Sect.~\ref{sec:sb1_rv}).}
\tablefoottext{c}{Velocities measured with a template broadened to the $v\sin i$ of the star (Sect.~\ref{sec:tess}).}}
\end{table*}
\clearpage

\section{The turn-off/subgiant reference sample}
\label{app:msto_table}
\begin{figure}[h]
\centering
\includegraphics[width=0.55\textwidth]{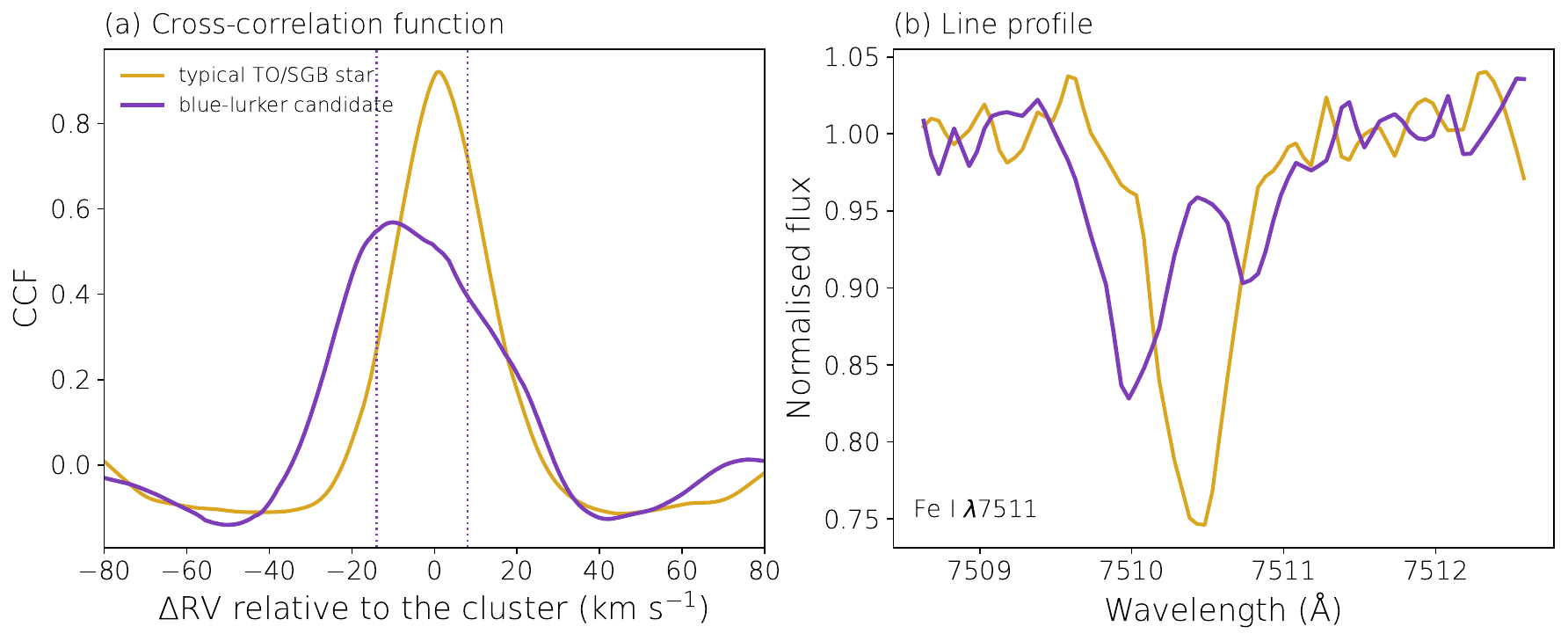}
\caption{Reference star \textit{Gaia}\,DR3 5856530891339792000 (purple) compared with a typical TO/SGB star (gold). (a) Cross-correlation function of the HR18 spectrum, with velocities relative to the cluster. (b) The Fe\,\textsc{i}~$\lambda$7511 line. The broad, asymmetric CCF and the split line show that the star is a double-lined binary.}
\label{fig:bl_sb2}
\end{figure}

Table~\ref{tab:msto} lists the adopted atmospheric parameters and
projected rotational velocities of the 42 TO/SGB reference stars used
throughout this work, derived with the same CMD projection onto
PARSEC evolutionary tracks applied to the BSS
(Sect.~\ref{sec:stellar_params}). Figure~\ref{fig:bl_sb2} shows the spectrum of the only reference star with a $v\sin i$ detection (Sect.~\ref{sec:bss_msto}).
\begin{table*}[h]
\centering
\caption{Stellar parameters and projected rotational velocities of the TO/SGB reference stars.}
\label{tab:msto}
\footnotesize
\begin{tabular}{lcccccl}
\hline\hline
Gaia DR3 ID & $T_\mathrm{eff}$ & $\log g$ & $M$ & $v\sin i$ & $G$ & Class \\
 & (K) & & (\msun) & (\kms) & (mag) & \\
\hline
5856437398493024640 & $5957\pm68$ & $3.92\pm0.03$ & $1.09\pm0.01$ & $<5$ & 16.07 & UL \\
5856438188766988672 & $5625\pm59$ & $3.82\pm0.02$ & $1.09\pm0.01$ & $<6$ & 16.10 & UL \\
5856438291846229632 & $5558\pm57$ & $3.82\pm0.02$ & $1.08\pm0.01$ & $<4$ & 16.17 & UL \\
5856510447294208256 & $5917\pm56$ & $3.94\pm0.03$ & $1.06\pm0.01$ & $<7$ & 16.23 & UL \\
5856510447294216704 & $5917\pm66$ & $3.94\pm0.02$ & $1.06\pm0.01$ & $<5$ & 16.23 & UL \\
5856512543238354944 & $5544\pm61$ & $3.83\pm0.02$ & $1.07\pm0.01$ & $<6$ & 16.23 & UL \\
5856515395096754560 & $5785\pm66$ & $3.89\pm0.02$ & $1.06\pm0.01$ & $<6$ & 16.23 & UL \\
5856515807413559808 & $5544\pm63$ & $3.83\pm0.02$ & $1.07\pm0.01$ & $<3$ & 16.18 & UL \\
5856516013572020352 & $5785\pm66$ & $3.89\pm0.02$ & $1.06\pm0.01$ & $<7$ & 16.21 & UL \\
5856523744513624832 & $5658\pm59$ & $3.85\pm0.02$ & $1.07\pm0.01$ & $<8$ & 16.14 & UL \\
5856526768171301760 & $5658\pm63$ & $3.85\pm0.02$ & $1.07\pm0.01$ & $<8$ & 16.19 & UL \\
5856527558444953472 & $5937\pm20$ & $3.93\pm0.01$ & $1.07\pm0.01$ & $<4$ & 16.11 & UL \\
5856527627164428928 & $5511\pm80$ & $3.80\pm0.02$ & $1.10\pm0.01$ & $<5$ & 16.14 & UL \\
5856527627164435584 & $5937\pm76$ & $3.93\pm0.03$ & $1.07\pm0.01$ & $<4$ & 16.11 & UL \\
5856527661524190720 & $5785\pm63$ & $3.89\pm0.02$ & $1.06\pm0.01$ & $<5$ & 16.24 & UL \\
5856527691559243648 & $5544\pm57$ & $3.83\pm0.02$ & $1.07\pm0.01$ & $<4$ & 16.22 & UL \\
5856527695883932288 & $5785\pm66$ & $3.89\pm0.03$ & $1.06\pm0.01$ & $<7$ & 16.18 & UL \\
5856528035184266240 & $5658\pm63$ & $3.85\pm0.02$ & $1.07\pm0.01$ & $<5$ & 16.15 & UL \\
5856528275672955264 & $5544\pm56$ & $3.83\pm0.02$ & $1.07\pm0.01$ & $<5$ & 16.21 & UL \\
5856528348679920384 & $5957\pm69$ & $3.92\pm0.03$ & $1.09\pm0.01$ & $<10$ & 16.11 & UL \\
5856528417438843008 & $5804\pm76$ & $3.88\pm0.03$ & $1.08\pm0.01$ & $<8$ & 16.10 & UL \\
5856528486157924864 & $5544\pm57$ & $3.83\pm0.02$ & $1.07\pm0.01$ & $<6$ & 16.20 & UL \\
5856528520517675520 & $5785\pm66$ & $3.89\pm0.02$ & $1.06\pm0.01$ & $<5$ & 16.24 & UL \\
5856528623597266176 & $5821\pm57$ & $3.87\pm0.02$ & $1.09\pm0.01$ & $<13$ & 16.06 & UL \\
5856528623597414016 & $5917\pm66$ & $3.94\pm0.02$ & $1.06\pm0.01$ & $<41$ & 16.22 & UL \\
5856528657956641024 & $5785\pm129$ & $3.89\pm0.04$ & $1.06\pm0.01$ & $<5$ & 16.20 & UL \\
5856528726676523648 & $6076\pm60$ & $3.98\pm0.03$ & $1.08\pm0.01$ & $<5$ & 16.07 & UL \\
5856528829755357824 & $5658\pm121$ & $3.85\pm0.03$ & $1.07\pm0.01$ & $<6$ & 16.23 & UL \\
5856529203385974656 & $5917\pm56$ & $3.94\pm0.03$ & $1.06\pm0.01$ & $<4$ & 16.21 & UL \\
5856529242072201728 & $5544\pm61$ & $3.83\pm0.02$ & $1.07\pm0.01$ & $<4$ & 16.22 & UL \\
5856529997986467712 & $5917\pm68$ & $3.94\pm0.03$ & $1.06\pm0.01$ & $<6$ & 16.18 & UL \\
5856529997986476800 & $6052\pm79$ & $3.99\pm0.03$ & $1.06\pm0.01$ & $<5$ & 16.19 & UL \\
5856530066705954944 & $5707\pm64$ & $3.84\pm0.02$ & $1.09\pm0.01$ & $<4$ & 16.09 & UL \\
5856530375943652864 & $5707\pm64$ & $3.84\pm0.02$ & $1.09\pm0.01$ & $<6$ & 16.09 & UL \\
5856530616462211968 & $5957\pm60$ & $3.92\pm0.03$ & $1.09\pm0.01$ & $<6$ & 16.02 & UL \\
5856530891339792000 & $6052\pm79$ & $3.99\pm0.03$ & $1.06\pm0.01$ & $26\pm24$ & 16.17 & D \\
5856531338016611840 & $5658\pm63$ & $3.85\pm0.02$ & $1.07\pm0.01$ & $<8$ & 16.16 & UL \\
5856531647254279296 & $6030\pm57$ & $4.00\pm0.02$ & $1.05\pm0.01$ & $<7$ & 16.24 & UL \\
5856531681614034688 & $5785\pm63$ & $3.89\pm0.02$ & $1.06\pm0.01$ & $<6$ & 16.17 & UL \\
5856531819052637184 & $5917\pm66$ & $3.94\pm0.02$ & $1.06\pm0.01$ & $<5$ & 16.23 & UL \\
5856531956491962496 & $5658\pm57$ & $3.85\pm0.01$ & $1.07\pm0.01$ & $<7$ & 16.20 & UL \\
5856532368808782464 & $5917\pm66$ & $3.94\pm0.03$ & $1.06\pm0.01$ & $<6$ & 16.17 & UL \\
\hline
\end{tabular}
\tablefoot{$T_{\rm eff}$, $\log g$ and mass from the same CMD projection onto PARSEC tracks used for the BSS (Table~\ref{tab:bss}). $v\sin i$ as in Sect.~\ref{sec:vsini_method}; UL = $1\sigma$ upper limit; D = detection ($\Delta\chi^2=1$).}
\end{table*}

\clearpage
\section{Differential reddening}
\label{app:diffred}
We tested the effect of differential reddening with two independent
methods. First, we used the dust map of \citet{Schlegel1998} through
the \texttt{dustmaps} package \citep{Green2018}. The three-dimensional
maps of the package do not cover this direction, so this map gives
the total dust column along the line of sight, with a resolution of
about 6\arcmin. To keep only the differences across the field, we
assumed that the dust in front of the cluster is a constant fraction
of the total, and we scaled each value to the mean $A_V=0.93$.
Second, we built an empirical map following \citet{Milone2012}. We
selected 1887 astrometric members from \textit{Gaia}\,DR3 and used
about 1340 main-sequence stars with $16.4<G<18.0$ as reference, after
removing the binaries. For each star we measured the shift along the
reddening vector ($A_G=0.83\,A_V$, $E(G_{\rm BP}-G_{\rm RP})=0.43\,A_V$)
between its position and the fiducial line of the main sequence. The
correction for each target is the median shift of its 40 nearest
reference stars, and we iterated the procedure twice.

\begin{figure*}[h]
\centering
\includegraphics[width=0.9\textwidth]{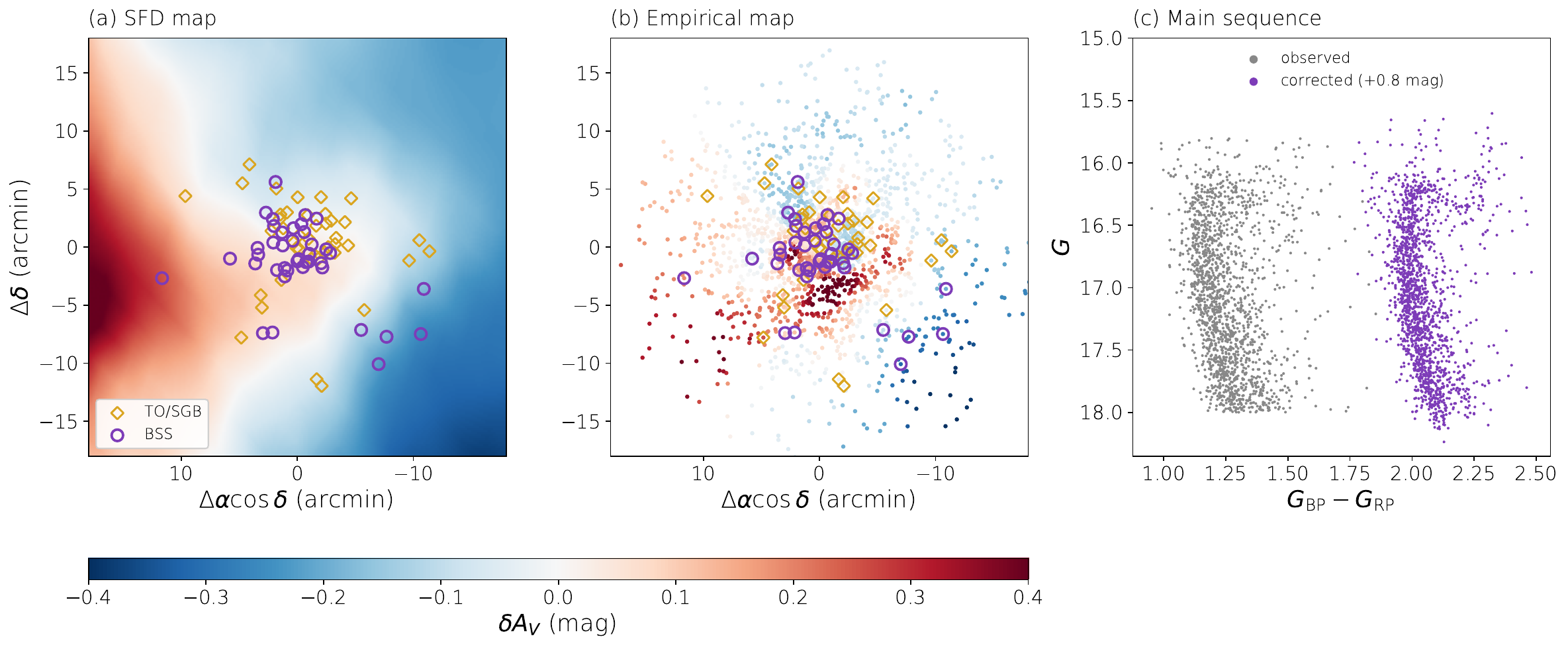}
\caption{Differential reddening in \Cr. (a) SFD map \citep{Schlegel1998}, as $\delta A_V$ relative to the mean $A_V=0.93$. (b) Empirical map from \textit{Gaia}\,DR3 main-sequence members \citep{Milone2012}. Open circles are the BSS and diamonds the TO/SGB stars; east is to the left. (c) Main sequence before (grey) and after (purple, shifted by $+0.8$~mag) the correction.}
\label{fig:diffred}
\end{figure*}

Figure~\ref{fig:diffred} shows the result. The correction is real:
the colour spread of the main sequence goes down from 0.22 to 0.13~mag
in units of $A_V$, whereas it does not change if we repeat the procedure
after shuffling the positions of the stars. The map has a scatter of
0.16~mag in $A_V$. It follows the same pattern as the SFD map
(Spearman $\rho=0.57$), with a larger amplitude, as expected for a
map with better resolution. The noise of the map is about 0.04~mag.
In the centre of the cluster the 40 neighbours lie within about
1\arcmin\ of each BSS. At the edges of the field there are fewer
reference stars, the radius grows to 3--5\arcmin\ (BSS~1, 2, 3, 4, 7
and 16), and the correction is less precise. For these stars the two
maps give the same sign and a similar size.

For the BSS the correction goes from $-0.31$ to $+0.21$~mag in $A_V$.
We converted the colour shift to temperature with the relation of
\citet{Mucciarelli2021}. The median change is 260~K, and 11 BSS
change by more than 300~K, most of them at the edges of the field.
For the TO/SGB reference stars the scatter is 120~K.

The effect on our results is small. The rotational velocities do not
depend on changes of this size (Sect.~\ref{sec:vsini_method}), so the
$v\sin i$ values and the fast/slow classification do not change. The
abundances change by 0.10--0.13~dex per 100~K
(Sect.~\ref{sec:chem_method}). For the slow rotators the temperature
changes are 10--450~K, with a median of 190~K, so the abundances move
by about 0.2~dex for a typical star and up to 0.5~dex at the edges of
the field. This is smaller than the 1~dex gap in [Ba/Fe] between
Groups~A and B, and smaller than the C and O excesses that define the
groups. A few stars close to the limits of the classification, mainly
in the (B) and marginal classes, could change class, but the two
groups and the gap between them do not depend on this correction.
The correction also explains part of the iron excesses of
Sect.~\ref{sec:feh}: for BSS~1, 2, 4 and 22 the corrected colour gives
a temperature 280--450~K lower, which brings their $\Delta$[Fe/H]
close to zero.

\end{appendix}
\end{document}